\documentclass[aps,prx,amsmath,floats,floatfix,twocolumn,
  superscriptaddress,nofootinbib,showpacs]{revtex4-1}

\pdfoutput=1 % Specifiy PDFLaTeX as per arXiv request. This has to be in the first 5 lines!
\usepackage{diagbox}
\usepackage{multirow}
\usepackage{braket}
\usepackage{amsfonts}
\usepackage{amsmath}
\usepackage{amssymb}
\usepackage{amsthm}
\usepackage{bm}
\usepackage{dcolumn}
\usepackage{epsfig}
\usepackage{graphicx}
\usepackage{graphics}
\usepackage[latin1]{inputenc}
\usepackage{latexsym}
\usepackage{rotating}
\usepackage[dvipsnames]{xcolor}
\usepackage{mathrsfs}
\usepackage{microtype}
\usepackage{verbatim}
\usepackage{url}

\usepackage[caption=false]{subfig}
\usepackage[normalem]{ulem}

\usepackage[breaklinks=true]{hyperref}
\hypersetup{
    colorlinks=true,
    linkcolor=NavyBlue,
    filecolor=Magenta,
    urlcolor=NavyBlue,
    citecolor=NavyBlue
}

\usepackage{yfonts}
\usepackage{float}
\usepackage{xspace} % Sensible space treatment at end of simple macros
\usepackage{mathrsfs}
\usepackage[toc,page]{appendix}
\usepackage{siunitx}
\usepackage{array}
\usepackage[normalem]{ulem}

\newcommand{\be}{\begin{equation}}
\newcommand{\ee}{\end{equation}}

\newcommand{\dCS}{{\mbox{\tiny dCS}}}

\newcommand{\GR}{{\mbox{\tiny GR}}}
\newcommand{\bGR}{{\mbox{\tiny bGR}}}

\newcommand{\geo}{{\mbox{\tiny geo}}}
\newcommand{\typeD}{{\mbox{\tiny D}}}
\newcommand{\nonD}{{\mbox{\tiny non-D}}}
\newcommand{\EdGB}{{\mbox{\tiny EdGB}}}

\newcommand{\HD}{{\mbox{\tiny HD}}}
\newcommand{\matter}{{\mbox{\tiny matter}}}
\newcommand{\field}{{\mbox{\tiny field}}}
\newcommand{\even}{{\mbox{\tiny even}}}
\newcommand{\odd}{{\mbox{\tiny odd}}}
\newcommand{\M}{{\mbox{\tiny M}}}

\newcommand{\ba}{\begin{align}}
\newcommand{\ea}{\end{align}}

\makeatletter
\newcommand*{\rom}[1]{\expandafter\@slowromancap\romannumeral #1@}
\makeatother

\AtBeginDocument{%
    \newwrite\bibnotes
    \def\bibnotesext{Notes.bib}
    \immediate\openout\bibnotes=\jobname\bibnotesext
    \immediate\write\bibnotes{@CONTROL{REVTEX41Control}}
    \immediate\write\bibnotes{@CONTROL{%
    apsrev41Control,author="08",editor="1",pages="1",title="0",year="1"}}
    \if@filesw
    \immediate\write\@auxout{\string\citation{apsrev41Control}}%
    \fi
}
\newcommand{\UIUC}{Illinois  Center  for  Advanced  Studies  of  the  Universe \&
Department of Physics, University of Illinois at Urbana-Champaign, Urbana, Illinois 61801, USA}
\newcommand{\CAL}{Theoretical Astrophysics 350-17, California Institute of Technology, Pasadena, CA 91125, USA}

\begin{document}
\title{Perturbations of spinning black holes beyond General Relativity: \\ Modified Teukolsky equation}

\author{Dongjun Li}
\affiliation{\CAL}
\email{dlli@caltech.edu}

\author{Pratik Wagle}
\affiliation{\UIUC}
\email{wagle2@illinois.edu}

\author{Yanbei Chen}
\affiliation{\CAL}

\author{Nicol\'as Yunes}
\affiliation{\UIUC}

\date{\today}

\begin{abstract}
    The detection of gravitational waves from compact binary mergers by the LIGO/Virgo collaboration has, for the first time, allowed for tests of relativistic gravity in the strong, dynamical and nonlinear regime. Outside Einstein's relativity, spinning black holes may be different from their general relativistic counterparts, and their merger may then lead to a modified ringdown. We study the latter and, for the first time, derive a modified Teukolsky equation, i.e., a set of linear, decoupled differential equations that describe dynamical perturbations of non-Kerr black holes for the radiative Newman-Penrose scalars $\Psi_0$ and $\Psi_4$. We first focus on non-Ricci-flat, Petrov type D black hole backgrounds in modified gravity, and derive the modified Teukolsky equation through direct decoupling and through a new approach, proposed by Chandrasekhar, that uses certain gauge conditions. We then extend this analysis to non-Ricci-flat, Petrov type I black hole backgrounds in modified gravity, assuming they can be treated as a linear perturbation of Petrov type D, black hole backgrounds in GR by generalizing Chandrasekhar's approach, and derive the decoupled modified Teukolsky equation. We further show that our formalism can be extended beyond linear order in both modified gravity corrections and gravitational wave perturbations. Our work lays the foundation to study the gravitational waves emitted in the ringdown phase of black hole coalescence in modified gravity for black holes of any spin. Our work can also be extended to compute gravitational waves emitted by extreme mass-ratio binary inspirals in modified gravity.
    
\end{abstract}

\maketitle
%%%%%%%%%%%%%%%%%%%%%%%%%%%%%%%%%%%%%%

\section{Introduction}
\label{sec:Intro}

General relativity (GR) has passed a plethora of experimental tests in the Solar system~\cite{Will2014} and in binary pulsars systems~\cite{Stairs2003,Wex:2020ald}, making it the most successful theory of gravity to date. With the detection of gravitational waves (GWs) by the LIGO/Virgo/Kagra (LVK) collaboration~\cite{TheLIGOScientific:2016src}, tests in the extreme gravity regime, where gravity is simultaneously strong, dynamical and non-linear, have gained prominence in the last decade~\cite{Yunes:2013dva,Will2014,Yagi:2016jml,Berti:2018cxi,Nair:2019iur}. Such tests will become only stronger with the next generation of ground-based~\cite{Abramovici:1992ah,Reitze:2019iox} and space-based detectors~\cite{Baker:2019nia}, allowing for even more stringent constraints on modifications to GR (see e.g., Refs.~\cite{Nair:2019iur,Perkins:2022fhr,Perkins:2020tra,Perkins:2021mhb,LIGOScientific:2020tif,LIGOScientific:2021sio,Gnocchi:2019jzp}).

Einstein's theory, although very successful, can be interpreted as having difficulties explaining certain theoretical and observational anomalies, which has motivated the study of modified theories of gravity. For example, the incompatibility between GR and quantum mechanics has motivated efforts in a variety of unified theories, such as loop quantum gravity~\cite{Birrell:1982ix,Ashtekar:1997yu,Ashtekar:2004eh} and string theory~\cite{Damour:1994zq,Mukhi:2011zz}. Observational anomalies could include the late-time acceleration of the Universe~\cite{Perlmutter:1998np,Riess:1998cb} (without the inclusion of an ``unnaturally'' small cosmological constant~\cite{Nojiri:2006ri,Tsujikawa:2010zza}), the anomalous galaxy rotation curves~\cite{Sofue:2000jx,Bertone:2016nfn} (without the inclusion of dark matter~\cite{Petraki:2013wwa}), and the matter-antimatter asymmetry of the Universe~\cite{Canetti:2012zc} (without the inclusion of additional sources of parity violation required by the Sakharov conditions~\cite{Petraki:2013wwa,Gell-Mann:1991kdm,Alexander:2004us}). All of these perceived anomalies have resulted in a zoo of modifications to GR, which can be both consistent with all current tests, while still yield deviations in the extreme gravity regime. For this class of theories, GWs may be excellent probes to study and possibly constrain deviations from Einstein's theory. 

An important source of GWs is the coalescence of compact objects: the inspiral, merger and ringdown of a binary system composed of black holes (BHs) and/or neutron stars (NSs). All of these coalescence phases can be used to test GR and constrain deviations. For instance, the presence of extra (scalar or vector) radiative degrees of freedom can be constrained with the inspiral phase of GWs emitted in binary BH coalescence. These fields can increase the rate at which orbital energy is radiated away from the system, thus affecting the orbital dynamics~\cite{Yagi:2015oca,Yagi:2013mbt,Berti:2018cxi,Wagle:2018tyk,R:2022cwe,Loutrel:2022tbk}, which can be modeled with post-Newtonian methods. The GW observations made by the LVK collaboration in the inspiral regime can then be used to determine whether binary BHs spiral in at the expected GR rate or not, thus allowing for constraints on the existence of these additional radiative fields~\cite{Perkins:2021mhb,Perkins:2022fhr,Perkins:2020tra}. 

On the other hand, modifications to the exterior BH geometry as well as the dynamics of these modified gravity theories may be constrained with ringdown GWs, emitted as the BH remnant settles to its final, stationary configuration. These waves can be characterized as a sum of quasinormal modes (QNMs), whose complex frequency contains information about the remnant BH background~\cite{Vishveshwara:1970cc,Vishveshwara:1970zz,Chandrasekhar_1983,Regge:PhysRev.108.1063,Zerilli:1971wd,Moncrief:1974am,Teukolsky:1973ha,Leaver:1985ax,Maggiore:2018sht}. The LVK observation of ringdown GWs and the measurement of the complex frequencies of a set of QNMs can then be used to probe the exterior geometry of the remnant~\cite{Dreyer:2003bv,Dreyer_2004}. In particular, these observations can yield tests of the Kerr hypothesis (i.e., that all astrophysical BHs can be described by the Kerr metric)~\cite{Dreyer:2003bv,Bambi:2015kza,Cardoso:2016ryw}. 
The GWs emitted during ringdown can be studied by considering gravitational perturbations of a background BH spacetime, obtaining their evolution equations, and then solving the latter to find the spectrum of perturbations.
Additionally, depending on the theory, there might be additional degrees of freedom present, leading to additional or coupled evolution equations that can be solved to obtain the QNM frequency spectra~\cite{Cardoso:2009pk, Molina:2010fb, Blazquez-Salcedo:2016enn, Blazquez-Salcedo_Khoo_Kunz_2017, Wagle_Yunes_Silva_2021, Srivastava_Chen_Shankaranarayanan_2021, Pierini:2021jxd, Pierini:2022eim}. This forms the basis of BH perturbation theory, which has been used to study QNMs of non-rotating BHs in GR~\cite{Regge:PhysRev.108.1063,Zerilli:1971wd,Moncrief:1974am,Vishveshwara:1970cc,Vishveshwara:1970zz} and modified gravity~\cite{Cardoso:2009pk, Molina:2010fb, Pani_Cardoso_Gualtieri_2011, Blazquez-Salcedo:2016enn, Blazquez-Salcedo_Khoo_Kunz_2017}. When the background spacetime is that of a non-rotating BH, the background metric is static and spherically symmetric, so the time and angular dependence of the evolution equations of the perturbations can be easily separated. In GR, the resulting coupled radial equations can then be further reduced to two decoupled equations, one for odd parity perturbations and another for even parity perturbations \cite{Regge:PhysRev.108.1063, Zerilli:1971wd}. In modified gravity, however, one may not be able to decouple all the radial equations, so there can be more than one equation in each parity besides the equations of extra non-metric fields \cite{Cardoso:2009pk, Molina:2010fb, Blazquez-Salcedo:2016enn, Blazquez-Salcedo_Khoo_Kunz_2017, Wagle_Yunes_Silva_2021, Srivastava_Chen_Shankaranarayanan_2021, Pierini:2021jxd, Pierini:2022eim}.

When considering background spacetimes that represent spinning BHs, however, the situation is much more complicated. This is because such BHs are mathematically represented through a background metric that is stationary and axisymmetric. The lack of spherical symmetry renders the evolution equations for the metric perturbations non-separable. Fortunately, an alternate method, prescribed by Teukolsky in 1973~\cite{Teukolsky:1973ha}, allows for the separation of the perturbation equations when one works with curvature quantities (instead of metric quantities), characterized in the Newman-Penrose (NP) formalism~\cite{Newman-Penrose}. The latter arises naturally from the introduction of spinor calculus into GR and is a special type of tetrad calculus. Using the NP formalism, the perturbations of a Schwarzschild BH in GR were studied by Price~\cite{Price:1972pw} and extended later in~\cite{Bardeen:1973xb}. Combining these results with Teukolsky's~\cite{Teukolsky:1973ha}, a separable decoupled equation for each of the two components of the perturbed Weyl tensor ($\Psi_0$ and $\Psi_4$) can be obtained. These decoupled equations paved the way for QNM studies in GR, allowing for the accurate computation of the QNM frequencies of Kerr BHs~\cite{Berti:2009kk,BertiRingdown}.

The Teukolsky formalism~\cite{Teukolsky:1973ha}, however, is not generally applicable in modified theories of gravity. In particular, this formalism applies only when the Einstein equations hold and when the background spacetime is of Petrov type D~\cite{Petrov:2000bs,Chandrasekhar_1983}, i.e., when all Weyl scalars except $\Psi_2$ vanish on the background spacetime. However, modified theories of gravity do not necessarily satisfy the Einstein equations, and the background BH solutions in these theories need not be of Petrov type D in general. This is the case, for instance, in quadratic theories of gravity (such as dynamical Chern-Simons (dCS) gravity~\cite{Jackiw:2003pm,Alexander:2009tp} or scalar-Gauss-Bonnet (sGB) gravity~\cite{Alty:1994xj,PhysRevD.98.021503}), where a dynamical field is non-minimally coupled to a quadratic curvature invariant. In these theories, the field equations are not Einstein's, and isolated, rotating BHs are of the algebraically general Petrov type \rom{1}~\cite{Owen:2021eez}, i.e., only the $\Psi_0$ and $\Psi_4$ background Weyl scalars vanish. Therefore, the Teukolsky formalism cannot be used directly to prescribe master equations for the evolution of curvature perturbations in such beyond GR BH backgrounds.

The study of BH perturbations and their QNMs in modified gravity has gained prominence in the recent decade. However, for the most part, these calculations have been limited to the non-rotating and the slowly-rotating case. In the spherically-symmetric, non-rotating case, QNMs have been calculated using metric perturbation theory, e.g., in dCS gravity~\cite{Cardoso:2009pk, Molina:2010fb, Pani_Cardoso_Gualtieri_2011}, Einstein-dilaton-Gauss-Bonnet (EdGB) gravity~\cite{Blazquez-Salcedo:2016enn, Blazquez-Salcedo_Khoo_Kunz_2017}, Einstein-Aether theory~\cite{Konoplya_Zhidenko_2006, Konoplya_Zhidenko_2007, Ding_2017, Ding_2019, Churilova_2020}, higher-derivative gravity (quadratic~\cite{Cardoso_Kimura_Maselli_Senatore_2018},~cubic \cite{deRham_Francfort_Zhang_2020}, and more generically~\cite{Cardoso_Kimura_Maselli_Berti_Macedo_McManus_2019, McManus_Berti_Macedo_Kimura_Maselli_Cardoso_2019}), and Horndeski gravity~\cite{Tattersall_Ferreira_2018}. In the axisymmetric, rotating case, reducing all the metric perturbations into a single perturbation function (e.g., Regge Wheeler function or a Zerilli-Moncrief function) is difficult, so studies have resorted to the slow-rotation approximation at leading order, e.g., in EdGB gravity~\cite{Pierini:2021jxd,Pierini:2022eim}, dCS gravity~\cite{Wagle_Yunes_Silva_2021,Srivastava_Chen_Shankaranarayanan_2021}, and higher-derivative gravity~\cite{Cano:2020cao, Cano_Fransen_Hertog_Maenaut_2021}. Purely numerical studies of perturbed spinning BHs, resulting from the merger of two other BHs, have also been done in dCS gravity, but they typically suffer from secularly-growing uncontrolled remainders~\cite{Okounkova:2019dfo, Okounkova_Stein_Moxon_Scheel_Teukolsky_2020}.

\begin{figure*}[t]
	\centering
	\includegraphics[width=\linewidth]{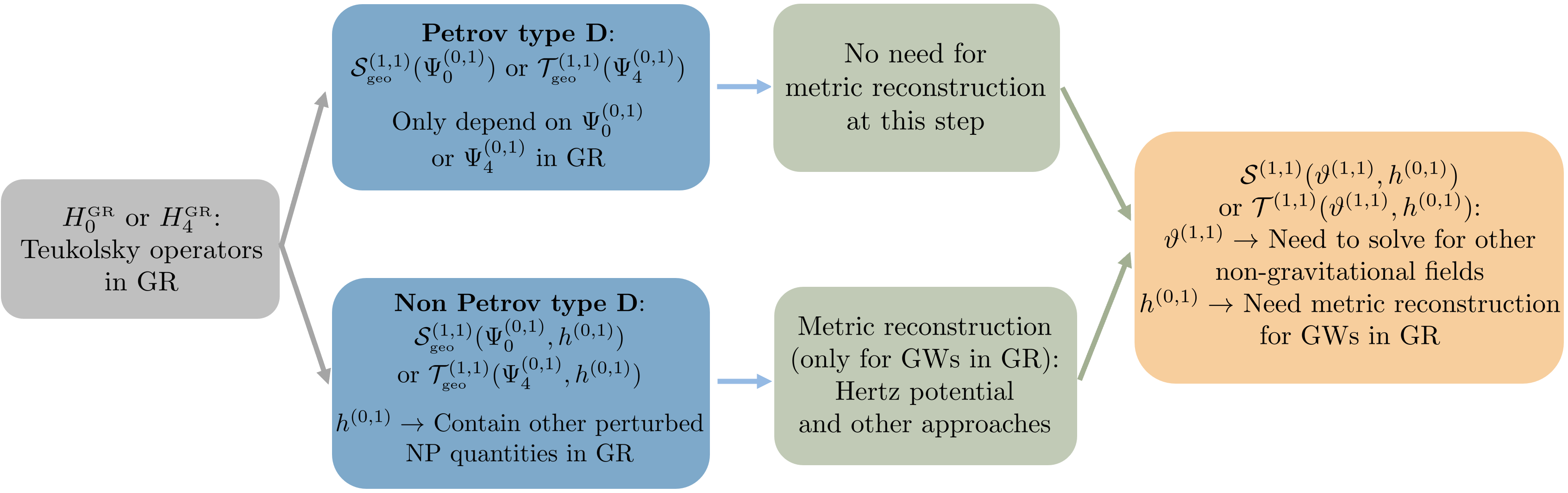}
	\caption{Schematic flow chart of the different possible terms that may arise in the modified Teukolsky equation for any Petrov type \rom{1} spacetime in modified gravity, where the background can be treated as a linear perturbation of a Petrov type D spacetime in GR. The origin of these correction terms and the strategies to evaluate them are outlined here and discussed in detail in Sec.~\ref{sec:The_modified_Teukolsky_equation}. For comparison, the corresponding procedures for any Petrov type D spacetime in modified gravity theory is also shown.}
	\label{fig:flowchart}
\end{figure*}

One can in principle extend the slow-rotation approximation to the QNM spectrum of rotating BHs in modified gravity to higher order in rotation, but this can be a daunting task. This is because the GWs emitted during ringdown are produced by BH remnants that typically spin at about $65\%$ of their maximum or higher~\cite{LIGOScientific:2021djp}. The accurate calculation of the QNM spectrum of such BHs then requires one to go to at least fifth order in a slow-rotation expansion or higher~\cite{Pani:2011vy}. Nonetheless, it has been shown in \cite{Pierini:2022eim} (see also \cite{Hatsuda:2020egs,Julie:2022huo}) that one can improve the convergence of the slow-rotation expansion using the Pad\'{e} approximation. In EdGB, one may then only consider up to second order in the slow-rotation expansion to deal with BHs spinning at about $70\%$ of their maximum. Additionally, going to higher order in spin leads to mode coupling between the $\ell$ modes, the $\ell \pm 1$ modes and higher modes~\cite{Pani:2013pma,Pani:2012bp,Wagle_Yunes_Silva_2021}, where $\ell$ is the orbital number of the spherical harmonic decomposition. Therefore, instead of extending the slow-rotation approximation, we here focus on developing a new formalism, motivated from the work of Teukolsky and Chandrasekhar, to understand the evolution of curvature perturbations and therefore the QNM spectrum of rotating BHs of arbitrary spin in modified gravity. 

\subsection*{Executive summary}

We here develop and apply a method to find the evolution equations of gravitational perturbations around non-Ricci-flat and Petrov type I BH backgrounds in modified gravity, where the BH background can be treated as linear perturbations of a Petrov type D background in GR. We begin by focusing on backgrounds that are still Petrov type D, but are not described by the Kerr metric because they satisfy field equations that are not Einstein's, i.e., the background spacetime is not Ricci flat. In this context, we extend the usual Teukolsky formalism, and also develop a new approach to find the curvature perturbation equations in a particular gauge, following Chandrasekhar~\cite{Chandrasekhar_1983}. We show that these two approaches yield the same perturbation equations.

Let us describe both of these approaches in more detail, beginning first with a brief refresher of how these approaches are applied in GR. In the traditional Teukolsky's approach, one begins by considering two Bianchi identities and one Ricci identity in the NP formalism. Using these equations along with the GR vacuum field equations and imposing the requirement that the background is Ricci-flat (i.e., the Ricci tensor vanishes on the background) and Petrov type D, one can in principle generate a commutator relation that eliminates the coupling between the perturbed Weyl scalars $\Psi_0^{(1)}$ and $\Psi_1^{(1)}$ and between $\Psi_4^{(1)}$ and $\Psi_3^{(1)}$. However, in the process of obtaining the commutation relation, one has to make use of additional Bianchi identities. This procedure is not tedious in GR because many NP scalars and spin coefficients vanish identically, but it can be non-trivial in modified gravity.

In Chandrasekhar's approach~\cite{Chandrasekhar_1983}, one makes use of suitable gauge conditions to simplify the perturbed equations without the need to use additional Bianchi identities. In this special gauge, the background and perturbed Weyl scalar $\Psi_1$ and $\Psi_3$ vanish, so the two Bianchi identities and the Ricci identity mentioned above simplify and depend now only on three unknown quantities. Decoupling these equations, one then obtains a master equation for the perturbed Weyl scalars $\Psi_i^{(1)}$ of the form,
\begin{equation}
    H_{i}^{\GR}\Psi_i^{(1)} = 0 \,, \quad i \in \{0,4\} \,,
\end{equation}
where $H_{i}^{\GR}$ are the Teukolsky differential operators~\cite{Teukolsky:1973ha}.

As mentioned earlier, we begin our analysis by modifying both of these approaches so that they are applicable in modified gravity for curvature perturbations of non-Ricci-flat BHs that are still Petrov type D. In the traditional Teukolsky's approach, we first develop a commutator relation by using additional Bianchi identities. Due to the complicated nature of the field equations in modified gravity, there are more non-vanishing NP quantities, thereby leading to more terms in the perturbation equations. To leading order in the perturbation and in deformations from GR, however, only the Bianchi identities and the commutator relations of GR are required since all additional terms vanish. In the Chandrasekhar's approach, we first show that even in modified gravity, a gauge still exists in which the perturbed $\Psi_1^{(1)}$ and $\Psi_3^{(1)}$ vanish. Using this gauge, the curvature perturbations can be easily decoupled. 

To derive the master equation, we find a two-parameter expansion useful. We use $\epsilon$ to denote the size of the GW perturbations and $\zeta$ the strength of the modified gravity correction. With this at hand, we show that any NP quantities $\Psi$ can be expanded as
\begin{align}
	\Psi
	=\Psi^{(0,0)}+\zeta\Psi^{(1,0)}+\epsilon\Psi^{(0,1)}
	+\zeta\epsilon\Psi^{(1,1)}\,.
\end{align}
We then show that both approaches lead to a modified evolution equation for the curvature perturbations of the form
\begin{align} \label{eq:sch-teuk2}
    & H_{0}^{\GR}\Psi_{0}^{(1,1)} =\mathcal{S}_{\geo}^{(1,1)}(\Psi_0^{(0,1)})
    +\mathcal{S}^{(1,1)}(\vartheta^{(1,1)}, h^{(0,1)})\,, \nonumber\\
    & H_{4}^{\GR}\Psi_{4}^{(1,1)} =\mathcal{T}_{\geo}^{(1,1)}(\Psi_4^{(0,1)})
    +\mathcal{T}^{(1,1)}(\vartheta^{(1,1)}, h^{(0,1)})\,,
\end{align}
where the $H_{i}^{\GR}$ differential operators are the same as the Teukolsky ones in GR~\cite{Teukolsky:1973ha}. Here, we have listed the dynamical quantities [i.e., $\mathcal{O}(\epsilon)$ terms] inside the parentheses. The source terms $\mathcal{S}^{(1,1)}$ and $\mathcal{T}^{(1,1)}$ arise from the perturbed and modified field equations, and they are functionals of any additional dynamical scalar, vector or tensor field in the theory (denoted as $\vartheta^{(1,1)}$ above) and the GW metric perturbation (denoted as $h^{(0,1)}$). The source terms $\mathcal{S}_{\geo}^{(1,1)}$ and $\mathcal{T}_{\geo}^{(1,1)}$ arise from the homogeneous part of the two Bianchi identities due to the correction to the background spacetime in modified gravity, and they are functionals of the dynamical $\Psi_{0,4}^{(0,1)}$ in GR. 

The evaluation of the source terms, which is required to evaluate the curvature perturbation evolution equations, requires knowledge of $h^{(0,1)}$ and $\vartheta^{(1,1)}$. The source terms $\mathcal{S}^{(1,1)}$ and $\mathcal{T}^{(1,1)}$ depend on $h^{(0,1)}$, so the evaluation of the right-hand side of Eq.~\eqref{eq:sch-teuk2} requires the reconstruction of the GW metric perturbation in GR $h^{(0,1)}$. This can be accomplished with the well-developed methods of Chrzanowski~\cite{Chrzanowski:1975wv} and others~\cite{Kegeles_Cohen_1979,Chandrasekhar_1983,Loutrel_Ripley_Giorgi_Pretorius_2020}. Moreover, the source terms $\mathcal{S}^{(1,1)}$ and $\mathcal{T}^{(1,1)}$ also depend on the evolution of the perturbed scalar, vector, or tensor degrees of freedom that the theory may admit $\vartheta$. The evolution of these degrees of freedom has to be solved simultaneously with the solution to the curvature perturbations. 

With this at hand, we then apply Chandrasekhar's approach to modified gravity theories for non-Ricci-flat and Petrov type \rom{1} BH backgrounds. In such spacetimes, the biggest challenge is that many background NP quantities are non-vanishing. Working perturbatively (i.e., treating the BH background as a deformation of the Petrov type D background in GR), one can eliminate the perturbed $\Psi_1$ and $\Psi_3$ from the evolution equations and obtain a separated and decoupled equation for $\Psi_0$ and $\Psi_4$. Schematically, these equations look a lot like the decoupled equations when dealing with non-Ricci-flat and Petrov type D backgrounds, except that now the source terms $\mathcal{S}_{\geo}^{(1,1)}$ and $\mathcal{T}_{\geo}^{(1,1)}$ can also be functionals of the GW metric perturbation, namely,
\begin{align} \label{eq:sch-teuk3}
    & H_{0}^{\GR}\Psi_{0}^{(1,1)} =\mathcal{S}_{\geo}^{(1,1)}(\Psi_0^{(0,1)}, h^{(0,1)})
    +\mathcal{S}^{(1,1)}(\vartheta^{(1,1)}, h^{(0,1)})\,, \nonumber\\
    & H_{4}^{\GR}\Psi_{4}^{(1,1)} =\mathcal{T}_{\geo}^{(1,1)}(\Psi_4^{(0,1)}, h^{(0,1)})
    +\mathcal{T}^{(1,1)}(\vartheta^{(1,1)}, h^{(0,1)})\,,
\end{align}
This time we see that both source terms to the curvature perturbation evolution equations require the reconstruction of the GW metric perturbation in GR. As in the Petrov type D case, we also see that the source terms $\mathcal{S}^{(1,1)}$ and $\mathcal{T}^{(1,1)}$ require knowledge of the evolution of the perturbed scalar, vector, or tensor degrees of freedom that the modified theory may admit $\vartheta$.
Figure~\ref{fig:flowchart} shows schematically the structure of the master equations for $\Psi_0$ and $\Psi_4$. 

In the rest of the paper, we derive and present the results summarized above in detail. In Sec.~\ref{sec:GRintro}, we present a brief review of the NP formalism and relevant NP equations. We also review the analysis presented by Teukolsky (i.e., the Teukolsky formalism) and by Chandrasekhar (using a gauge choice) for Petrov type D spacetimes in GR. In Sec.~\ref{sec:perturbation_framework}, we discuss a subset of modified gravity theories that our work can be applied to and prescribe a perturbation scheme for them. We then extend both Teukolsky's and Chandrasekhar's approaches to Petrov type D spacetimes in these modified gravity theories in Sec.~\ref{sec:nonGR}. In Sec.~\ref{sec:teuknonD}, we prescribe and discuss in detail the formalism to study perturbations of an algebraically general: Petrov type I spacetime in modified gravity theories which can be treated as a linear perturbation of a Petrov type D spacetimes in GR. In Sec.~\ref{sec:higher_order_perturbations}, we discuss the connection of the formalism developed in Sec.~\ref{sec:teuknonD} to the second-order Teukolsky formalism in GR. We further show that our formalism can be generalized to higher order in both $\zeta$ and $\epsilon$, which is thus a bGR extension of the higher-order Teukolsky formalism in GR developed in \cite{Campanelli:1998jv}. Finally, in Sec.~\ref{sec:discussions}, we summarize our work and discuss some avenues for future work. Henceforth, we adopt the following conventions unless stated otherwise: we work in 4-dimensions with metric signature $(-,+,+,+)$ as in~\cite{Misner:1974qy}. For all NP quantities except the metric signature, we use the notation adapted by Chandrasekhar in~\cite{Chandrasekhar_1983}.

%%%%%%%%%%%%%%%%%%%%%%%%%%%%%%%%%%%%%%
\section{NP formalism and perturbations of BHs in GR} \label{sec:GRintro}

With the study of GWs using tetrad and spinor calculus gaining prominence in the 1960s, Ezra Newman and Roger Penrose presented a formalism that combines these two techniques to derive a very compact and useful set of equations that are equivalent to the field equations~\cite{Newman:1961qr}. This set of equations consists of a linear combination of equations for the Riemann tensor in terms of the Ricci rotation coefficients or spinor affine connections~\cite{Newman:1961qr}. The different possible components of the Riemann tensor or the Weyl tensor in a null tetrad or a null basis were then associated with certain quantities, called the NP coefficients or NP scalars. This formalism provided a new tool to understand GW properties, such as polarizations and ringdown modes in more detail~\cite{Eardley:1973br,Eardley:1973zuo,Wagle:2019mdq,Teukolsky:1973ha,Press:1973zz}. Using the NP framework, Teukolsky presented a formalism to study the ringdown phase of spinning BHs in GR~\cite{Teukolsky:1973ha,Press:1973zz,Teukolsky:1974yv} and to study the dynamical perturbations of Kerr BHs, or more generally, Petrov type D spacetimes in GR.

In this section, we provide a quick refresher of the NP formalism and discuss the necessary equations for developing a formalism to obtain master equations for GW perturbations in GR. Using these equations, we present in brief the approach prescribed by Teukolsky~\cite{Teukolsky:1973ha} and by Chandrasekhar~\cite{Chandrasekhar_1983} to obtain separable decoupled differential equations for perturbations of BHs in GR. For a reader familiar with these topics, we recommend starting from Sec.~\ref{sec:perturbation_framework}, where we extend the aforementioned formalism to BHs in modified gravity.

\subsection{NP formalism: A quick review} \label{sec:NPformalism}

In this subsection, we present a quick overview of the relevant equations under the NP formalism required for our work. For an in-depth overview, we provide further details of the NP formalism~\cite{Newman:1961qr,Chandrasekhar_1983} in Appendix~\ref{appendix:NPformalism_addition}. In the NP formalism, a null tetrad $(l^\mu,n^\mu,m^\mu,\Bar{m}^\mu)$ is introduced at every point of a four-dimensional pseudo-Riemannian manifold of signature $+2$ and metric $g_{\mu\nu}$. The vectors $l^\mu$ and $n^\mu$ are real, whereas $m^\mu$ and $\Bar{m}^\mu$ are complex, with a overhead bar denoting complex conjugation. The tetrad 4-vectors must also satisfy the following orthogonality properties:
\begin{align}
\label{eq:tetrad_ortho}
l_\mu l^\mu = n_\mu n^\mu &= m_\mu m^\mu = \Bar{m}_\mu \Bar{m}^\mu = 0  \nonumber \,, \\ 
l_\mu n^\mu &= - m_\mu \Bar{m}^\mu = -1 \,, \nonumber \\ 
l_\mu m^\mu = l_\mu \Bar{m}^\mu &= n_\mu m^\mu = n_\mu \Bar{m}^\mu = 0 \,.
\end{align}
Given such a null tetrad, the metric can be expressed as
\be
\label{eq:tetrad_metric}
g_{\mu\nu} =-l_\mu n_\nu -n_\mu l_\nu + m_\mu \Bar{m}_\nu + \Bar{m}_\mu m_\nu \,.
\ee
Intrinsic derivatives in the NP formalism are defined as,
\begin{align}
\label{eq:derivative}
    D \phi \equiv \phi_{;\mu} l^\mu \,, ~~~~~~~ & ~~~~~~ 
    \Delta\phi \equiv \phi_{;\mu}n^\mu \,, \nonumber \\
    \delta \phi \equiv \phi_{;\mu} m^\mu \,, ~~~~~~~ & ~~~~~~ \delta^{*}\phi \equiv \phi_{;\mu}\Bar{m}^\mu \,.
\end{align}
For any tetrad, we can also perform Lorentz transformations on it, i.e., three rotations and three boosts. These transformations can be mapped to three types of tetrad rotations, which are characterized by six real variables on the tetrad basis vectors, such that the orthogonality properties in Eq.~\eqref{eq:tetrad_ortho} are preserved \cite{Chandrasekhar_1983}. These three types of tetrad rotations are discussed in detail in Appendix~\ref{appendix:NPformalism_addition}. 

In the NP formalism, the fundamental variables are 5 Weyl scalars ($\Psi_1,\Psi_2,...$), 12 spin coefficients ($\kappa,\pi,\varepsilon,...$), and 10 NP Ricci scalars ($\Phi_{00},\Phi_{01},..,\Lambda$), which are generally complex quantities. The mathematical form of all these quantities is presented in Appendix~\ref{appendix:NPformalism_addition}. These quantities allow one to construct certain fundamental relations of the NP formalism: $18$ complex Ricci identities [Eq.~\eqref{eq:Ricci_ids_all}] and $9$ complex plus $2$ real Bianchi identities [Eqs.~\eqref{eq:bianchi_iden}] \cite{Chandrasekhar_1983}. The Ricci identities are derived from appropriate linear combinations of Eq.~\eqref{eq:riemann_ricci_rot} and \eqref{eq:Weyl_Riemann}, while the Bianchi identities come from Eq.~\eqref{eq:Bianchi_gen}. Some Ricci identities relevant for this work are
\begin{subequations} \label{eq:RicciId}
\begin{align}
\label{eq:RicciId_Psi0}  \left(D-\rho-\rho^{*}\right.
&-\left.3\varepsilon+\varepsilon^{*}\right)\sigma \nonumber \\
-&(\delta-\tau+\pi^{*}  -\alpha^{*}-3\beta)\kappa-\Psi_{0}=0\,, \\
\label{eq:RicciId_Psi4}  \left(\Delta+\mu+\mu^{*} \right.
&+\left.3\gamma-\gamma^{*}\right)\lambda\nonumber \\
-&(\delta^{*}+3\alpha + \beta^{*} +\pi-\tau^{*})\nu+\Psi_{4}=0\,, \\
\label{eq:Riccieq_Dtau}  \left(D-\varepsilon+\varepsilon^* \right.
&-\left.\rho\right)\tau -\left(\Delta-3\gamma+\gamma^*\right)\kappa\nonumber \\
-& \pi^* \rho-\left(\tau^*+\pi\right)\sigma-\Psi_1-\Phi_{01} =0 \,,\\
\label{eq:Riccieq_Dbeta} \left(D -\rho^*+\varepsilon^* \right.
&\left.\right)\beta-\left(\delta+\alpha^*-\pi^*\right)\varepsilon \nonumber \\
-&\left(\alpha +\pi\right)\sigma-\left(\mu+\gamma\right)\kappa-\Psi_1=0 \,,\\  
\label{eq:Riccieq_Drho}  \left(D-\rho^*+\varepsilon^*\right.
&\left.\right)\beta-\left(\delta+\alpha^*-\pi^*\right)\varepsilon\nonumber \\
-& \left(\alpha+\pi\right)\sigma-\left(\mu+\gamma\right)\kappa-\Psi_1  =0 \,,
\end{align}
\end{subequations}
while the Bianchi identities useful for this work are
\begin{subequations} \label{eq:bianchiid}
\begin{align}
\label{eq:BianchiId_Psi0_1}
& \left(\delta^{*}-4\alpha+\pi\right)\Psi_{0} - (D-2\varepsilon-4\rho) \Psi_{1}-3\kappa\Psi_{2}=S_1\,, 
\\
\label{eq:BianchiId_Psi0_2}
& (\Delta-4\gamma+\mu)\Psi_{0} - (\delta-4\tau-2\beta)\Psi_{1}-3\sigma\Psi_{2}=S_2\,, 
\\
\label{eq:BianchiId_Psi4_1}
& (\delta+4\beta-\tau)\Psi_{4} - (\Delta+2\gamma+4\mu)\Psi_{3}+3\nu\Psi_{2}=S_3 \,, 
\\
\label{eq:BianchiId_Psi4_2}
& (D+4\varepsilon-\rho)\Psi_{4}	-\left(\delta^{*}+4\pi+2\alpha\right)\Psi_{3}+3\lambda\Psi_{2}=S_4\,,
\end{align}
\end{subequations}
where we have defined
\begin{subequations}
\label{eq:source_bianchi}
\begin{align}
    \label{eq:source_bianchi_1}
	\begin{split} 
		S_1\equiv& \;\left(\delta+\pi^{*}-2\alpha^{*}-2\beta\right)\Phi_{00}
		-\left(D-2\varepsilon-2\rho^{*}\right)\Phi_{01} \\
		& \;+2\sigma\Phi_{10}-2\kappa\Phi_{11}-\kappa^{*}\Phi_{02}\,,
	\end{split} \\
	\label{eq:source_bianchi_2}
	\begin{split} 
		S_2\equiv& \;\left(\delta+2\pi^{*}-2\beta\right)\Phi_{01}
		-\left(D-2\varepsilon+2\varepsilon^{*}-\rho^{*}\right)\Phi_{02} \\
		& \;-\lambda^{*}\Phi_{00}+2\sigma\Phi_{11}-2\kappa\Phi_{12}\,, 
	\end{split} \\
	\label{eq:source_bianchi_3}
	\begin{split}
		S_3\equiv& \;-\left(\Delta+2\mu^{*}+2\gamma\right)\Phi_{21}
		+\left(\delta^{*}-\tau^{*}+2\alpha+2\beta^{*}\right)\Phi_{22} \\
		& \;+2\nu\Phi_{11}+\nu^{*}\Phi_{20}-2\lambda\Phi_{12}\,,
	\end{split} \\
	\label{eq:source_bianchi_4}
	\begin{split} 
		S_4\equiv& \;-\left(\Delta+\mu^{*}+2\gamma-2\gamma^{*}\right)\Phi_{20}
		+\left(\delta^{*}+2\alpha-2\tau^{*}\right)\Phi_{21} \\
		& \;+2\nu\Phi_{10}-2\lambda\Phi_{11}+\sigma^{*}\Phi_{22}\,.
	\end{split}
\end{align}
\end{subequations}
The remaining equations are presented in Appendix~\ref{appendix:NPformalism_addition}.

The above equations can be recast in a simpler form if we define the following operators: 
\begin{equation} \label{eq:simplify_operators_0}
    \begin{aligned}
        & F_1\equiv\delta^{*}-4\alpha+\pi\,,\quad
	    F_2\equiv\Delta-4\gamma+\mu\,, \\
	    & J_1\equiv D-2\varepsilon-4\rho\,,\quad
		J_2\equiv\delta-4\tau-2\beta\,, \\
        & E_1\equiv\delta-\tau+\pi^{*}-\alpha^{*}-3\beta\,, \\
		& E_2\equiv D-\rho-\rho^{*}-3\varepsilon+\varepsilon^{*}\,, \\
    \end{aligned}
\end{equation}
\begin{equation} \label{eq:simplify_operators_4}
    \begin{aligned}
        & F_3\equiv \delta+4\beta-\tau \,,\quad
	    F_4\equiv D+4\varepsilon-\rho \,, \\
	    & J_3\equiv \Delta+2\gamma+4\mu \,,\quad
		J_4\equiv \delta^{*}+4\pi+2\alpha \,, \\
		& E_3\equiv \delta^{*}+3\alpha + \beta^{*} +\pi-\tau^{*}\,, \\
		& E_4\equiv \Delta+\mu+\mu^{*} + 3 \gamma-\gamma^{*}\,,
    \end{aligned}
\end{equation}
so we can rewrite Eqs.~\eqref{eq:BianchiId_Psi0_1}-\eqref{eq:BianchiId_Psi0_2} and Eq.~\eqref{eq:RicciId_Psi0} as
\begin{subequations}
\label{eq:Bianchi_simplified}
\begin{align} 
    & F_1\Psi_0-J_1\Psi_1-3\kappa\Psi_2=S_1\,, \label{eq:BianchiId_Psi0_1_simplify} \\
    & F_2\Psi_0-J_2\Psi_1-3\sigma\Psi_2=S_2\,, \label{eq:BianchiId_Psi0_2_simplify} \\
    & E_2\sigma-E_1\kappa-\Psi_0=0 \label{eq:RicciId_Psi0_simplify}\,,
\end{align}
\end{subequations}
while Eqs.~\eqref{eq:BianchiId_Psi4_1}-\eqref{eq:BianchiId_Psi4_2} and Eq.~\eqref{eq:RicciId_Psi4} can be written as
\begin{subequations}
\label{eq:Bianchi_simplified_psi4}
\begin{align} 
    & F_3\Psi_4-J_3\Psi_3+3\nu\Psi_2=S_3\,, \label{eq:BianchiId_Psi4_1_simplify} \\
    & F_4\Psi_4-J_4\Psi_3+3\lambda\Psi_2=S_4\,, \label{eq:BianchiId_Psi4_2_simplify} \\
    & E_4\lambda-E_3\nu+\Psi_4=0 \label{eq:RicciId_Psi4_simplify}\,.
\end{align}
\end{subequations}
For this work, we also need a commutator of the intrinsic derivatives introduced in Eq.~\eqref{eq:derivative}, namely,
\begin{align}
\label{eq:commutator}
    [\delta,D]
    =& \;\left(\alpha^*+\beta-\pi^*\right)D
    +\kappa\Delta-\left(\rho^*+\varepsilon-\varepsilon^*\right)\delta
    \nonumber\\
    & \;-\sigma\delta^*\,.
\end{align}
The other commutators of intrinsic derivatives can be found in Appendix~\ref{appendix:NPformalism_addition}. 

Let us conclude with a brief discussion of the Petrov classification~\cite{Petrov:2000bs,Chandrasekhar_1983}. The Petrov classification is an organizational scheme based on the examination of the algebraic structure of the Weyl curvature tensor. Since the Weyl scalars in the NP formalism depend on the Weyl tensor [see e.g., Eq.~\eqref{eq:Weylscalar_NP}], one can classify solutions in a given theory based on the vanishing of the Weyl scalars for the given solution. The classification is as follows:
\begin{enumerate}
    \item Type I: $\Psi_0 = \Psi_4 = 0$.
    \item Type II: $\Psi_0 = \Psi_1 = \Psi_4 = 0$.
    \item Type D: $\Psi_0 = \Psi_1 = \Psi_3 = \Psi_4 = 0$.
    \item Type III: $\Psi_0 = \Psi_1 = \Psi_2 = \Psi_4 = 0$.
    \item Type N: $\Psi_0 = \Psi_1 = \Psi_2 = \Psi_3 = 0$.
\end{enumerate}
Isolated stationary BHs in GR are of Petrov type D, while these BHs in modified gravity theories, such as in dCS gravity or EdGB gravity, are of Petrov type I~\cite{Owen:2021eez,Alexander:2009tp}. Since Petrov type I spacetimes are the most general type of spacetime in the Petrov classification, they are also called algebraically general. The rest of the spacetimes in the Petrov classification, including Petrov type D, are classified as algebraically special.

\subsection{Teukolsky formalism for Petrov type D spacetimes in GR} \label{sec:TeukGR}

In this subsection, we present the formalism first prescribed by Teukolsky in 1972~\cite{Teukolsky:1973ha}, where using the NP formalism, he obtained a set of separable, decoupled gravitational perturbation equations for Kerr BHs in GR. More specifically, Teukolsky expanded all curvature quantities into a background plus a perturbation; for example, the Weyl scalars are expanded into
\begin{equation}
    \Psi_i = \Psi_i^{(0)} + \epsilon \; \Psi_i^{(1)}
\end{equation}
for $i \in (0,1,2,3,4)$, where the superscript ${(0)}$ means that these quantities are computed from the background metric, while the superscript $(1)$ stands for a perturbation from this background with $\epsilon$ an order-counting parameter. With this in hand, Teukolsky was then able to derive separable and decoupled equations for the curvature perturbations $\Psi_0^{(1)}$ and $\Psi_4^{(1)}$ of a Kerr BH.

The following derivation, which follows closely that of~\cite{Teukolsky:1973ha}, applies to any Petrov type D vacuum background metric in GR, which includes the Schwarzschild and Kerr metrics. Let us then choose the $l^\mu$ and $n^\mu$ vectors of the unperturbed tetrad along the repeated principal null directions of the Weyl tensor. Thus, for a Petrov type D vacuum GR spacetime, we have
\begin{align}
\label{eq:bg_values_typeD}
    \Psi_0^{(0)} = \Psi_1^{(0)} =& \Psi_3^{(0)} = \Psi_4^{(0)} = 0 \,, \nonumber \\
    \kappa^{(0)} = \sigma^{(0)} =& \nu^{(0)} = \lambda^{(0)} = 0 \,.
\end{align}
The result on the second line of Eq.~\eqref{eq:bg_values_typeD} can also be seen to come from the Bianchi identities in Eq.~\eqref{eq:bianchi_iden}.

The GR field equations in trace-reversed form can be expressed as
\be
\label{eq:fieldeq-GR}
R^{\mu\nu} = 8 \pi \left( T^{\mu\nu} -\frac{1}{2}T g^{\mu\nu} \right) \,,
\ee
where $T^{\mu\nu}$ is the stress-energy tensor and $T$ is its trace. Since we are working with vacuum spacetimes, $T^{\mu\nu} = 0$, and thus $R^{\mu\nu} = 0 $. Using this in Eq.~\eqref{eq:Ricci_NP}, we can see that all background and perturbed values of $\Phi_{ij}$ for $i,j \in \{0,1,2\}$ vanish. For instance,
\be \label{eq:eg_phi00}
\Phi_{00} \equiv -\frac{1}{2}R_{11} = -\frac{1}{2}R_{\mu\nu} l^\mu l^\nu = 4 \pi T_{ll} = 0 \,.
\ee
Thus, using Eq.~\eqref{eq:source_bianchi}, we see that $S_1, S_2, S_3 ~\rm{and}~ S_4$ vanish identically for vacuum GR spacetimes. 

To study the perturbations of BHs, we require differential equations for $\Psi_0^{(1)}$ and $\Psi_4^{(1)}$ since these represent curvature perturbations associated with propagating metric perturbations. We first present the formalism to obtain a differential equation for $\Psi_0^{(1)}$, and later, we apply the same to $\Psi_4^{(1)}$. Consider then the vacuum Ricci identity of Eq.~\eqref{eq:RicciId_Psi0_simplify} and the Bianchi identities in Eqs.~\eqref{eq:BianchiId_Psi0_1_simplify} and~\eqref{eq:BianchiId_Psi0_2_simplify}. As mentioned previously, in vacuum GR spacetimes, the right-hand side of these equations vanish. Furthermore, using Eq.~\eqref{eq:bg_values_typeD}, the corresponding perturbation equations to leading order in the perturbation take the form
\begin{subequations} \label{eq:teukpert}
\begin{align}
\begin{split} \label{eq:teuk1pert}
    F_1^{(0)}\Psi_{0}^{(1)} - J_1^{(0)} \Psi_{1}^{(1)} -3\kappa^{(1)}\Psi_{2}^{(0)}&=0\,, 
\end{split} \\
\begin{split} \label{eq:teuk2pert}
    F_2^{(0)}\Psi_{0}^{(1)} - J_2^{(0)}\Psi_{1}^{(1)}  -3\sigma^{(1)} \Psi_{2}^{(0)}&=0\,, 
\end{split} \\ 
\begin{split} \label{eq:teuk3pert}
    E_2^{(0)}\sigma^{(1)} - E_1^{(0)}\kappa^{(1)}-\Psi_{0}^{(1)}&=0\,.
\end{split}
\end{align}
\end{subequations}
In order to simplify the notation, we will drop the superscript $(0)$ for all background quantities for the remainder of this section. Multiplying Eq.~\eqref{eq:teuk3pert} by the background $\Psi_2$ Weyl scalar and plugging in for $E_1$ and $E_2$ using Eq.~\eqref{eq:simplify_operators_0}, one finds
\begin{equation}
\begin{split} \label{eq:teuk3apert}
    \left(D-4\rho-\rho^{*} - 3\varepsilon+\varepsilon^{*}\right)\left(\Psi_2\sigma^{(1)}\right) -(\delta-4\tau+\pi^{*}   \\
    -\alpha^{*} -3\beta) \left(\Psi_2\kappa^{(1)}\right)-\Psi_2\Psi_{0}^{(1)}=0\,,
\end{split}
\end{equation}
where we have used Eqs.~\eqref{eq:bianchi_dpsi2} and \eqref{eq:bianchi_delpsi2}, which for the background $\Psi_2$ reduce to
\begin{equation} \label{eq:GRbianchi}
    D\Psi_2 = 3 \rho \Psi_2, \qquad 
    \delta\Psi_2 = 3 \tau \Psi_2 \,.
\end{equation}
In order to be consistent with the simplified notation, we introduce
\begin{subequations} \label{eq:commutator_simplify}
\begin{align} 
    E_1^{\rm{\GR}} &= \delta-4\tau+\pi^{*} -\alpha^{*} -3\beta \,, \\
    E_2^{\rm{\GR}} &= D-4\rho-\rho^{*} - 3\varepsilon+\varepsilon^{*} \,,
\end{align}
\end{subequations}
so Eq.~\eqref{eq:teuk3apert} can be written more compactly as
\begin{equation}
    E_2^{\rm{\GR}} \left(\Psi_2 \sigma^{(1)}\right) - E_1^{\rm{\GR}} \left( \Psi_2 \kappa^{(1)}\right) = \Psi_2 \Psi_0^{(1)}\,.
\end{equation}

To obtain a differential equation for $\Psi_0^{(1)}$, we need to eliminate $\Psi_1^{(1)}$ from Eqs.~\eqref{eq:teuk1pert} and \eqref{eq:teuk2pert}. This can be done by making use of the following commutation relation. 
\begin{align} \label{eq:commutatorgr}
        E_2^{\rm{\GR}} J_2 - E_1^{\rm{\GR}} J_1 = 0 \,.
\end{align}
This relation can be shown to hold for any Petrov type D spacetime in GR by using Eqs.~\eqref{eq:Riccieq_Dtau}-\eqref{eq:Riccieq_Drho} and Eq.~\eqref{eq:commutator}. 
On operating $E_2^{\rm{\GR}}$ on Eq.~\eqref{eq:teuk2pert}, $E_1^{\rm{\GR}}$ on Eq.~\eqref{eq:teuk1pert}, and subtracting one equation from the other, $\Psi_1^{(1)}$ vanishes identically. Using Eq.~\eqref{eq:teuk3apert}, we finally have
\be \label{eq:teukpsi0}
\left( E_2^{\rm{\GR}}F_2 - E_1^{\rm{\GR}} F_1 - 3 \Psi_2 \right) \Psi_0^{(1)} = 0 \,.
\ee
This is the decoupled equation for $\Psi_0^{(1)}$ for any Petrov type D vacuum spacetime in GR. As shown by Geroch, Held, and Penrose (GHP)~\cite{Geroch_Held_Penrose_1973}, the NP equations are invariant under the exchange $l^\mu \leftrightarrow n^\mu$ and $m^\mu \leftrightarrow \Bar{m}^\mu $, where the choice of $l^\mu$ and $n^\mu$ has no effect on this symmetry. Applying this transformation to Eq.~\eqref{eq:teukpsi0}, one finds the decoupled differential equation for $\Psi_4^{(1)}$ for a Petrov type D vacuum spacetime in GR, namely,
\begin{align} \label{eq:teukpsi4}
    \left(E_4^{\rm{\GR}} F_4 - E_3^{\rm{\GR}} F_3 
	-3\Psi_2 \right) \Psi_{4}^{(1)}=0 \,,
\end{align}
where we have introduced
\begin{align} \label{eq:teukpsi4-new}
    E_3^{\rm{\GR}} &\equiv \delta^{*}+3\alpha +\beta^{*}+4\pi-\tau^{*}\,,
\nonumber \\
    E_4^{\rm{\GR}} &\equiv \Delta +4\mu+\mu^{*} + 3 \gamma-\gamma^{*}\,.
\end{align}
An alternate derivation using the GHP formalism was provided by Stewart~\cite{Stewart:1974uz}. However, for the purpose of this section, we will stick with the formalism laid down by Teukolsky.

%%%%%%%%%%%%%%%%%%%%%%%%%%%%%%%%%%%%%%

\subsection{Chandrasekhar's approach for Petrov type D spacetimes in GR} 
\label{sec:gauge_GR}

Chandrasekhar introduced another way to derive the Teukolsky equation in~\cite{Chandrasekhar_1983} by utilizing the gauge freedom of the tetrad. As briefly mentioned in Sec.~\ref{sec:NPformalism} and discussed in detail in Appendix~\ref{appendix:NPformalism_addition}, one is free to rotate the tetrad following Eq.~\eqref{eq:tetrad_rotations} such that all the normalization and orthogonality conditions in Eq.~\eqref{eq:tetrad_ortho} are preserved.

Let us then consider a type \rom{2} rotation, which is given by 
\begin{align} 
\label{eq:rotate2-first}
    & \;n\rightarrow n\,,\; m\rightarrow m+bn\,,\;
    \bar{m}\rightarrow \bar{m}+b^*n\,,\; \nonumber\\
    & \;l\rightarrow l+b^*m+b\bar{m}+bb^*n
\end{align} 
[see also Eq.~\eqref{eq:rotate2}], and set the rotation parameter $b$ to be of leading order in the perturbation, i.e., $b=b^{(1)}$. Ignoring all higher-order terms, the perturbed Weyl scalars transform into [see e.g., Eq.~\eqref{eq:change2}]
\begin{equation} \label{eq:rotate2_reduce_typeD}
	\begin{array}{l}
		\Psi_{0}^{(1)}\rightarrow\Psi_{0}^{(1)}+4b^{(1)}\Psi_{1}^{(0)}\,,\;
		\Psi_{1}^{(1)}\rightarrow\Psi_{1}^{(1)}+3b^{(1)}\Psi_{2}^{(0)}\,, \\
		\Psi_{2}^{(1)}\rightarrow\Psi_{2}^{(1)}+2b^{(1)}\Psi_{3}^{(0)}\,,\; 
		\Psi_{3}^{(1)}\rightarrow\Psi_{3}^{(1)}+b^{(1)}\Psi_{4}^{(0)}\,, \\ 
		\Psi_{4}^{(1)}\rightarrow\Psi_{4}^{(1)}\,.
	\end{array} 
\end{equation}
Since for a Petrov type D spacetime, $\Psi_{i\neq2}^{(0)}=0$, all the $\Psi_{i\neq1}^{(1)}$ remain invariant under such a rotation. 
By choosing $b^{(1)}=-\Psi_1^{(1)}/\left(3\Psi_2^{(0)}\right)$, the perturbed Weyl scalar $\Psi_1^{(1)}$ can be removed directly without the use of any additional Bianchi identities and commutation relations used in Sec.~\ref{sec:TeukGR}. Another way to understand this gauge choice is that we have three equations for four unknowns in Eqs.~\eqref{eq:teukpert}, so there is one arbitrary function to be determined. 

Using this gauge freedom to set $\Psi_1^{(1)}=0$ through a tetrad rotation, one can now easily derive the Teukolsky equation. First, use this gauge freedom to set $\Psi_1^{(1)}=0$ in Eqs.~\eqref{eq:teuk1pert}-\eqref{eq:teuk2pert}, and then solve for $\kappa^{(1)}$ and $\sigma^{(1)}$. Now insert these solutions back into Eq.~\eqref{eq:teuk3pert} to find 
\begin{equation} \label{eq:3_to_1_typeD}
    \begin{aligned}
        & \left(\mathcal{E}_2F_2 - \mathcal{E}_1 F_1
        -3\Psi_2\right)\Psi_0^{(1)}=0\,,
    \end{aligned}
\end{equation}
where we have defined 
\begin{equation} \label{eq:def_tile_E_operators}
     \mathcal{E}_i\equiv\Psi_2 E_i\Psi_2^{-1}\,.
\end{equation}
Here, we have dropped the superscript $(0)$ for all unperturbed quantities. Applying the GHP transformation explained below Eq.~\eqref{eq:teukpsi0}, one finds an equation for $\Psi_4^{(1)}$, namely,
\begin{equation} \label{eq:3_to_1_typeD_Psi4}
    \begin{aligned}
        & \left(\mathcal{E}_4F_4-\mathcal{E}_3F_3
        -3\Psi_2\right)\Psi_4^{(1)}=0\,.
    \end{aligned}
\end{equation}
The $\mathcal{E}_{i}$ operators can be simplified using the product rule. Doing so, one finds
\begin{subequations} \label{eq:tile_E_operators}
    \begin{align}
        & \mathcal{E}_1=\delta-\tau+\pi^{*}-\alpha^{*}
        -3\beta-\frac{1}{\Psi_2}\delta\Psi_2 \,,  \\
        & \mathcal{E}_2=D-\rho-\rho^{*}-3\varepsilon
        +\varepsilon^{*}-\frac{1}{\Psi_2}D\Psi_2\,, \\
        & \mathcal{E}_3=\delta^{*}+3\alpha+\beta^{*} +\pi-\tau^{*}-\frac{1}{\Psi_2}\delta^*\Psi_2\,, \\
        & \mathcal{E}_4=\Delta+\mu+\mu^{*}+3\gamma-\gamma^*
        -\frac{1}{\Psi_2}\Delta\Psi_2\,,
    \end{align}
\end{subequations}
In deriving  Eqs.~\eqref{eq:3_to_1_typeD}-\eqref{eq:3_to_1_typeD_Psi4}, we have also multiplied the whole equation by $3\Psi_2$.
We will see in Sec.~\ref{sec:gauge_nonGR} that this makes Eqs.~\eqref{eq:3_to_1_typeD}-\eqref{eq:3_to_1_typeD_Psi4} exactly the same as Eqs.~\eqref{eq:teukpsi0}-\eqref{eq:teukpsi4}, so $\mathcal{E}_i=E_i^{\rm{\GR}}$. Note that one can also derive the equation for $\Psi_4$ in the same way we derived an equation for $\Psi_0$ (i.e., without the GHP transformation), using the fact that a type \rom{1} rotation at $\mathcal{O}(\epsilon)$ can be used to set $\Psi_3^{(1)}$ to zero. 

It should not be surprising that one obtains the same equation following the traditional Teukolsky's approach and Chandrasekhar's approach. From Eq.~\eqref{eq:rotate2_reduce_typeD} and other tetrad rotations discussed in Appendix~\ref{appendix:NPformalism_addition} that one can perform in Eqs.~\eqref{eq:tetrad_rotations}, one can see that $\Psi_0^{(1)}$ and $\Psi_4^{(1)}$ are gauge-invariant quantities under linear perturbations. In Chandrasekhar's approach, since one does not need to use any additional Bianchi identities and commutation relations to cancel off $\Psi_1^{(1)}$, there are fewer equations one needs to worry about, and this will be helpful when dealing with the more complicated non-Petrov-type-D spacetime backgrounds of modified gravity theories. However, to convince ourselves that the equivalence between these two approaches is not broken when considering beyond GR theories, in Sec.~\ref{sec:nonGR} we will find a modified master equation using both approaches and show that the two methods are equivalent in modified gravity theories.

%%%%%%%%%%%%%%%%%%%%%%%%%%%%%%%%%%%%%%

\section{Framework of Perturbation in Modified Gravity Theories}
\label{sec:perturbation_framework}
In this section, we discuss a subset of modified gravity theories that the formalism developed in this work can be applied to. We classify these theories into two classes based on the presence of additional non-metric fields in the action that define these theories. For both classes, we provide some examples by explicitly writing down the Lagrangian, the equations of motion for all the fields, and the properties of BH spacetimes, which serve as the background to our perturbation analysis. We then prescribe a perturbation scheme using a two-parameter expansion for both classes of modified gravity theories.

\subsection{Theories of gravity beyond GR}
\label{sec:beyondGR-theory}

In this subsection, we provide a quick overview of certain modified theories of gravity relevant for this work and discuss the BH spacetimes in these theories, which serve as a background for our perturbation scheme. Consider then a class of theories defined through the following beyond GR Lagrangian:
\be
\label{eq:lagrangian}
\mathcal{L} = \mathcal{L}_{\rm{\GR}} + \ell^p \mathcal{L}_{\rm{\bGR}} + \mathcal{L}_{\rm{\matter}} + \mathcal{L}_{\rm{\field}} \,,
\ee
where $\mathcal{L}_{\rm{\GR}}$ is the Einstein-Hilbert Lagrangian, $\mathcal{L}_{\rm{\matter}}$ is the matter Lagrangian, $\mathcal{L}_{\rm{\field}}$ is the Lagrangian for all other (non-metric) dynamical fields (including all kinetic and potential terms of these fields) that the theory may permit, and $\mathcal{L}_{\rm{\bGR}}$ is a Lagrangian that contains non-Einstein-Hilbert curvature terms and can, in principle, include non-minimal couplings to the non-metric dynamical fields of the theory. The quantity $\ell$ in Eq.~\eqref{eq:lagrangian} is a dimension-full scale that characterizes the strength of the GR correction, and $p$ is a number to ensure that $\ell^p {\cal{L}}_{\rm{\bGR}}$ has the right dimensions. We can classify the beyond GR theories described by the Lagrangian in Eq.~\eqref{eq:lagrangian} based on the presence or absence of additional non-metric dynamical fields, i.e., based on whether $\mathcal{L}_{\rm{\field}}$ vanishes. Note that we here do not consider theories with non-dynamical, prior or ``fixed'' fields that couple to the metric tensor. In this work then, we define this classification as:
\begin{itemize}
\centering
    \item $ \mathcal{L}_{\rm{\field}} \neq 0 \Longrightarrow$ Class A,
    \item $ \mathcal{L}_{\rm{\field}} = 0 \Longrightarrow$ Class B. 
\end{itemize}

An example of beyond GR theories of class A that we will consider is dCS gravity. This theory is defined by the Lagrangian in Eq.~\eqref{eq:lagrangian} with the choices
\begin{align}
    &\mathcal{L}_{\rm{\GR}}=(16\pi)^{-1}R\,, \nonumber \\  
    &\mathcal{L}_{\rm{\bGR}}^\dCS=\frac{1}{4}\vartheta ~^*\! R^{\mu}{}_{\nu}{}^{\kappa\delta}R^\nu{}_{\mu\kappa\delta}\,, \nonumber \\
    &\mathcal{L}_{\rm{\field}}^\dCS
    =-\frac{1}{2}g^{\mu\nu}(\nabla_\mu\vartheta)(\nabla_\nu\vartheta)\,,
\end{align}
and $\ell = \ell_{\dCS}$ is the dCS coupling constant with $p = 2$. $R$ is the Ricci scalar, $g_{\mu\nu}$ is the metric, and $\vartheta$ is a massless, pseudoscalar, axion-like field that non-minimally couples to the Pontryagin curvature invariant $^*\! R^{\mu}{}_{\nu}{}^{\kappa\delta} R^\nu{}_{\mu \kappa \delta}$, where 
\begin{equation}
    ^*\! R^{\mu}{}_{\nu}{}^{\kappa\delta} = \frac{1}{2}\epsilon^\mu{}_{\nu\alpha\beta} R^{\alpha\beta\kappa\delta}
\end{equation}
is the dual of the Riemann tensor.
The field equations in dCS gravity are
\begin{align}
\label{eq:tracericci}
R_{\mu \nu} &= 8\pi \Big\{ (T_{\mu\nu}^{\rm{\M}}-\frac{1}{2}g_{\mu\nu}T^{\rm{\M}}) + (\nabla_\mu \vartheta)(\nabla_\nu \vartheta) \nonumber \\
& - 2\alpha_\dCS \left[ (\nabla_\sigma \vartheta)\epsilon^{\sigma \delta \alpha}{}_{(\mu}\nabla_\alpha R_{\nu) \delta} 
+ (\nabla_\sigma \nabla_\delta \vartheta)^*\!R^\delta{}_{(\mu \nu)}{}^\sigma \right] \Big\} \,, \\
\label{eq:sfeqdcs}
\Box \vartheta &= -\frac{\alpha_\dCS}{4}  ~^*\! R^{\mu}{}_{\nu}{}^{\kappa\delta} R^\nu{}_{\mu \kappa \delta}\,,
\end{align}
where Eq.~\eqref{eq:tracericci} is the trace-reversed metric field equation, and Eq.~\eqref{eq:sfeqdcs} is the scalar field equation. The dCS coupling constant $\alpha_\dCS \equiv \ell_\dCS^2$ determines the strength of the Chern-Simons (CS) modification and has dimensions of $[\rm{Length}]^2$. Stationary and vacuum BH solutions in this theory are not Ricci-flat, so they are obviously not represented by the Kerr metric \cite{Yunes_Pretorius_2009,Yagi:2012ya,Delsate:2018ome}. Instead, spinning BHs in dCS gravity have a corrected event horizon location, ergosphere, and different exterior multipole moments~\cite{Yunes_Pretorius_2009} to name a few corrected quantities. Moreover, dCS BHs are of non-Ricci-flat Petrov type \rom{1} spacetimes in the Petrov classification given in Sec.~\ref{sec:NPformalism}. To leading order in spin, however, the BHs in this theory remain non-Ricci-flat and of Petrov type D~\cite{Owen:2021eez,Yunes:2007ss,Yunes_Pretorius_2009}.

Another example of a class A beyond GR theory is EdGB gravity \cite{Kanti:1995vq}, which is a special case of sGB gravity \cite{Witek:2018dmd}. Using Eq.~\eqref{eq:lagrangian} and the conventions in \cite{Witek:2018dmd, Pierini:2021jxd}, EdGB theory is defined via
\begin{align} \label{eq:lag-sgb}
    & \mathcal{L}_{\rm{\GR}}=(16\pi)^{-1} R\,, \nonumber \\  
    & \mathcal{L}_{\rm{\bGR}}^\EdGB
    =(64\pi)^{-1}e^{\theta}\mathcal{G}\,, \nonumber \\
    & \mathcal{L}_{\rm{\field}}^\EdGB
    =-(32\pi)^{-1}g^{\mu\nu}(\nabla_\mu\theta) (\nabla_\nu\theta)\,,
\end{align}
where 
\begin{equation}
    \mathcal{G}
    =R^{\mu\nu\rho\sigma}R_{\mu\nu\rho\sigma}
    -4R^{\mu\nu}R_{\mu\nu}+R^2
\end{equation}
is the Gauss-Bonnet curvature invariant, and $\ell = \ell_{\EdGB}$ is the EdGB coupling constant with $p=2$. The quantity $\theta$ is a massless dilaton-like scalar field that non-minimally couples to the Gauss-Bonnet invariant $\mathcal{G}$. The metric field equation for EdGB gravity in trace-reversed form is then given by \cite{Pierini:2021jxd}
\begin{align}
\label{eq:tracericci-sgb}
    R_{\mu \nu} 
    &= 8\pi(T_{\mu\nu}^{\rm{\M}}-\frac{1}{2}g_{\mu\nu}T^{\rm{\M}})
    +\frac{1}{2}(\nabla_\mu\theta)(\nabla_\nu\theta) \nonumber \\
    & -\alpha_\EdGB\left(\mathcal{K}_{\mu\nu}
    -\frac{1}{2}g_{\mu\nu}\mathcal{K}\right)\,, \nonumber\\
    \mathcal{K}_{\mu\nu}
    &=\frac{1}{8}\left(g_{\mu\rho}g_{\nu\sigma}
    +g_{\mu\sigma}g_{\nu\rho}\right)\epsilon^{\delta\sigma\gamma\alpha} \nabla_\beta\left(^*\!R^{\rho\beta}{}_{\gamma\alpha}
    e^\theta\nabla_\delta\theta\right)\,, \nonumber\\
    \mathcal{K}&=g_{\mu\nu}\mathcal{K}^{\mu\nu}\,, 
\end{align}
whereas the scalar field equation is
\begin{equation} \label{eq:EdGB_scalar_EOM}
    \Box\theta=-\frac{\alpha_\EdGB}{4}e^{\theta}\mathcal{G} \,.
\end{equation}
The quantity $\alpha_\EdGB \equiv l_\EdGB^2$ is the coupling constant of EdGB theory and has dimensions of $[\rm{Length}]^2$. Stationary and vacuum BH solutions in this theory, just like in dCS gravity, are non-Ricci-flat and are not represented by the Kerr metric~\cite{Kleihaus:2015aje,Yunes:2011we,Kleihaus:2011tg,Maselli:2015tta,Sullivan:2020zpf}. Rotating BHs in EdGB theory are described by non-Ricci-flat Petrov type \rom{1} spacetimes in general, but to leading order in spin, they are described by non-Ricci-flat Petrov type D spacetimes~\cite{Owen:2021eez}.

An example of class B beyond GR theories is higher-derivative gravity~\cite{Cano_Fransen_Hertog_Maenaut_2021} because this theory contains no non-metric dynamical fields. Following Eq.~\eqref{eq:lagrangian}, the Lagrangian of this theory can be represented by
\begin{align}
    \mathcal{L}_{\GR}
    =& \;(16\pi)^{-1}R\,, \nonumber\\
    \mathcal{L}^{\HD}_{\bGR}
    =& \;(16\pi)^{-1}(\lambda_{\even}R_{\mu\nu}{ }^{\rho\sigma}
    R_{\rho\sigma}{ }^{\delta\gamma}R_{\delta\gamma}{ }^{\mu\nu} \nonumber\\
    & \;+\lambda_{\odd}R_{\mu\nu}{ }^{\rho\sigma}
    R_{\rho\sigma}{ }^{\delta\gamma}~^*\!R_{\delta\gamma}{ }^{\mu\nu})\,, \nonumber\\
    \mathcal{L}^{\HD}_{\field}=&\;0\,,
\end{align}
where we have only kept terms with up to six derivatives of the metric (a more general discussion can be found in \cite{Cano_Fransen_Hertog_Maenaut_2021}). $\ell=\ell_{\HD}$ is the higher-derivative gravity coupling constant with $p=4$. The quantities $\lambda_{\even}$ and $\lambda_{\odd}$ are dimensionless coupling constants that are introduced to distinguish terms that preserve or break parity. The field equation in trace-reversed form is \cite{Cano_Fransen_Hertog_Maenaut_2021}
\begin{align} \label{eq:tracericci-highder}
    R_{\mu\nu}
    =& \;8\pi(T_{\mu\nu}^{\rm{\M}}
    -\frac{1}{2}g_{\mu\nu}T^{\rm{\M}})
    -\mathcal{E}_{\mu\nu}^{(6)}\,, \nonumber \\
    \mathcal{E}_{\mu\nu}^{(n)}
    =& \;P^{(n)}{ }_{(\mu}{ }^{\rho\sigma\gamma} R_{\nu)\rho\sigma\gamma}
    -\frac{1}{2}g_{\mu\nu}\mathcal{L}_{(n)}
    +2\nabla^{\sigma}\nabla^{\rho}
    P_{(\mu|\sigma|\nu)\rho}^{(n)}\,, \nonumber\\
    P_{\mu\nu\rho\sigma}^{(6)} 
    =& \;3\alpha^{\even}_{\HD}R_{\mu\rho}^{\alpha\beta} R_{\alpha\beta\rho\sigma} \nonumber\\
    +& \;\frac{3\alpha^{\odd}_{\HD}}{2}
    \left(R_{\mu\rho}^{\alpha\beta}~^*\!R_{\alpha\beta\rho\sigma}
    +R_{\mu\rho}^{\alpha\beta}
    ~^*\!R_{\rho\sigma\alpha\beta}\right)\,, 
\end{align}
where $\alpha^{\even}_{\HD}\equiv\ell_{\HD}^4\lambda_{\even}$ and $\alpha^{\odd}_{\HD}\equiv\ell_{\HD}^4\lambda_{\odd}$ are coupling constants that determine the strength of the parity-preserving and the parity-breaking higher-derivative gravity corrections. The quantity $\mathcal{L}_{(n)}$ refers to the Lagrangian with $n$ derivatives of the metric in higher-derivative gravity, so $\mathcal{L}_{(6)}=\mathcal{L}_{\bGR}^{\HD}$. Rotating BHs in higher-derivative gravity are non-Ricci-flat~\cite{Cano_Fransen_Hertog_Maenaut_2021}, but their Petrov type has not yet been studied in detail.

Theories described by the Lagrangian given in Eq.~\eqref{eq:lagrangian} only form a subset of all possible theories. This subset does not just include dCS gravity~\cite{Jackiw:2003pm,Alexander:2009tp}, EdGB gravity~\cite{Gross_Sloan_1987,Kanti:1995vq,Moura:2006pz,Nojiri:2010wj,Kleihaus:2011tg}, and higher-derivative theories of gravity~\cite{Burgess_2004, Donoghue_2012, Endlich_Gorbenko_Huang_Senatore_2017, Cano_Ruiperez_2019, Cano:2020cao, Cano_Fransen_Hertog_Maenaut_2021}, but it also includes, for example, sGB gravity in general \cite{Sotiriou:2013qea}, quadratic gravity theories without additionally coupled fields~\cite{Sotiriou:2008rp,Clifton:2011jh}, and higher dimensional gravity theories~\cite{Dvali:2000hr,deRham:2010kj} to name a few. These theories can also be classified based on whether their stationary and vacuum (i.e., no matter) BH solutions are Ricci-flat or non-Ricci-flat. For a beyond GR theory that admits Ricci-flat, Petrov type D BH spacetimes, perturbations can be studied within the standard Teukolsky formalism presented in Sec.~\ref{sec:TeukGR}, so we do not focus on these theories here. In this work, instead, we focus on the dynamical perturbations of BHs that are non-Ricci-flat and either Petrov type D or Petrov type \rom{1}. Therefore, our work applies to dCS gravity~\cite{Jackiw:2003pm,Alexander:2009tp,Owen:2021eez,Yunes:2007ss}, EdGB and sGB gravity~\cite{Owen:2021eez}, and higher-derivative gravity \cite{Burgess_2004, Donoghue_2012, Endlich_Gorbenko_Huang_Senatore_2017, Cano_Ruiperez_2019, Cano:2020cao, Cano_Fransen_Hertog_Maenaut_2021}.

\subsection{Perturbation Scheme}
\label{sec:perturbation_scheme}
In this subsection, we discuss the perturbation scheme that is applicable to the modified gravity theories discussed in Sec.~\ref{sec:beyondGR-theory}. To solve for the dynamical gravitational perturbations of a BH background in any such modified gravity theory perturbatively, we need a multi-variable expansion of all NP quantities. Generalizing the discussion in \cite{Okounkova:2019dfo} for dCS gravity to any modified gravity theory that can be studied perturbatively (in an effective field theory approach), we need at least two expansion parameters\footnote{Note that in \cite{Okounkova:2019dfo}, $\epsilon$ is used for the strength of the correction to GR, and $\zeta$ is used for the size of GW perturbations, which is opposite to our choices here. We here choose to remain consistent with previous literature in GR~\cite{Regge:PhysRev.108.1063,Chandrasekhar_1983} and in dCS gravity~\cite{Alexander:2009tp,Yunes_Pretorius_2009}}: 
\begin{itemize}
    \item[(i)] $\zeta$, a dimensionless parameter that characterizes the strength of the correction to GR (which typically will depend on the ratio of the scale $\ell$ to the BH mass), and 
    \item[(ii)] $\epsilon$, a dimensionless parameter that describes the size of the GW perturbations, which also appears in GR. 
\end{itemize}
In this work, we additionally impose that $\zeta$ is the leading order at which beyond GR corrections to the metric field $h_{\mu\nu}^{\bGR}$ appear, while the leading-order correction to other non-metric fields may enter with other (possibly lower) powers of $\zeta$. 

In order to understand the coupling constant $\zeta$ better, let us first relate it to the coupling constants of the different modified gravity theories we used as examples in Sec.~\ref{sec:beyondGR-theory}. For class A beyond GR theories with non-minimal coupling, the extra non-metric fields $\vartheta_{\bGR}$, e.g., $\vartheta$ in dCS gravity and $\theta$ in EdGB gravity, are sourced by the metric field and are proportional to terms of $\mathcal{O}(\alpha_{\bGR})$, where $\alpha_{\bGR}$ is the coupling constant associated with $\mathcal{L}_{\bGR}$ in Eq.~\eqref{eq:lagrangian}, e.g., $\alpha_{\dCS}$ in dCS gravity and $\alpha_{\EdGB}$ in EdGB gravity. The field $\vartheta_{\bGR}$ then back-reacts onto the metric and sources the metric perturbations $h_{\mu\nu}^{\bGR}$, which are also multiplied by a factor of $\alpha_\bGR$. Thus, to leading order, $\vartheta_{\bGR}\sim\alpha_\bGR$ and $h_{\mu\nu}^{\bGR}\sim \alpha_\bGR\vartheta_{\bGR}$, so $\zeta\sim\alpha_\bGR^2$. This is evident from Eqs.~\eqref{eq:tracericci} and ~\eqref{eq:tracericci-sgb}, where $\zeta\sim\alpha_{\dCS}^2$ for dCS gravity and $\zeta\sim\alpha_{\EdGB}^2$ for EdGB gravity. For class B beyond GR theories, the metric perturbations are driven by the metric fields at lower order and are proportional to $\alpha_\bGR$, so $\zeta\sim\alpha_\bGR$. For example, from Eq.~\eqref{eq:tracericci-highder}, one can see that $\zeta\sim\alpha_{\HD}^{\even,\odd}$. 

By requiring that $h_{\mu\nu}^{\bGR}$ enters at $\mathcal{O}(\zeta)$, $R_{\mu\nu}$ must also enter at $\mathcal{O}(\zeta)$ since we are focusing on background spacetimes that are perturbed from the vacuum solutions in GR. This can be seen in Eqs.~\eqref{eq:tracericci}, \eqref{eq:tracericci-sgb}, and \eqref{eq:tracericci-highder}. In addition, for both classes of beyond GR theories, since metric perturbations in modified gravity are sourced by the metric field in GR either indirectly via extra non-metric fields (class A) or directly (class B), the leading-order terms of the metric field in $R_{\mu\nu}$ must be of $\mathcal{O}(\zeta^0)$. Thus, when computing $R_{\mu\nu}$, we only need the metric at $\mathcal{O}(\zeta^0,\epsilon^0)$ or $\mathcal{O}(\zeta^0,\epsilon^1)$. The perturbative order of $R_{\mu\nu}$ and the metric field in it will be important when we discuss the decoupling of the modified Teukolsky equation in Sec.~\ref{sec:TeuknonGR} and Sec.~\ref{sec:The_modified_Teukolsky_equation}.

Besides the metric field, we also have the NP quantities (i.e., tetrad basis vectors, Weyl scalars, spin coefficients, and NP Ricci scalars) generated from it. Although the beyond GR correction to the metric field enters at $\mathcal{O}(\zeta)$, the beyond GR correction to the NP quantities does not necessarily enter at $\mathcal{O}(\zeta)$ if we make certain gauge choices on some NP quantities, which will be discussed in detail in Sec.~\ref{sec:gauge_background}. For simplicity, we want all the NP quantities to have the same expansion pattern as the metric field, so here we construct a NP tetrad which is corrected by beyond GR theories at $\mathcal{O}(\zeta)$ to leading order. Thus, all the other NP quantities are naturally corrected by modified gravity at $\mathcal{O}(\zeta)$ to leading order.

In order to ensure that all the NP quantities will be corrected at ${\cal{O}}(\zeta)$, we must find a tetrad that shared this same property, namely,
\begin{equation}
    \label{eq:pert-tetrad}
    e_{a\mu} = e_{a\mu}^{(0,0)}+\zeta\delta e_{a\mu}^{(1,0)}\,,
\end{equation}
where $\delta e_{a\mu}^{(1,0)}$ is a perturbation of ${\cal{O}}(\zeta^1,\epsilon^0)$ of the Kinnersley tetrad 
$e_{a\mu}^{(0,0)}$. Here, we have used the superscript $(n,m)$ to denote terms at $\mathcal{O}(\zeta^n,\epsilon^m)$. The only constraint on a NP tetrad is the orthogonality condition in Eq.~\eqref{eq:tetrad_ortho}. Let us expand the correction to the Kinnersley tetrad $\delta e_{a\mu}^{(1,0)}$ in terms of the original tetrad $e_{a\mu}^{(0,0)}$ in GR,
\begin{equation} \label{eq:expansion_tetrad}
	\delta e_{a\mu}^{(1,0)}
	=A_{ab}^{(1,0)}e^{\;b(0,0)}_{\mu}\,.
\end{equation}
To satisfy Eq.~\eqref{eq:tetrad_ortho}, we need to have that
\begin{align} \label{eq:tetrad_conditions}
	& \left(e_{a\mu}^{(0,0)}+\zeta\delta e_{a\mu}^{(1,0)}\right)
	\left(e_{b\nu}^{(0,0)}+\zeta\delta e_{b\nu}^{(1,0)}\right)
	\left(g^{\mu\nu(0,0)}+\zeta h^{\mu\nu(1,0)}\right) \nonumber\\ 
	& =\eta_{ab}\,,
\end{align} 
where $\eta_{ab}$ is the metric defined in Eq.~\eqref{eq:ztetrad}, $g^{\mu\nu(0,0)}$ is the metric of the GR background, and $h^{\mu\nu(1,0)}$ represents the modification to the metric due to deviation from GR. Up to $\mathcal{O}(\zeta)$, we can equivalently require that
\begin{align} \label{eq:tetrad_conditions_reduce}
	\delta e_{a\mu}^{(1,0)}e_{b\nu}^{(0,0)}g^{\mu\nu(0,0)}
	+e_{a\mu}^{(0,0)}&\delta e_{b\nu}^{(1,0)}g^{\mu\nu(0,0)}  \nonumber \\
	=&-e_{a\mu}^{(0,0)}e_{b\nu}^{(0,0)}h^{\mu\nu(1,0)}\,,
\end{align}
where we have used the condition $g^{\mu\nu(0,0)}e_{a\mu}^{(0,0)}e_{b\nu}^{(0,0)}=\eta_{ab}$. Inserting the expansion of Eq.~\eqref{eq:expansion_tetrad} in the above condition and using the condition $g^{\mu\nu(0,0)}e_{a\mu}^{(0,0)}e_{b\nu}^{(0,0)}=\eta_{ab}$ again, one finds
\begin{equation} \label{eq:tetrad_conditions_final}
	A_{ab}^{(1,0)}+A_{ba}^{(1,0)}=2A_{(ab)}^{(1,0)}
	=-h_{ab}^{(1,0)}\,,
\end{equation}
where $h_{ab}^{(1,0)}=e_{a\mu}^{(0,0)}e_{b\nu}^{(0,0)}h^{\mu\nu(1,0)}$, and thus $A_{(ab)}^{(1,0)}=-\frac{1}{2}h_{ab}^{(1,0)}$. In general, $A_{ab}^{(1,0)}$ can have $16$ independent components, which can be separated into a symmetric tensor $A_{(ab)}^{(1,0)}$ with $10$ independent components and an antisymmetric tensor $A_{[ab]}^{(1,0)}$ with $6$ independent components. Since Eq.~\eqref{eq:tetrad_conditions_final} does not impose any constraints on $A_{[ab]}^{(1,0)}$, the components of $A_{[ab]}^{(1,0)}$ correspond to $6$ degrees of gauge freedom to further rotate the tetrad. We can choose $A_{[ab]}^{(1,0)}=0$, so the perturbed tetrad is
\begin{equation} \label{eq:A_ab}
	A_{ab}^{(1,0)}=-\frac{1}{2}h_{ab}^{(1,0)}\,,\quad
	\delta e_{a\mu}^{(1,0)}
	=-\frac{1}{2}e_{a\nu}^{(0,0)}h^{\;\nu(1,0)}_{\mu}\,.
\end{equation}

Using the tetrad in Eqs.~\eqref{eq:pert-tetrad} and~\eqref{eq:A_ab}, we are able to expand the metric field and all the NP quantities generated from it with the same perturbative scheme. In this paper, we are interested in linear dynamical perturbations of any Petrov type I stationary spacetime, which itself is a linear deformation of the Kerr metric, so all terms beyond $\mathcal{O}(\zeta^1,\epsilon^1)$ will be ignored. Up to $\mathcal{O}(\zeta^1,\epsilon^1)$, if we use the tetrad in Eqs.~\eqref{eq:pert-tetrad} and \eqref{eq:A_ab}, the Weyl scalars can be expanded as
\begin{align} \label{eq:expansion_Weyl}
	\Psi_i
	&=\Psi_i^{(0)}+\epsilon\Psi_i^{(1)} \nonumber\\
	&=\Psi_i^{(0,0)}+\zeta\Psi_{i}^{(1,0)}+\epsilon\Psi_{i}^{(0,1)}
	+\zeta\epsilon\Psi_{i}^{(1,1)}\,,
\end{align}
and the same expansion applies to the metric field and all the other NP quantities.For the beyond GR theories of class A mentioned in Sec.~\ref{sec:beyondGR-theory}, additional fields may be present. For the examples presented, the pseudoscalar field in dCS gravity can be perturbatively expanded as
\begin{align}
    \vartheta
	=\vartheta^{(0)}+\epsilon\vartheta^{(1)}  = \zeta\vartheta^{(1,0)} + \zeta\epsilon\vartheta^{(1,1)}\,.
\end{align}
A scalar field $\theta$ in EdGB gravity can also be expanded perturbatively in a similar manner. For both $\vartheta$ and $\theta$, the background and perturbed GR pieces vanish. Notice that other work sometimes chooses to expand extra fields starting at $\zeta^0$ \cite{Okounkova_Stein_Scheel_Hemberger_2017, Okounkova_Scheel_Teukolsky_2019, Okounkova:2019dfo, Okounkova_Stein_Moxon_Scheel_Teukolsky_2020} or $\zeta^{\frac{1}{2}}$ \cite{Yagi:2012ya}, since these extra fields usually enter at lower order than the metric field as explained above. In our case, we choose to absorb the coupling constant into the expansion of the extra fields for convenience in the order counting, so our expansion starts at $\zeta$. In latter sections, we may also rotate the tetrad in Eqs.~\eqref{eq:pert-tetrad} and \eqref{eq:A_ab} using Eqs.~\eqref{eq:tetrad_rotations} such that certain NP quantities vanish on the background. If the expansion in Eq.~\eqref{eq:expansion_Weyl} is not broken, we will use the rotated tetrad for the convenience of calculations. In the case that Eq.~\eqref{eq:expansion_Weyl} is violated due to those rotations, we will use Eqs.~\eqref{eq:pert-tetrad} and \eqref{eq:A_ab} as our background tetrad.

Besides $\zeta$ and $\epsilon$, one may have to deal with additional expansion parameters, such as the dimensionless spin $\chi$ in the slow-rotation expansion, but an expansion in $\zeta$ and $\epsilon$ is necessary and sufficient to demonstrate how the Teukolsky equation in modified gravity can be derived. Below, we may write some quantities with only one superscript, e.g., $\Psi^{(n)}$, which represents the $n$-th order term in the expansion of $\Psi$ in $\epsilon$, as shown in the first line of Eq.~\eqref{eq:expansion_Weyl}, so all the other expansions are hidden for simplicity.

%%%%%%%%%%%%%%%%%%%%%%%%%%%%%%%%%%%%%%%

\section{Perturbations of Petrov type D spacetimes in theories beyond GR} 
\label{sec:nonGR}

In this section, we present a method to extend the formalism shown in Sec.~\ref{sec:GRintro} for obtaining the perturbation equations for Petrov type D BHs in modified theories of gravity discussed in Sec.~\ref{sec:beyondGR-theory}. We particularly focus on spacetimes that are stationary and vacuum solutions to modified gravity theories, and although they may not be Ricci flat, they remain of Petrov type D. As discussed in Sec.~\ref{sec:beyondGR-theory}, an example of such a spacetime is BH solutions in dCS gravity, expanded to leading order in the dimensionless spin parameter~\cite{Owen:2021eez,Alexander:2009tp} and obtained in an effective field theory (EFT) approach. We will use the perturbation scheme introduced in Sec.~\ref{sec:perturbation_scheme}. Extending the formalism developed for Petrov type D spacetimes in GR (either the traditional Teukolsky's approach or the Chandrasekhar's approach) to include Petrov type D spacetimes that are non-Ricci-flat in modified gravity is a stepping stone in developing a formalism that is applicable to algebraically general Petrov type I spacetimes in beyond GR theories. 

\subsection{Extending the Teukolsky formalism beyond GR: Non-Ricci-flat and Petrov type D backgrounds} 
\label{sec:TeuknonGR}

In this subsection, we present an extension to the Teukolsky formalism presented in Sec.~\ref{sec:TeukGR} for non-GR non-Ricci-flat Petrov type D spacetimes. We follow a procedure similar to that presented in Sec.~\ref{sec:TeukGR} with the aim of developing a formalism to obtain the decoupled differential equation describing the dynamical pieces of $\Psi_0$ and $\Psi_4$. This subsection along with the next one form the backbone of the development of a formalism for the algebraically general Petrov type \rom{1} spacetimes in beyond GR theories.

We begin by considering modified theories of gravity whose isolated (stationary and vacuum) BH solutions are non-Ricci-flat, i.e., the Ricci tensor obtained from trace-reversed vacuum field equations (i.e., no matter present) no longer vanish. For instance, in theories such as dCS or EdGB, where a scalar field is non-minimally coupled to a quadratic term in curvature~\cite{Jackiw:2003pm,Alexander:2009tp,Wagle_Yunes_Silva_2021,Blazquez-Salcedo:2016enn}, cubic, or higher-order theories of gravity~\cite{Burgess_2004, Sotiriou:2006hs, Donoghue_2012, Endlich_Gorbenko_Huang_Senatore_2017, Cano_Ruiperez_2019, Cano:2020cao, Cano_Fransen_Hertog_Maenaut_2021}, the metric field equations lead to a non-vanishing Ricci tensor and are therefore non-Ricci-flat. This can easily be seen in the dCS gravity example with the trace-reversed field equation~\eqref{eq:tracericci}, where the Ricci tensor clearly does not vanish even in vacuum due to the non-vanishing of the Riemann tensor and a non-trivial pseudo-scalar field.

When the background is non-Ricci-flat, the unperturbed Bianchi identities acquire sources. In the NP language, the non-vanishing of the Ricci tensor implies that NP Ricci scalars $\Phi_{ij}$ for $i,j \in (0,1,2)$ also do not vanish [see e.g., Eq.~\eqref{eq:eg_phi00}]. Consequently, the source terms of Eqs.~\eqref{eq:BianchiId_Psi0_1_simplify}-\eqref{eq:BianchiId_Psi0_2_simplify} are non-vanishing for non-Ricci-flat, non-GR BH background. But if we require that the non-Ricci-flat background be of Petrov type D, then the background Weyl scalars
\begin{align}
\label{eq:bg_values_nonGR}
    \Psi_0^{(0)} = \Psi_1^{(0)} =& \Psi_3^{(0)} = \Psi_4^{(0)} = 0 \,.
\end{align}
Unlike in the GR case, however, the background spin coefficients no longer vanish in general, as one can verify explicitly by inserting Eq.~\eqref{eq:bg_values_nonGR} in Eqs.~\eqref{eq:bianchi_iden}. Consequently, we still have additional terms that are non-vanishing in the equations presented in Sec.~\ref{sec:TeukGR}. More specifically, the full Bianchi identities recast in the form of Eqs.~\eqref{eq:Bianchi_simplified} now take the form
\begin{subequations}
\begin{align}
\begin{split} \label{eq:nongr-teuk1pert}
    F_1\Psi_{0} - J_1 \Psi_{1} -3\kappa\Psi_{2}  &=S_1\,, 
\end{split} \\
\begin{split} \label{eq:nongr-teuk2pert}
    F_2\Psi_{0} - J_2\Psi_{1}  -3\sigma \Psi_{2} &=S_2\,, 
\end{split} \\ 
\begin{split} \label{eq:nongr-teuk3pert}
     E_2\sigma - E_1\kappa -\Psi_{0}&=0\,,
\end{split}
\end{align}
\end{subequations}
where $S_1$ and $S_2$ are given in Eq.~\eqref{eq:source_bianchi}, $(E_{1,2},F_{1,2},J_{1,2})$ are defined in Eq.~\eqref{eq:simplify_operators_0}, and $(\kappa,\sigma)$ are spin coefficients presented in Appendix~\ref{appendix:NPformalism_addition}. Notice that we have not yet performed a perturbative expansion to separate the background from the perturbed Weyl scalars.

Adapting a method similar to that presented in Sec.~\ref{sec:TeukGR} to obtain a differential equation for $\Psi_0$, we need to eliminate the $\Psi_1$ dependence from the above equations by developing an appropriate commutation relation for this type of beyond GR theories. While eliminating the $\Psi_1$ dependence, we will also naturally decouple $\Psi_0$ from the $\kappa$ and $\sigma$ dependence in the above equations, as shown below. To decouple the equations, we prescribe the following steps:
\begin{enumerate}
    \item Multiply Eq.~\eqref{eq:nongr-teuk3pert} by $\Psi_2$.
    \item Use the chain rule such that the intrinsic derivatives act on the product of $\Psi_2$ with either $\sigma$ or $\kappa$. For instance,
    \begin{equation} \label{eq:psidsigma}
        \Psi_2 (D\sigma) = D(\Psi_2 \sigma) - \sigma (D\Psi_2) \,.
    \end{equation}
    For modified theories of gravity, the second term above is different from Eq.~\eqref{eq:GRbianchi} because it is modified due to the non-vanishing of the NP Ricci scalars. For instance, when looking at Eq.~\eqref{eq:bianchi_dpsi2},
    \begin{equation}
        D\Psi_2 = 3\rho\Psi_2 - P_1 \,,
    \end{equation}
    where $P_1$ are all the non-vanishing terms from the Bianchi identity in Eq.~\eqref{eq:bianchi_dpsi2}.
    However, when working with this approach, we have more algebraic complications involved in decoupling all curvature perturbations. Therefore, for the purpose of this subsection, we continue to work with Eq.~\eqref{eq:psidsigma}.
    \item Using Eq.~\eqref{eq:psidsigma}, we can rewrite the operators in Eq.~\eqref{eq:nongr-teuk3pert} as $\mathcal{E}_1$ and $\mathcal{E}_2$, as defined in Eqs.~\eqref{eq:tile_E_operators}.
    \item The commutator acting on $\Psi_1$ is then given by
    \begin{equation} \label{eq:comm}
       \left(\mathcal{E}_2J_2-\mathcal{E}_1J_1\right) \Psi_1\,.
    \end{equation}
    \item Now expand $\Psi_1$ as shown in Eq.~\eqref{eq:expansion_Weyl}, i.e.,
    \begin{equation}
        \Psi_1 = \Psi_1^{(0,0)} + \zeta \Psi_1^{(1,0)} + \epsilon \Psi_1^{(0,1)} + \zeta \epsilon \Psi_1^{(1,1)} \,.
    \end{equation}
    Since the BH background is Petrov type D, the background $\Psi_1^{(0,0)}$ and $\Psi_1^{(1,0)}$ vanish. The quantity $\Psi_1^{(0,1)}$ is generated by the perturbed (GW) metric in GR, which can be set to zero through a convenient choice of gauge, as we have shown in Sec.~\ref{sec:gauge_GR}. Therefore, to leading order in $\zeta$ and $\epsilon$, the terms inside the parenthesis of Eq.~\eqref{eq:comm} must be evaluated on the GR BH background as in Eq.~\eqref{eq:commutatorgr}. 
    \emph{Following these arguments, the commutator given by Eq.~\eqref{eq:comm} vanishes for non-Ricci-flat and Petrov type D BH backgrounds in the class of modified gravity theories we considered.}
\end{enumerate}

Multiplying Eqs.~\eqref{eq:nongr-teuk1pert} and \eqref{eq:nongr-teuk2pert} by $\mathcal{E}_1$ and $\mathcal{E}_2$, respectively, subtracting one from the other, and expanding to leading order in $\epsilon$, we find
\begin{align} \label{eq:nongr-teuk}
   H_{0}^{(0)}\Psi_0^{(1)}=\mathcal{S}^{(1)}\,,
\end{align}
where we have defined
\begin{subequations} \label{eq:operators_typeD}
    \begin{align}
        H_{0}
        & =\mathcal{E}_2F_2-\mathcal{E}_1F_1-3\Psi_2\,, 
        \label{eq:H0} \\
        \mathcal{S}
        & =\mathcal{E}_2S_2-\mathcal{E}_1S_1\,.
        \label{eq:source1}
    \end{align}
\end{subequations}
Expanding Eq.~\eqref{eq:nongr-teuk} using the two parameter expansion in Eq.~\eqref{eq:expansion_Weyl}, at leading orders in $\zeta$ and $\epsilon$, we have
\begin{equation} \label{eq:nongr-teuk-expanded}
    H_{0}^{(0,0)}\Psi_0^{(1,1)}+H_{0}^{(1,0)}\Psi_0^{(0,1)}
   =\mathcal{S}^{(1,1)}\,.
\end{equation}
Notice, similar to the case in GR, the expansion in $\epsilon$ is sufficient to derive Eq.~\eqref{eq:nongr-teuk}, and an expansion in $\zeta$ is imposed at the end to get the equation at $\mathcal{O}(\zeta^1,\epsilon^1)$.

We can now use the GHP transformation to derive an analogous modified Teukolsky equation for the perturbed $\Psi_4$. Let us then apply the exchange transformation $l^\mu \leftrightarrow n^\mu,\, m^\mu \leftrightarrow \Bar{m}^\mu $ to Eq.~\eqref{eq:nongr-teuk} and use the definitions given in Eq.~\eqref{eq:simplify_operators_0} to find
\begin{align} \label{eq:nongr-teuk-psi4}
    H_{4}^{(0)}\Psi_4^{(1)}=\mathcal{T}^{(1)}\,,
\end{align}
which, expanded in $\zeta$, becomes
\begin{align} \label{eq:nongr-teuk-psi4-expanded}
   H_{4}^{(0,0)}\Psi_4^{(1,1)}+H_{4}^{(1,0)}\Psi_4^{(0,1)} =\mathcal{T}^{(1,1)}\,,
\end{align}
where we have defined
\begin{subequations}
    \begin{align}
        H_{4}
        & =\mathcal{E}_4F_4-\mathcal{E}_3F_3-3\Psi_2\,, 
        \label{eq:H4} \\ 
        \mathcal{T}
        & =\mathcal{E}_4S_4-\mathcal{E}_3S_3\,.
        \label{eq:source2}
    \end{align}
\end{subequations}
$S_3$ and $S_4$ are defined in Eq.~\eqref{eq:source_bianchi}, while $\mathcal{E}_3$ and $\mathcal{E}_4$ are defined in Eq.~\eqref{eq:tile_E_operators}.

Equations~\eqref{eq:nongr-teuk} and~\eqref{eq:nongr-teuk-psi4} therefore represent a modified Teukolsky equation. The differential operators acting on $\Psi_{0,4}^{(1)}$ are similar in functional form to those of the standard Teukolsky equation in GR. Notice, however, that these operators are not the same as their GR counterparts [i.e., corrected by $H_{0,4}^{(1,0)}$ in Eqs.~\eqref{eq:nongr-teuk-expanded} and \eqref{eq:nongr-teuk-psi4-expanded}] because the Bianchi identities are modified. In the GR limit, one can of course show that they are equivalent to each other because the Bianchi identities no longer depend on NP Ricci scalars, so they reduce to Eq.~\eqref{eq:GRbianchi}. Note importantly that the left-hand side of Eqs.~\eqref{eq:nongr-teuk} and \eqref{eq:nongr-teuk-psi4} describe all GW perturbations since they are not expanded in power of $\zeta$.

The modified Teukolsky equations~\eqref{eq:nongr-teuk} and~\eqref{eq:nongr-teuk-psi4} contain source terms that are of ${\cal{O}}(\zeta)$ and thus absent in GR. After an expansion in $\zeta$ in Eqs.~\eqref{eq:nongr-teuk-expanded} and \eqref{eq:nongr-teuk-psi4-expanded}, we notice that the source terms ${\cal{S}}^{(1)}$ and ${\cal{T}}^{(1)}$ depend on dynamical NP quantities at ${\cal{O}}(\zeta^1,\epsilon^1)$ [i.e., ${\cal{S}}^{(1,1)}$ and ${\cal{T}}^{(1,1)}$]. These sources terms depend on the $S_i$ terms in Eqs.~\eqref{eq:source1} and~\eqref{eq:source2}, which are products of differential operators constructed from the tetrad and the NP Ricci scalars $\Phi_{ij}$. As discussed in Sec.~\ref{sec:perturbation_scheme}, since $R_{\mu\nu}$ is $\mathcal{O}(\zeta)$, $\Phi_{ij}$ is always of ${\cal{O}}(\zeta^1,\epsilon^0)$ or ${\cal{O}}(\zeta^1,\epsilon^1)$, which then means the tetrad that is needed to compute the differential operators must be of ${\cal{O}}(\zeta^0,\epsilon^0)$ and ${\cal{O}}(\zeta^0,\epsilon^1)$. In addition, all the metric fields in $R_{\mu\nu}$ must also be of ${\cal{O}}(\zeta^0,\epsilon^0)$ and ${\cal{O}}(\zeta^0,\epsilon^1)$. \textit{We therefore conclude that curvature perturbations of a non-Ricci-flat, Petrov type D BH background satisfy a decoupled equation.}

The tetrad at ${\cal{O}}(\zeta^0,\epsilon^0)$ is just the Kinnersley tetrad of Eq.~\eqref{eq:teukpsi0}, but the tetrad at ${\cal{O}}(\zeta^0,\epsilon^1)$ must be reconstructed from the metric perturbation at ${\cal{O}}(\zeta^0,\epsilon^1)$. That is, one needs to first solve the Teukolsky equation in GR for the GR Weyl scalars $\Psi_{0,4}^{(0,1)}$ and then reconstruct the GR GW metric perturbation to build the perturbed tetrad at ${\cal{O}}(\zeta^0,\epsilon^1)$.
This is in stark contrast to the GR case since for a Ricci-flat Petrov type D BH background in GR, metric reconstruction is not required to study GW perturbations. 
Metric reconstruction in GR has already been worked out in the vacuum case by Chrzanowski~\cite{Chrzanowski:1975wv} and Cohen and Kegeles~\cite{Kegeles_Cohen_1979} (see e.g., \cite{Whiting_Price_2005,Yunes_Gonzalez_2006} for a short review) using Hertz potential. There are also approaches that avoid using Hertz potential by solving the remaining Bianchi identities, Ricci identities, and commutation relations, for example in \cite{Chandrasekhar_1983, Loutrel_Ripley_Giorgi_Pretorius_2020}. Clearly then, such metric reconstruction in GR is possible, and we leave a further analysis of their implementation in our decoupled equations to future work.

%%%%%%%%%%%%%%%%%%%%%%%%%%%%%%%%%%%%%%
\subsection{Extending Chandrasekhar's approach beyond GR: Non-Ricci-Flat and Petrov type D backgrounds} 
\label{sec:gauge_nonGR}

Similar to the Petrov type D vacuum GR case, we can also follow Chandrasekhar's approach to remove $\Psi_1^{(1)}$ directly. By doing the same type \rom{2} rotation of Sec.~\ref{sec:gauge_GR} with the rotation parameter $b^{(1)}=-{\Psi_1^{(1)}}/({3\Psi_2^{(0)}})$, we can set $\Psi_1^{(1)}=0$. Then, from Eqs.~\eqref{eq:nongr-teuk1pert}-\eqref{eq:nongr-teuk2pert}, we again solve for $\kappa$ and $\sigma$ first. Notice that the $\kappa$ and $\sigma$ we have solved for may also contain $\mathcal{O}(\epsilon^0)$ terms since they do not necessarily vanish in a non-Ricci-flat Petrov type D background. We then insert the solutions for $\kappa$ and $\sigma$ in terms of $\Psi_0^{(1)}$ and $S_i^{(1)}$ back into Eq.~\eqref{eq:nongr-teuk3pert} to obtain a single equation for $\Psi_0^{(1)}$. We have verified explicitly that this equation is exactly the same as Eq.~\eqref{eq:nongr-teuk}. Applying the GHP transformation, one again finds Eq.~\eqref{eq:nongr-teuk-psi4} for $\Psi_4^{(1)}$. 

As shown above, the final modified Teukolsky equation obtained using the two approaches (i.e., the Teukolsky's approach and Chandrasekhar's approach) are equivalent for both Ricci-flat and non-Ricci-flat, Petrov type D BH backgrounds. A main difference between the two methods is in how the equations for the curvature perturbations $\Psi_0$ and $\Psi_4$ are decoupled from $\Psi_1$ and $\Psi_3$, respectively. Chandrasekhar's approach has a significant algebraic advantage over Teukolsky's original formalism, as the former utilizes available gauge freedom to make convenient gauge choices to eliminate $\Psi_1$ and $\Psi_3$ dependence. For non-Ricci-flat, Petrov type D backgrounds in modified gravity, Teukolsky's approach is not significantly more complicated than in GR, but this is no longer true when considering non-Ricci-flat, Petrov type \rom{1} backgrounds. In the latter case, Teukolsky's approach is more involved because of the non-vanishing of additional NP quantities leading to more non-vanishing terms in these equations. In Chandrasekhar's approach, however, one can continue to leverage gauge freedom to eliminate certain NP quantities without the need for developing a commutator relation like that of Eqs.~\eqref{eq:commutatorgr} and \eqref{eq:comm} or using additional Bianchi identities. Because of this, we will employ Chandrasekhar's approach in what follows to develop a formalism to study perturbations of non-Ricci-flat, Petrov type \rom{1} spacetimes in modified theories of gravity.

%%%%%%%%%%%%%%%%%%%%%%%%%%%%%%%%%%%%%%%
\section{Extension of the Teukolsky formalism beyond GR: \\ Non-Ricci-flat and non-Petrov-type-D backgrounds}
\label{sec:teuknonD}

In this section, we extend Chandrasekhar's approach to non-Ricci-flat backgrounds that are algebraically general. As seen in Sec.~\ref{sec:gauge_GR} and \ref{sec:gauge_nonGR}, choosing a convenient gauge for the background and for the perturbed NP quantities, certain NP quantities can be eliminated from the NP equations when deriving the (modified) Teukolsky equation to obtain a single decoupled equation for $\Psi_0$ and $\Psi_4$. In this section, we first explore these gauge choices for background and perturbed NP quantities in more detail while treating the Petrov type \rom{1} spacetime as a linear perturbation of a Petrov type D spacetime in GR. We then derive the master equations for dynamical Weyl scalars $\Psi_0$ and $\Psi_4$, discuss the modifications introduced due to non-GR effects, and provide a brief discussion on how to evaluate this equation for beyond GR theories.

Before proceeding with this section, it is important to distinguish between two background concepts that we introduce in this work. In general, the line element of a BH background spacetime for theories beyond GR discussed in Sec.~\ref{sec:beyondGR-theory} can be expressed as
\be \label{eq:bg-metric}
ds^2 = ds^2_{\GR} + \zeta \Tilde{ds}^{2}_{\bGR} \,.
\ee
Here, we have introduced the following symbols:
\begin{itemize}
    \item[(i)] $ds^2$ is the line element of the \textit{background spacetime} or the \textit{background} for short, which is the stationary part of the full spacetime. 
    \item[(ii)] $ds^2_{\GR}$ is the line element of the \textit{original background}, which is the background \textit{all} the perturbations, including the stationary ones (e.g., $\Tilde{ds}^{2}_{\bGR}$), are built on top of.
\end{itemize}
For instance, the line element of a slowly rotating BH in dCS gravity to leading order in spin takes the form of Eq.~\eqref{eq:bg-metric} with~\cite{Yunes_Pretorius_2009}
\begin{align} 
    \Tilde{ds}^{2}_{\dCS} 
    &=\frac{5M^4}{4}\frac{a}{r^4}\left(1+\frac{12}{7}\frac{M}{r}
    +\frac{27}{10}\frac{M^2}{r^2}\right)\sin^2\theta dtd\phi\,, \label{eq:ds2}\\ 
    ds^2_{\GR} 
    &=-f(r)dt^2-\frac{4Ma\sin^2\theta}{r}dtd\phi+f(r)^{-1}dr^2 \nonumber\\
    & \quad +r^2 d\theta^2+r^2\sin^2\theta d\phi^2\,. \label{eq:ds3}
\end{align}
Here, in our notation, the original background is given by Eq.~\eqref{eq:ds3} whereas the background spacetime is given by the sum of Eqs.~\eqref{eq:ds2} and~\eqref{eq:ds3}. This is of course just a simple example of our notation, which holds true for theories that can be described using the Lagrangian given in Eq.~\eqref{eq:lagrangian}.
In general, the background spacetime includes ${\cal{O}}(\zeta^0,\epsilon^0)$ and ${\cal{O}}(\zeta^1,\epsilon^0)$ parts, while the original background is just of ${\cal{O}}(\zeta^0,\epsilon^0)$ (i.e., it is the Kerr BH spacetime for arbitrarily spinning BHs).

\begin{figure*}[t]
	\centering
	\includegraphics[width=0.8\linewidth]{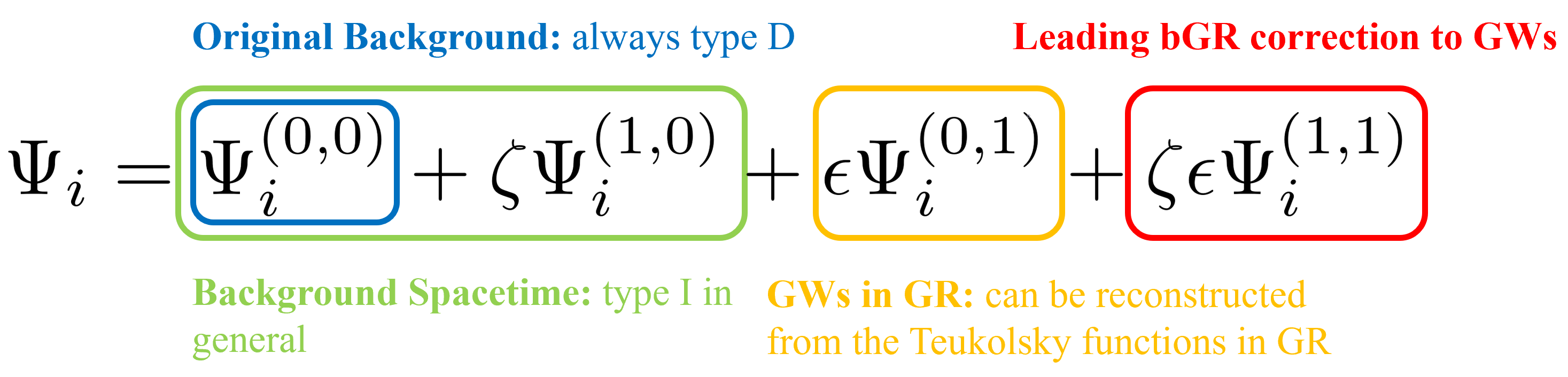}
	\caption{A diagram to illustrate the meaning of different terms in the expansion of NP quantities in Eq.~\eqref{eq:expansion_Weyl}.}
	\label{fig:expansion}
\end{figure*}

Although the concepts of a background and an original background spacetime may sometimes correspond to the same thing (e.g., to the Kerr BH spacetime in GR), these concepts can sometimes be different in modified gravity theories. For example, in the theories discussed in Sec.~\ref{sec:beyondGR-theory}, the Kerr metric is not a solution for all stationary and axisymmetric BHs. Rather these BHs are represented by spacetimes that are non-Ricci-flat and non-Petrov-type-D when not expanded in spin. In such cases, the background of the dynamical gravitational perturbation we study would be such a non-Ricci-flat and non-Petrov-type-D spacetime, but the original background would still be the Kerr spacetime. In Fig.~\ref{fig:expansion}, we present the relation between these two different background concepts and the terms in the expansion of NP quantities in Eq.~\eqref{eq:expansion_Weyl}.

\subsection{Gauge choice for the background spacetime: $\mathcal{O}(\zeta^0,\epsilon^0)$ and $\mathcal{O}(\zeta^1,\epsilon^0)$} \label{sec:gauge_background}

For a non-Petrov-type-D modified background spacetime, the gauge choice in Eq.~\eqref{eq:bg_values_nonGR} is not possible. For example, as found in \cite{Yagi:2012ya}, the metric describing a rotating BH in dCS gravity need not be of Petrov type D once one incorporates second-order and higher in rotation effects; in that case, the metric is now of Petrov type \rom{1}, which is the most general type in the Petrov classification. However, we can still set $\Psi_0^{(0)}=\Psi_4^{(0)}=0$ for a Petrov type \rom{1} spacetime as discussed in \cite{Chandrasekhar_1983} and shown for dCS gravity in \cite{Owen:2021eez}, so we {\it could} use a gauge such that
\begin{equation} \label{eq:gauge_background_typeI}
	\Psi_{0,1,3,4}^{(0,0)}=0\,,\quad \Psi_0^{(1,0)}=\Psi_4^{(1,0)}=0\,,
\end{equation}
but we will {\it not} for the following reasons. 

Although the gauge defined by requiring that Eq.~\eqref{eq:gauge_background_typeI} holds simplifies Eqs.~\eqref{eq:Bianchi_simplified} and \eqref{eq:Bianchi_simplified_psi4}, it may spoil our assumption that the leading correction to the tetrad enters at $\mathcal{O}(\zeta^1)$. As shown in \cite{Owen:2021eez}, for dCS gravity in the slow-rotation approximation, in order to impose that $\Psi_0^{(1,0)}=\Psi_4^{(1,0)}=0$ at $\mathcal{O}(\chi^2)$, we need to modify the tetrad at $\mathcal{O}(\zeta^{\frac{1}{2}},\chi^2)$, and this induces a nonzero $\Psi_1^{(\frac{1}{2},0)}$ and $\Psi_3^{(\frac{1}{2},0)}$. These $\mathcal{O}(\zeta^{\frac{1}2})$ terms are not covered by our expansion strategy in Eq.~\eqref{eq:expansion_Weyl}, which only contains terms of $\mathcal{O}(\zeta^0,\epsilon^0)$, $\mathcal{O}(\zeta^1,\epsilon^0)$, $\mathcal{O}(\zeta^0,\epsilon^1)$, and $\mathcal{O}(\zeta^1,\epsilon^1)$ for all quantities. For this reason, we only impose
\begin{equation} \label{eq:gauge_background_general}
	\Psi_{0,1,3,4}^{(0,0)}=0
\end{equation}
and leave all $\mathcal{O}(\zeta^1,\epsilon^0)$ perturbations general. These properties are summarized on the left two columns of Table~\ref{tab:psi_table}. In this case, we will use the background tetrad in Eqs.~\eqref{eq:pert-tetrad} and \eqref{eq:A_ab} such that Eq.~\eqref{eq:gauge_background_general} is satisfied, and the expansion in Eq.~\eqref{eq:expansion_Weyl} is not broken.
	
\subsection{Gauge choice for the dynamical perturbations: $\mathcal{O}(\zeta^0,\epsilon^1)$ and $\mathcal{O}(\zeta^1,\epsilon^1)$} \label{sec:gauge_dynamical}

Different gauge choices can be made separately at different perturbative orders. Sec.~\ref{sec:gauge_background} fixed the gauge for the background spacetime at $\mathcal{O}(\zeta^0,\epsilon^0)$ and $\mathcal{O}(\zeta^1,\epsilon^0)$, but we still have gauge freedom at $\mathcal{O}(\zeta^0,\epsilon^1)$ and $\mathcal{O}(\zeta^1,\epsilon^1)$. As in Secs.~\ref{sec:gauge_GR} and ~\ref{sec:gauge_nonGR}, we shall impose 
\begin{equation} \label{eq:gauge_perturbation}
	\Psi_1^{(0,1)}=\Psi_3^{(0,1)}=\Psi_1^{(1,1)}=\Psi_3^{(1,1)}=0\,.
\end{equation}
In this gauge, Eqs.~\eqref{eq:BianchiId_Psi0_1_simplify}-\eqref{eq:BianchiId_Psi0_2_simplify} for the dynamical part of $\Psi_0$ and $\Psi_1$ decouple directly, and so do Eqs.~\eqref{eq:BianchiId_Psi4_1_simplify}-\eqref{eq:BianchiId_Psi4_2_simplify} for the dynamical part of $\Psi_3$ and $\Psi_4$.

As discussed in \cite{Chandrasekhar_1983}, in a Petrov type D spacetime, we can always make a gauge choice such that the linear perturbations to $\Psi_1$ and $\Psi_3$ vanish without affecting $\Psi_0$ and $\Psi_4$, so only $\Psi_0$ and $\Psi_4$ are gauge invariant quantities in a linear perturbation theory. Since at $\mathcal{O}(\zeta^0)$, the background spacetime is the Petrov type D spacetime of GR, it then follows that we can always make the gauge choice in Eq.~\eqref{eq:gauge_perturbation} at $\mathcal{O}(\zeta^0,\epsilon^1)$.

Next, we need to show that Eq.~\eqref{eq:gauge_perturbation} holds at $\mathcal{O}(\zeta^1,\epsilon^1)$. If we treat $\Psi_{1,3}^{(1,1)}$ as the $\mathcal{O}(\zeta^0,\epsilon^1)$ perturbation to $\Psi_{1,3}^{(1,0)}$, it is not clear that we can make a gauge choice in Eq.~\eqref{eq:gauge_perturbation} since the \textit{background spacetime} at $\mathcal{O}(\zeta^1,\epsilon^0)$ is not necessarily Petrov type D. However, we can also treat $\Psi_{1,3}^{(1,1)}$ as the $\mathcal{O}(\zeta^1,\epsilon^1)$ perturbation to $\Psi_{1,3}^{(0,0)}$ in the \textit{original background}. Since the \textit{original background} is the Petrov type D spacetime in GR, Eq.~\eqref{eq:gauge_perturbation} should still hold.

Let us show that, at $\mathcal{O}(\zeta^1,\epsilon^1)$, $\Psi_{1,3}^{(1,1)}$ can be eliminated by a tetrad rotation at $\mathcal{O}(\zeta^1,\epsilon^1)$. Let us consider $\Psi_1^{(1,1)}$  explicitly and apply a type \rom{2} rotation [cf.~Eq.~\eqref{eq:rotate2}], with a parameter $b^{(1,1)}$ at $\mathcal{O}(\zeta^1,\epsilon^1)$. This leads to, at $\mathcal{O}(\zeta^1,\epsilon^1)$,
\begin{equation} \label{eq:rotate2_reduce}
	\begin{array}{l}
		\Psi_{0}^{(1,1)}\rightarrow\Psi_{0}^{(1,1)}+4b^{(1,1)}\Psi_{1}^{(0,0)}\,, \\
		\Psi_{1}^{(1,1)}\rightarrow\Psi_{1}^{(1,1)}+3b^{(1,1)}\Psi_{2}^{(0,0)}\,, \\
		\Psi_{2}^{(1,1)}\rightarrow\Psi_{2}^{(1,1)}+2b^{(1,1)}\Psi_{3}^{(0,0)}\,, \\
		\Psi_{3}^{(1,1)}\rightarrow\Psi_{3}^{(1,1)}+b^{(1,1)}\Psi_{4}^{(0,0)}\,, \\ 
		\Psi_{4}^{(1,1)}\rightarrow\Psi_{4}^{(1,1)}\,.
	\end{array} 
\end{equation}

We are motivated to require that $b = \mathcal{O}(\zeta^1,\epsilon^1)$ since we want to perturb about the \textit{original background}. By letting $b^{(1,1)}=-\Psi_1^{(1,1)}/(3\Psi_2^{(0,0)})$, we can set $\Psi_1^{(1,1)}=0$. With the background gauge choice that ensures Eq.~\eqref{eq:gauge_background_general} holds, we can easily see from Eq.~\eqref{eq:rotate2_reduce} that all the other Weyl scalars at $\mathcal{O}(\zeta^1,\epsilon^1)$ are unaffected such that
\begin{equation} \label{eq:rotate2_result}
	\begin{array}{l}
		\Psi_{1}^{(1,1)}\rightarrow0\,,\quad	
		\Psi_{0,2,3,4}^{(1,1)}\rightarrow\Psi_{0,2,3,4}^{(1,1)}\,.
	\end{array}
\end{equation}
Similarly, by applying a type \rom{1} rotation [Cf.~Eq.~\eqref{eq:rotate1}] and choosing the rotation parameter $a^{(1,1)}=[-\Psi_3^{(1,1)}/(3\Psi_2^{(0,0)})]^{*}$, we can set 
\begin{equation} \label{eq:rotate2_result_Psi3}
	\begin{array}{l}
		\Psi_{3}^{(1,1)}\rightarrow0\,,\quad	
		\Psi_{0,1,2,4}^{(1,1)}\rightarrow\Psi_{0,1,2,4}^{(1,1)}\,. 
	\end{array}
\end{equation}
Properties of the $\mathcal{O}(\zeta^0,\epsilon^1)$ and $\mathcal{O}(\zeta^1,\epsilon^1)$ contributions to the Weyl scalars are summarized on the right half of Table~\ref{tab:psi_table}.

\begin{table*}[]
    \centering
    \begin{tabular}{c|c|c||c|c|}
        \hline
        \multirow{3}{*}{Types of Terms}
        & \multicolumn{2}{c||}{Stationary Background} 
        & \multicolumn{2}{c|}{Dynamical GWs} \\
        \cline{2-5}
        & \begin{tabular}{c} 
            Original Background \\ (GR)\end{tabular} 
        & \begin{tabular}{c} 
            Stationary Modification \\ to Original Background
        \end{tabular} 
        & \begin{tabular}{c} 
            GWs on \\ Original Background
        \end{tabular}
        & \begin{tabular}{c}
            GW \\ Corrections 
        \end{tabular} \\
        \hline
        \diagbox{Weyl Scalar}{Orders}        
        & \begin{tabular}{cc}
            $\mathcal{O}(\zeta^0,\epsilon^0)$
        \end{tabular}
        & \begin{tabular}{cc}
            $\mathcal{O}(\zeta^1,\epsilon^0)$
        \end{tabular}
        & \begin{tabular}{cc}
            $\mathcal{O}(\zeta^0,\epsilon^1)$
        \end{tabular}  
        & \begin{tabular}{cc}
            $\mathcal{O}(\zeta^1,\epsilon^1)$
        \end{tabular} \\ 
        \hline
        $\Psi_0$ & 0 & $\neq 0$ & $\neq 0^{(a)}$ & $\neq 0^{(a)}$ \\
        $\Psi_1$ & $0$ & $\neq 0$ & $0^{(b)}$ & $0^{(b)}$ \\
        $\Psi_2$ & $\neq 0$ & $\neq 0$ & $\neq 0^{(c)}$ & $\neq 0^{(d)}$ \\
        $\Psi_3$ & 0 & $\neq 0$ & $0^{(b)}$ & $0^{(b)}$ \\
        $\Psi_4$ & 0 & $\neq 0$ & $\neq 0^{(a)}$ & $\neq 0^{(a)}$ \\
        \hline
    \end{tabular}
    \caption{Properties of Weyl scalars in the Chandrasekhar gauge for non-Petrov-type-D modified BH spacetimes with GWs. Quantities on the {\it stationary background} columns are already known.  For quantities on the {\it dynamical GWs} columns, items labeled as $(a)$ are scalars that need to be solved for, labeled as $(b)$ are set to zero by gauge, labeled as $(c)$ can be reconstructed from $\Psi_{0}^{(0,1)}$ or $\Psi_{4}^{(0,1)}$, while labeled by $(d)$ do not appear in the modified Teukolsky equation.}
    \label{tab:psi_table}
\end{table*}

\subsection{Modified Teukolsky equation in non-Ricci-flat and algebraically general backgrounds} \label{sec:The_modified_Teukolsky_equation}

We can now derive the modified Teukolsky equation for non-Ricci-flat and Petrov type \rom{1} spacetimes. Here, we only show how to obtain the equation for the dynamical perturbation to $\Psi_0$, but the same procedure can be applied to $\Psi_4$, or one can perform the GHP transformation $l^{\mu}\leftrightarrow n^{\mu}$, $m^{\mu}\leftrightarrow\bar{m}^{\mu}$ on the $\Psi_0$ equation to find the equation for $\Psi_4$~\cite{Geroch_Held_Penrose_1973}.

\subsubsection{Elimination of $\kappa$ and $\sigma$}
\label{sec:elimination_kappa_sigma}

From Eqs.~\eqref{eq:BianchiId_Psi0_1_simplify}-\eqref{eq:BianchiId_Psi0_2_simplify}, we can solve for $\kappa$ and $\sigma$ in terms of other NP quantities. Inserting $\kappa$ and $\sigma$ from Eqs.~\eqref{eq:BianchiId_Psi0_1_simplify}-\eqref{eq:BianchiId_Psi0_2_simplify} into Eq.~\eqref{eq:RicciId_Psi0_simplify} and multiplying the resulting equation by $3\Psi_2$ to match the form of the original Teukolsky equation~\cite{Teukolsky:1973ha} when $\zeta=0$, one finds
\begin{equation}
    \begin{aligned}
        & \Psi_2E_2\left[\Psi_2^{-1}\left(F_2\Psi_0
        -J_2\Psi_1-S_2\right)\right] \\
        - & \Psi_2E_1\left[\Psi_2^{-1}\left(F_1\Psi_0
        -J_1\Psi_1-S_1\right)\right]  -3\Psi_2\Psi_0=0\,.
    \end{aligned}
\end{equation}
Re-organizing this equation to extract the operators that act on $\Psi_0$, $\Psi_1$, $S_1$, and $S_2$, we find
\begin{equation} \label{eq:3_to_1}
    H_{0}\Psi_0-H_{1}\Psi_1= \mathcal{S}\,,
\end{equation}
where $H_0$ and $\mathcal{S}$ are defined in Eq.~\eqref{eq:operators_typeD}, and we have defined
\begin{equation} \label{eq:operators_non_typeD}
    H_{1}\equiv \mathcal{E}_2 J_2 - \mathcal{E}_1 J_1\,, 
\end{equation}
with $\mathcal{E}_i$ defined in Eq.~\eqref{eq:tile_E_operators}.

\subsubsection{Gauge choice and general strategy}

The derivation so far has combined the three equations in Eqs.~\eqref{eq:BianchiId_Psi0_1_simplify}-\eqref{eq:RicciId_Psi0_simplify} into a single equation \eqref{eq:3_to_1}.  Our next goal is to keep only $\Psi_0^{(1,1)}$ and no other $\mathcal{O}(\zeta^1,\epsilon^1)$ contributions of Weyl scalars, spin connection coefficients, or intrinsic derivatives. Note that $\mathcal{O}(\zeta^0,\epsilon^0)$ and $\mathcal{O}(\zeta^1,\epsilon^0)$ are known background components, while $\mathcal{O}(\zeta^0,\epsilon^1)$ can be reconstructed from linear perturbation of Kerr. 

For terms on the left-hand side of Eq.~\eqref{eq:3_to_1}, we will find the following pattern, where an operator $O$ operates on a field $\psi$, and we are interested in the $\mathcal{O}(\zeta^1,\epsilon^1)$ component, with 
\begin{align}
    (O\psi)^{(1,1)} 
    &=O^{(1,1)} \psi^{(0,0)}+O^{(0,1)} \psi^{(1,0)} \nonumber\\
    &+O^{(1,0)} \psi^{(0,1)}+O^{(0,0)} \psi^{(1,1)}\,.
\end{align}
As we shall see in Sec.~\ref{H0Psi0}, because of our gauge choice in Table~\ref{tab:psi_table}, the only non-vanishing $\mathcal{O}(\zeta^1,\epsilon^1)$ quantity we will encounter will be $\Psi_0^{(1,1)}$.

For terms on the right-hand side of Eq.~\eqref{eq:3_to_1}, we will argue in Sec.~\ref{subsubsec:source} that they can all be obtained from the background geometry and the $\mathcal{O}(\zeta^0,\epsilon^1)$ metric perturbation $h^{(0,1)}$ because GWs on the modified background $h^{(1,1)}$ do not contribute to the source term.

\subsubsection{Analysis of the General Modified Teukolsky Equation: \\ the $H_0\Psi_0$ and $H_1\Psi_1$ terms}
\label{H0Psi0}

For the first term on the left-hand side of Eq.~\eqref{eq:3_to_1}, expanding $H_0\Psi_0$ to $\mathcal{O}(\zeta^1,\epsilon^1)$, one finds the following three types of terms:
\begin{equation} \label{eq:H0_Psi0}
    (H_0\Psi_0)^{(1,1)}
    =H_0^{(0,0)}\Psi_0^{(1,1)}+H_0^{(1,0)}\Psi_0^{(0,1)}
    +H_0^{(0,1)}\Psi_0^{(1,0)}\,,
\end{equation}
Since at $\mathcal{O}(\zeta^0,\epsilon^0)$, Eq.~\eqref{eq:3_to_1} becomes $H_0^{(0,0)}\Psi_0^{(0,0)}=0$, $H_0^{(0,0)}$ is the Teukolsky differential operator that acts on $\Psi_0$ in GR, which was discussed in Sec.~\ref{sec:TeuknonGR}. Therefore, the first term in Eq.~\eqref{eq:H0_Psi0} is just the Teukolsky equation in GR but for $\Psi_0^{(1,1)}$. The second term vanishes in GR but is generically nonzero in modified gravity. This is because $\Psi_0^{(0,1)}$ is a solution to the Teukolsky equation presented in Eq.~\eqref{eq:teukpsi0}. As discussed in Sec.~\ref{sec:GRintro}, this is a gauge invariant quantity and thus non-vanishing in general. On the other hand, the operator $H_0^{(1,0)}$ can be evaluated using the background metric for the spacetime in the modified theory of gravity under consideration.

The third term only shows up for non-Petrov-type-D spacetime since $\Psi_0^{(1,0)}=0$ if the modified background spacetime is Petrov type D. The operator $H_0^{(0,1)}$ contains Weyl scalars, spin coefficients, and intrinsic derivatives at $\mathcal{O}(\zeta^0,\epsilon^1)$, so as discussed at the end of Sec.~\ref{sec:TeuknonGR}, we need to reconstruct the metric of GW perturbations in GR. By applying one of these metric reconstruction procedures and rotating the reconstructed tetrad to the gauge in Eq.~\eqref{eq:gauge_perturbation}, one is able to evaluate all the terms in $H_0^{(0,1)}$. 

The last two terms in Eq.~\eqref{eq:H0_Psi0} come from the homogeneous part of the Bianchi and Ricci identities. These terms are purely geometrical, and we can interpret them as source terms induced by stationary perturbations contained in the background geometry. We can then rewrite Eq.~\eqref{eq:H0_Psi0} as
\begin{align}
        & (H_0\Psi_0)^{(1,1)}=H_{0}^{\GR}\Psi_0^{(1,1)}
        -\mathcal{S}_{0,\typeD}^{(1,1)}
        -\mathcal{S}_{0,\nonD}^{(1,1)}\,, 
\end{align}
where we have defined
\begin{align}
        & H_{0}^{\GR}\equiv H_0^{(0,0)}\,, \\
        & \label{eq:S0typeD} \mathcal{S}_{0,\typeD}^{(1,1)}\equiv-H_0^{(1,0)}\Psi_0^{(0,1)}\,, \\
        & \label{eq:S0typenonD}\mathcal{S}_{0,\nonD}^{(1,1)}\equiv-H_0^{(0,1)}\Psi_0^{(1,0)}\,.
\end{align}

Moving on to the second term on the left-hand side of Eq.~\eqref{eq:3_to_1} and using properties in Table~\ref{tab:psi_table}, we obtain
\begin{equation} \label{eq:H1_Psi1}
    (H_1\Psi_1)^{(1,1)}=H_1^{(0,1)}\Psi_1^{(1,0)}\,.
\end{equation}
Similar to $H_0^{(0,1)}$, $H_1^{(0,1)}$ is also made up of Weyl scalars, spin coefficients, and intrinsic derivatives at $\mathcal{O}(\zeta^0,\epsilon^1)$, so we need metric reconstruction for this term as well. This term vanishes in any Petrov type D spacetime since $\Psi_1=0$ with an appropriate choice of gauge at the background level. Similar to $H_0^{(0,1)}\Psi_0^{(1,0)}$, we can effectively treat $H_1^{(0,1)}\Psi_1^{(1,0)}$ as a source term involving $\Psi_1^{(1,0)}$ and induced by the stationary perturbation of background geometry. Let us then define
\begin{equation} \label{eq:S1nonD}
    \mathcal{S}_{1,\nonD}^{(1,1)}\equiv H_1^{(0,1)}\Psi_1^{(1,0)}\,.
\end{equation}

The source term $\mathcal{S}_{1,\nonD}^{(1,1)}$ along with the source terms $\mathcal{S}_{0,\typeD}^{(1,1)}$ and $\mathcal{S}_{0,\nonD}^{(1,1)}$ given in Eqs.~\eqref{eq:S0typeD}-\eqref{eq:S0typenonD} come from the homogeneous part of the Bianchi and Ricci identities. Grouping these source terms together, we define
\begin{equation}
\label{eq:source_free_driving}
    \mathcal{S}_{\geo}^{(1,1)}\equiv \mathcal{S}_{0,\typeD}^{(1,1)}+\mathcal{S}_{0,\nonD}^{(1,1)}+\mathcal{S}_{1,\nonD}^{(1,1)}\,.
\end{equation}
	
\subsubsection{Analysis of the General Modified Teukolsky Equation: \\ the $\mathcal{S}$ term}
\label{subsubsec:source}

Besides the source terms generated by the correction to the background metric, we also have corrections to the Einstein-Hilbert action due to modified gravity theory, including extra fields not present in GR (i.e., class A beyond GR theories) or higher-order terms in curvature (i.e., class B beyond GR theories) as discussed in detail in Sec.~\ref{sec:beyondGR-theory}. In a perturbative treatment, all these corrections manifest as some source terms on the right-hand side of the Einstein equations, so we have a non-zero ``effective'' stress tensor, or in the trace reversed form, a non-zero Ricci tensor, even in the case without ordinary matter (see e.g., the discussion of dCS gravity, EdGB gravity, and higher-derivative gravity cases in Sec.~\ref{sec:beyondGR-theory}). 

Let us first look at class A beyond GR theories, where there are additional fields introduced by modified gravity, such as the pseudo scalar field coupled to the Pontryagin density in dCS gravity. Let us focus on one of these extra fields, which we represent generically as $\vartheta$. Since this field vanishes in GR, $\vartheta^{(0,0)}=\vartheta^{(0,1)}=0$ also in general. From Eqs.~\eqref{eq:source_bianchi_1}-\eqref{eq:source_bianchi_2}, we see that the terms in $S_i$ couple $\Phi_{ij}$ with either the directional derivatives or the spin coefficients. According to Eq.~\eqref{eq:Ricci_NP}, the $\Phi_{ij}$ are linear functions of $R_{\mu\nu}$ contracted with the tetrad basis,
\begin{equation}
	\Phi_{ij} \propto  R_{\mu\nu}e_{i}{}^{\mu}e_{j}{}^{\nu}\,, \quad \{i,j\} \in \{0,1,2\} \,.
\end{equation}
Since $\vartheta^{(0,0)}=\vartheta^{(0,1)}=0$, $\Phi_{ij}^{(1,1)}\sim\vartheta^{(1,0)}h^{(0,1)}+\vartheta^{(1,1)}g^{(0,0)}$, where $g^{(0,0)}$ represents the terms only involving background metric in GR. Then, $S_i$ in $\mathcal{S}$ can only enter at $\mathcal{O}(\zeta^1)$, so
\begin{equation} \label{eq:source1_11}
    \begin{aligned}
        \mathcal{S}^{(1,1)}
        =& \;\mathcal{E}^{(0,0)}_2 S_2^{(1,1)}
        -\mathcal{E}^{ (0,0)}_1 S_1^{(1,1)} \\
        & \;+\mathcal{E}^{(0,1)}_2 S_2^{(1,0)}
        -\mathcal{E}^{(0,1)}_1 S_1^{(1,0)} \\
        \sim& \;\vartheta^{(1,0)}h^{(0,1)}+\vartheta^{(1,1)}g^{(0,0)}\,.
    \end{aligned}
\end{equation}
The source ${\cal{S}}$ at ${\cal{O}}(\zeta^1,\epsilon^1)$, $\mathcal{S}^{(1,1)}$, couples the GWs in GR and the extra field $\vartheta$, so we need to solve the equations of motions of these non-gravitational fields to find their contributions to the stress tensor and $\mathcal{S}^{(1,1)}$ in the modified Teukolsky equation. In our notation, the modified Teukolsky equation describing the evolution of the GW perturbations due to the modification to GR can then be expressed as
\be \label{eq:masterbgr}
H_{0}^{\GR}\Psi_0^{(1,1)} = \mathcal{S}_{0,\typeD}^{(1,1)} + \mathcal{S}_{0,\nonD}^{(1,1)} + \mathcal{S}_{1,\nonD}^{(1,1)} + \mathcal{S}^{(1,1)} \,,
\ee
where all the quantities have been defined in Eqs.~\eqref{eq:S0typeD},~\eqref{eq:S0typenonD},~\eqref{eq:S1nonD} and~\eqref{eq:source1}. Notice that the differential operator acting on $\Psi_0^{(1,1)}$ is the same as the differential operator that appears in the Teukolsky equation for GR BH spacetimes discussed previously in Sec.~\ref{sec:TeukGR}.

One can find the solution to these extra fields in different ways. One way is to solve the equations of motions of these extra fields and the modified Teukolsky equation in parallel. Another way is to use the order-reduction scheme introduced in \cite{Okounkova_Stein_Scheel_Hemberger_2017}, in which one solves the equations of motions of these extra fields first and then insert them into the modified Teukolsky equation. Notice here that we have absorbed the coupling constant multiplying $\vartheta$ in $R_{\mu\nu}$ into the perturbative order of $\vartheta$. For example, as discussed in Sec.~\ref{sec:perturbation_scheme}, $\vartheta$ itself is usually of $\mathcal{O}(\alpha_{\bGR})$, where $\alpha_{\bGR}$ is the coupling constant in front of $\mathcal{L}_{\bGR}$ in Eq.~\eqref{eq:lagrangian}. The same coupling constant also shows up in front of these beyond GR corrections in $R_{\mu\nu}$, e.g., Eqs.~\eqref{eq:tracericci} and ~\eqref{eq:tracericci-sgb}, so the contribution of $\vartheta$ to $R_{\mu\nu}$ is of $\mathcal{O}(\alpha_{\bGR}^2)$ or $\mathcal{O}(\zeta)$. Thus, the equation of motion of $\vartheta$ is at lower order than the gravitational field equation, which allows us to follow the order-reduction scheme in \cite{Okounkova_Stein_Scheel_Hemberger_2017}, although this procedure is likely to introduce secularly-growing uncontrolled remainders. All these calculations depend on the details of the target modified gravity theory, so we will not discuss them in detail here, and instead, provide some examples in Sec.~\ref{sec:examples_non_metric_EOM} and leave the case-by-case study to future work. 

Another way to generate these source terms is due to corrections to the Einstein-Hilbert action that are only made up of gravitational fields, e.g., higher-derivative gravity \cite{Burgess_2004, Donoghue_2012, Endlich_Gorbenko_Huang_Senatore_2017, Cano_Ruiperez_2019}, which we classified as class B beyond GR theories in Sec.~\ref{sec:beyondGR-theory}. In this case, by pure order counting, the kind of terms that can appear are of the form $h^{(1,0)}h^{(0,1)}$. These terms are similar in form to $\mathcal{S}_{\geo}^{(1,1)}$, given in Eq.~\eqref{eq:source_free_driving}, and so have that $\mathcal{S}_{\geo}^{(1,1)} = {\cal{O}}(h^{(1,0)}h^{(0,1)})$. Therefore,
$\mathcal{S}_{\geo}^{(1,1)}$ takes the form of a coupling between the GWs in GR and the stationary modification to the background metric. Though $h^{(1,0)}$ can be generated by $\vartheta^{(1,0)}$, if we treat it as an arbitrary stationary correction to the background metric, the way it couples to GWs in GR is independent of the gravity theory, as we have discussed above. In contrast, the source terms coming from the non-vanishing stress tensor and made up of only gravitational fields depend on the details of the modified gravity theory, so they cannot be treated universally when only knowing the correction to the background metric. On the other hand, like these $\mathcal{S}_{\geo}^{(1,1)}$ terms, we do not need to solve the equations of motion of other non-gravitational fields, so these terms can be evaluated directly with the background metric and the reconstructed metric for GWs in GR when knowing the stress tensor in the target modified gravity theory.

One of the major successes of Teukolsky's formalism in GR, presented in Sec.~\ref{sec:TeukGR}, was the separation of the master equation into a radial and an angular equation, when written in a coordinate basis, such as in the Boyer-Lindquist coordinates of the Kerr BH spacetime. Each of these equations need then to be solved independently as an eigenvalue problem. Since the differential operator acting on the beyond GR, leading-order correction to GW perturbations remains unchanged from GR, the left-hand side of the beyond GR master equation in Eq.~\eqref{eq:masterbgr} is naturally separable into a radial and an angular part. Furthermore, one can separate the right-hand side of Eq.~\eqref{eq:masterbgr} by making use of the orthogonality properties of the spin-weighted spheroidal harmonics (which are the solution to the angular master equation for GR BH Petrov type D spacetimes) to project the source terms onto the original angular basis. Following this trick, the separability of the master equations into a radial and an angular equation must hold for beyond GR, Petrov type I, non-Ricci-flat spacetimes as well. When looking at the example theories presented in Sec.~\ref{sec:beyondGR-theory}, one may also encounter a mode coupling between different $\ell$ modes (e.g., between $\ell$ and $\ell \pm 1$ modes at leading order in the slow rotation expansion \cite{Wagle_Yunes_Silva_2021,Srivastava_Chen_Shankaranarayanan_2021,Pierini:2021jxd}). This is seen when coupling between different perturbation functions exist, both in GR~\cite{Pani:2013pma} and beyond GR theories~\cite{Wagle_Yunes_Silva_2021,Srivastava_Chen_Shankaranarayanan_2021,Pierini:2021jxd,Pierini:2022eim}.

\subsubsection{Examples of equations of motion of extra (non-metric) fields}
\label{sec:examples_non_metric_EOM}
In the previous section, we showed that to evaluate $\mathcal{S}^{(1,1)}$, one needs to solve the equations of motion of these non-metric extra fields. In this section, we provide the equations of motion of the pseudoscalar field $\vartheta$ in dCS gravity and the scalar field $\theta$ in EdGB gravity as a demonstration.

In dCS gravity, expanding the equation of motion of $\vartheta$ in Eq.~\eqref{eq:sfeqdcs} using the perturbation scheme in Eq.~\eqref{eq:expansion_Weyl}, we find, at $\mathcal{O}(\zeta^1,\epsilon^1)$,
\begin{align} \label{eq:dCS_EOM_scalar_11}
	\square^{(0,0)}\vartheta^{(1,1)}
	=-\pi^{-\frac{1}{2}}M^2\left[R^*\!R\right]^{(0,1)}
	-\square^{(0,1)}\vartheta^{(1,0)}\,,
\end{align}
where $R^*\!R$ is a shorthand for $~^*\!R^{\mu}{}_{\nu}{}^{\kappa\sigma}R^{\nu}{}_{\mu\kappa\sigma}$, and we follow \cite{Wagle_Yunes_Silva_2021} to use $\zeta_{\dCS}\equiv {16\pi\alpha_\dCS^2}/{M^4}$ as the dCS gravity expansion parameter. We have also absorbed a factor of $(\zeta_{\dCS})^{{1}/{2}}$ into the expansion of $\vartheta$. To solve Eq.~\eqref{eq:dCS_EOM_scalar_11} in the Teukolsky formalism, one first needs to project all quantities onto the NP tetrad. For example, the Pontryagin density and the wave operator decompose into
\begin{equation} \label{eq:R*R}
    R~^*\!R=8i\mathcal{E}(3\Psi_2^2-4\Psi_1\Psi_3+\Psi_0\Psi_4-c.c.)\,,
\end{equation}
\begin{equation} \label{eq:square_vartheta}
    \begin{aligned}
        \square\vartheta
        =& \;\left[\{\delta,\delta^{*}\}-\{D,\Delta\}
        +(\gamma+\gamma^{*}-\mu-\mu^{*})D\right. \\
        &\; \left.+(\rho+\rho^{*}-\varepsilon-\varepsilon^{*})\Delta
        +(\pi-\tau^{*}-\alpha+\beta^{*})\delta\right. \\
        &\; \left.+(\pi^{*}-\tau-\alpha^{*}+\beta)
        \delta^{*}\right]\vartheta\,,
    \end{aligned}
\end{equation}
where $i\mathcal{E}=\epsilon_{\mu\nu\rho\sigma}l^{\mu}n^{\nu}m^{\rho}\bar{m}^{\sigma}$, and $\mathcal{E}$ is a real function. These NP projected quantities now need to be expanded in the two-parameter scheme to properly evaluate Eq.~\eqref{eq:dCS_EOM_scalar_11} and then to solve it.

Similarly, in EdGB gravity, using $\zeta_{\EdGB}\equiv {16\pi\alpha_{\EdGB}^2}/{M^4}$ as the EdGB gravity expansion parameter and expanding Eq.~\eqref{eq:EdGB_scalar_EOM}, we find
\begin{equation} \label{eq:EdGB_EOM_scalar_11}
    \square^{(0,0)}\theta^{(1,1)}
    =-\pi^{-\frac{1}{2}}M^2\mathcal{G}^{(0,1)}-\square^{(0,1)}\theta^{(1,0)}\,.
\end{equation}
Now, the wave operator and the Gauss-Bonnet invariant must be projected onto the NP tetrad to find, once more that $\Box\theta$ is given by Eq.~\eqref{eq:square_vartheta} after replacing $\vartheta$ with $\theta$, and the NP projected $\mathcal{G}$ is
\begin{equation} \label{eq:R^2}
    \mathcal{G}=8(3\Psi_2^2-4\Psi_1\Psi_3+\Psi_0\Psi_4+c.c.)\,.
\end{equation}
Here, we also absorbed a factor of $(\zeta_{\EdGB})^{{1}/{2}}$ into the expansion of $\theta$. As before, to solve Eq.~\eqref{eq:EdGB_EOM_scalar_11}, one must now expand these NP projected quantities in our two-parameter scheme.

For both cases, we end up with a usual scalar field equation with source terms that depend on NP quantities at $\mathcal{O}(\zeta^{0},\epsilon^1)$. Thus, we can first reconstruct these NP quantities and then use Eqs.~\eqref{eq:R*R}, \eqref{eq:square_vartheta}, and Eq.~\eqref{eq:R^2} to express the source terms in terms of $\Psi_0^{(0,1)}$ or $\Psi_4^{(0,1)}$. After this, one can either solve the scalar field equation and the modified Teukolsky equation concurrently \cite{Wagle_Yunes_Silva_2021, Srivastava_Chen_Shankaranarayanan_2021, Pierini:2021jxd, Pierini:2022eim}, or use the order-reduction scheme to solve for the scalar field first and plug it into the modified Teukolsky equation.

\vspace{0.4cm}
To summarize, we have found the modified Teukolsky equation of $\Psi_0$ for any non-Ricci-flat and algebraically general background spacetime that can be treated as a linear perturbation of a Petrov type D spacetime, namely,
\begin{align} \label{eq:master_eqn_non_typeD_Psi0}
    H_{0}^{\GR}\Psi_0^{(1,1)}
    =\mathcal{S}_{\geo}^{(1,1)}+\mathcal{S}^{(1,1)}\,,
\end{align}
where we have defined
\begin{align}
    & \mathcal{S}_{\geo}^{(1,1)}=\mathcal{S}_{0,\typeD}^{(1,1)}
    +\mathcal{S}_{0,\nonD}^{(1,1)}+\mathcal{S}_{1,\nonD}^{(1,1)}\,, \nonumber\\
    & \mathcal{S}_{0,\typeD}^{(1,1)}=-H_0^{(1,0)}\Psi_0^{(0,1)}\,, \nonumber\\
    & \mathcal{S}_{0,\nonD}^{(1,1)}=-H_0^{(0,1)}\Psi_0^{(1,0)}\,, \nonumber\\
    & \mathcal{S}_{1,\nonD}^{(1,1)}=H_1^{(0,1)}\Psi_1^{(1,0)}\,,
\end{align}
where $H_0$ and $H_1$ are defined in Eqs.~\eqref{eq:operators_non_typeD}, and $\mathcal{S}$ is defined in Eq.~\eqref{eq:source1}. The equation for $\Psi_4$ can be derived by performing a GHP transformation on Eq.~\eqref{eq:master_eqn_non_typeD_Psi0},
\begin{align} \label{eq:master_eqn_non_typeD_Psi4}
    & H_{4}^{\GR}\Psi_4^{(1,1)}
    =\mathcal{T}_{\geo}^{(1,1)}+\mathcal{T}^{(1,1)}\,,
\end{align}
where we have defined
\begin{align} \label{eq:source_non_typeD_Psi4}
    & \mathcal{T}_{\geo}^{(1,1)}=\mathcal{T}_{4,\typeD}^{(1,1)}
        +\mathcal{T}_{4,\nonD}^{(1,1)}+\mathcal{T}_{3,\nonD}^{(1,1)}\,, \nonumber \\
        & \mathcal{T}_{4,\typeD}^{(1,1)}=-H_4^{(1,0)}\Psi_4^{(0,1)}\,, \nonumber\\
        & \mathcal{T}_{4,\nonD}^{(1,1)}=-H_4^{(0,1)}\Psi_4^{(1,0)}\,, \nonumber\\
        & \mathcal{T}_{3,\nonD}^{(1,1)}=H_3^{(0,1)}\Psi_3^{(1,0)}\,,
\end{align}
where $H_{4}^{\GR}$ is the Teukolsky operator for $\Psi_4$ in GR [see Eq.~\eqref{eq:H4}], and
\begin{equation} \label{eq:operators_non_typeD_Psi4}
    H_{3}\equiv \mathcal{E}_4 J_4 - \mathcal{E}_3 J_3\,. 
\end{equation}
For the source terms $\mathcal{S}_{\geo}^{(1,1)}$ or $\mathcal{T}_{\geo}^{(1,1)}$, they can be computed from the modified background metric, the solutions to the Teukolsky equation in GR, and the reconstructed metric for GWs in GR. For $\mathcal{S}^{(1,1)}$ or $\mathcal{T}^{(1,1)}$, we may need to solve the equations of motion of other non-gravitational fields and evaluate the stress tensor. We have collected the full expressions of all the terms in the modified Teukolsky equation above in Appendix~\ref{appendix:eqns_in_one_place}. In addition, the equations above are presented in an abstract form using NP symbols; they can be further simplified when considering perturbations of specific background spacetimes in specific coordinates and tetrads, e.g., Kerr in Boyer-Lindquist coordinates and in the Kinnersley tetrad.

%---------------------------------

\section{Extension of Framework to Higher Order in the Coupling}
\label{sec:higher_order_perturbations}

One important observation about Eqs.~\eqref{eq:master_eqn_non_typeD_Psi0} and \eqref{eq:master_eqn_non_typeD_Psi4} is that they are in a very similar format to the second-order Teukolsky equation in GR \cite{Campanelli:1998jv}. In this section, we discuss the connection between the leading-order modified Teukolsky formalism and the second-order Teukolsky formalism in GR, which demonstrates that many techniques well-developed (in different contexts) in GR can be directly reused in modified gravity. Moreover, we show that our formalism can be generalized to higher orders [i.e., $\mathcal{O}(\zeta^m,\epsilon^n)$, $m\geq0$, $n\geq1$], which is then a beyond GR extension of the higher-order Teukolsky formalism developed in \cite{Campanelli:1998jv} for GR. For a general discussion of non-linear multiple-parameter perturbation theory in relativity, we refer the reader to \cite{Sonego:1997np, Bruni:1996im, Bruni:2002sma, Sopuerta:2003rg}.

\subsection{Connection to the second-order Teukolsky formalism in GR}
\label{sec:second_order_Teuk_GR}
Since Teukolsky presented the linear-order perturbation equation in \cite{Teukolsky:1973ha}, higher-order Teukolsky equations have been of great interest to the community. On the one hand, the inability of the linear-order Teukolsky equation to estimate the errors due to the use of a perturbative expansion makes the study of higher-order Teukolsky equations necessary~\cite{Campanelli:1998jv}. On the other hand, higher-order perturbations enable the study of certain physical systems that cannot be studied sufficiently accurately within the linear-order scheme, such as head-on collisions in the close-limit approximation~\cite{Campanelli:1998jv, Abrahams:1995wd, Gleiser:1998rw}, self-force in EMRIs~\cite{Lousto:2008vw, Shah:2010bi, Keidl:2010pm, Pound:2012nt, Gralla:2012db, vandeMeent:2017zgy, Pound:2021qin, Loutrel_Ripley_Giorgi_Pretorius_2020}, etc. On the observational side, recent studies of non-linearities that show up in numerical relativity suggest that second- and higher-order perturbations may be important for the analysis of gravitational wave data  \cite{Ma:2022wpv, Mitman:2022qdl, Cheung:2022rbm}.

In~\cite{Campanelli:1998jv}, the Teukolsky equation was successfully extended to second- and higher-order, so let us show now that these higher-order equations are very similar to what we obtained in this paper. Comparing our Eq.~\eqref{eq:master_eqn_non_typeD_Psi4} to the vacuum case ($T^{\mu\nu}_{\matter}=0$) of Eqs.~(7)-(10) in \cite{Campanelli:1998jv}, these equations take a very similar format if we replace all the terms proportional to $h^{(0,1)}h^{(1,0)}$ with $h^{(0,1)}h^{(0,1)}$ and set the source term due to $\mathcal{L}_{\bGR}$ in Eq.~\eqref{eq:lagrangian} to zero, $\mathcal{T}^{(1,1)}=0$. More precisely, if we follow the approach in this work to derive the Teukolsky equation at $\mathcal{O}(\zeta^0,\epsilon^2)$, we find
\begin{equation} \label{eq:second_order_GR_Psi4_eqn}
    H_{4}^{\GR}\Psi_4^{(0,2)}=\mathcal{T}_{\geo}^{(0,2)}\,,\quad
    \mathcal{T}_{\geo}^{(0,2)}=-H_4^{(0,1)}\Psi_4^{(0,1)}\,.
\end{equation}
These are the equations that ought to be compared to the work in GR at second order in perturbation theory.

Equation~\eqref{eq:second_order_GR_Psi4_eqn} and Eqs.~(7)-(10) from~\cite{Campanelli:1998jv} are similar in form,
as expected in perturbation theory, where the principal part of the equation remains unchanged at each order and is driven by lower order perturbations. Nonetheless, our Eq.~\eqref{eq:second_order_GR_Psi4_eqn} is simpler. First, there are no terms in $\Psi_3^{(0,1)}$ since they are removed by our gauge choice in Eq.~\eqref{eq:gauge_perturbation}. Second, there are no terms that depend on $\lambda^{(0,1)}$ and $\nu^{(0,1)}$, since $\lambda$ and $\nu$, just like $\kappa$ and $\sigma$, are eliminated from the equations from the beginning, as shown in Sec.~\ref{sec:elimination_kappa_sigma}. To compare Eq.~\eqref{eq:second_order_GR_Psi4_eqn} with Eqs.~(7)-(10) from~\cite{Campanelli:1998jv}, we choose the same gauge given in Eq.~\eqref{eq:gauge_perturbation}. In this case, $\Psi_3^{(0,1)}=0$, and one can solve for $\lambda^{(0,1)}$ and $\nu^{(0,1)}$ in terms of $\Psi_4^{(0,1)}$~\cite{Chandrasekhar_1983}, so all the $\lambda^{(0,1)}$ and $\nu^{(0,1)}$ related terms become additional operators acting on $\Psi_4^{(0,1)}$ in Eq.~\eqref{eq:second_order_GR_Psi4_eqn}. In Appendix~\ref{appendix:consistency_check}, we have shown this consistency explicitly following this prescription.

Further, we notice that Eq.~\eqref{eq:second_order_GR_Psi4_eqn} and Eqs.~\eqref{eq:master_eqn_non_typeD_Psi4}-\eqref{eq:source_non_typeD_Psi4}, are also similar. When studying Petrov type \rom{1} spacetimes in modified gravity, we did not make any assumptions about what NP quantities vanish at $\mathcal{O}(\zeta^1,\epsilon^0)$ to avoid sabotaging our perturbation scheme, as discussed in Sec.~\ref{sec:gauge_background}. For the second-order Teukolsky formalism in GR, the stationary Petrov type I background at $\mathcal{O}(\zeta^1,\epsilon^0)$ is replaced by the ``dynamical background," driven by GW perturbations at $\mathcal{O}(\zeta^0,\epsilon^1)$, where most NP quantities also do not vanish. Due to this connection, many challenges shared by these two situations have been solved in the second-order Teukolsky formalism in GR, such as metric reconstruction at $\mathcal{O}(\zeta^0,\epsilon^1)$. The success of applying the second-order Teukolsky formalism to the study of self-force in \cite{Lousto:2008vw, Shah:2010bi, Keidl:2010pm, Pound:2012nt, Gralla:2012db, vandeMeent:2017zgy, Pound:2021qin, Loutrel_Ripley_Giorgi_Pretorius_2020} strongly suggests that our modified Teukolsky formalism is feasible numerically.

Despite these similarities, there are also differences between these two efforts. One major difference is the presence of extra non-metric fields in class A beyond GR theories. Unlike in GR, even without matter, one needs to evaluate the effective stress-energy tensor driven by these intrinsic extra fields, and thus, solve their equations of motion concurrently. Nonetheless, as discussed in Sec.~\ref{sec:examples_non_metric_EOM}, this issue was already dealt with in the studies of slowly-rotating BHs using metric perturbations in dCS \cite{Wagle_Yunes_Silva_2021, Srivastava_Chen_Shankaranarayanan_2021} and EdGB \cite{Pierini:2021jxd, Pierini:2022eim}. Besides the issue of extra fields, one also has to be careful when constructing the background tetrad in these non-Ricci-flat backgrounds, as shown in Sec.~\ref{sec:perturbation_scheme}.

\subsection{Modified Teukolsky formalism beyond $\mathcal{O}(\zeta^1,\epsilon^1)$}
\label{sec:teuknonD_nonlinear}
As illustrated in the previous section, second- and higher-order BH perturbation theory in GR has been of great interest due to its importance in constraining the first-order perturbations and its need when dealing with certain physical systems. In the case of modified gravity, one does not just have to deal with non-linear terms in $\epsilon$, but also with non-linear terms in the dimensionless coupling constant $\zeta$. When the beyond GR theory itself is known at higher order, these higher-order corrections due to modified gravity might be interesting, since there might be non-linear phenomena that is not described by the linear theory. For these reasons, we follow~\cite{Campanelli:1998jv} to extend our formalism beyond $\mathcal{O}(\zeta^1,\epsilon^1)$.

Let us consider some perturbations at $\mathcal{O}(\zeta^M,\epsilon^N)$, $M\geq0$, $N\geq1$. First, we need to find a tetrad with terms up to $\mathcal{O}(\zeta^M,\epsilon^N)$, such that the orthogonality condition in Eq.~\eqref{eq:tetrad_ortho} is satisfied while our perturbation scheme is preserved, similar to what we did in Sec.~\ref{sec:perturbation_scheme}. For $1\leq m\leq M$, expanding the correction to the tetrad at $\mathcal{O}(\zeta^m,\epsilon^0)$, we have
\begin{equation}
    \delta e_{a\mu}^{(m,0)}=A_{ab}^{(m,0)}\delta e_{b\mu}^{(0,0)}\,.
\end{equation}
Through induction, one can easily show that we can solve for all $A_{ab}^{(m,0)}$ iteratively, where $1\leq m\leq M$. Let us assume $\delta e_{b\mu}^{(1,0)}\,,\cdots\,,\delta e_{b\mu}^{(M-1,0)}$ are known, and the base case $\delta e_{b\mu}^{(1,0)}$ was shown in Sec.~\ref{sec:perturbation_scheme}. We also assume that the corrections to the background metric $h_{\mu\nu}^{(1,0)}\,,\cdots\,, h_{\mu\nu}^{(M,0)}$ are known. Then, to satisfy Eq.~\eqref{eq:tetrad_ortho}, we need 
\begin{equation}
    \begin{aligned}
        & \left(e_{a\mu}^{(0,0)}
        +\sum_{m=1}^{M}\zeta^{m}\delta e_{a\mu}^{(m,0)}\right)
        \left(e_{b\nu}^{(0,0)}
        +\sum_{m=1}^{M}\zeta^{m}\delta e_{b\nu}^{(m,0)}\right) \\
        & \left(g^{\mu\nu(0,0)}
        +\sum_{m=1}^{M}\zeta^{m}h^{\mu\nu(m,0)}\right)
        =\eta_{ab}\,.
    \end{aligned}
\end{equation}
For convenience, let us introduce
\begin{equation}
    \mathcal{U}^{(M,0)}
    \equiv\sum_{\substack{i+j+k=M,\\M>i,j,k>0}}
    \delta e_{a\mu}^{(i,0)}\delta e_{b\nu}^{(j,0)}h^{\mu\nu(k,0)}\,,
\end{equation}
where every term on the right-hand side is assumed to be known, and  $\mathcal{U}^{(0,0)}=0$ when $M=1$. Then, following the same procedure as in Sec.~\ref{sec:perturbation_scheme}, at $\mathcal{O}(\zeta^M,\epsilon^0)$ we have
\begin{equation}
    2A_{(ab)}^{(M,0)}=-h_{ab}^{(M,0)}-\mathcal{U}^{(M,0)}\,,
\end{equation}
where $\mathcal{U}^{(M,0)}$ contains $A_{(ab)}^{(m,0)}$, with $1\leq m<M$ solved in the previous steps. If we pick the same gauge as in Sec.~\ref{sec:perturbation_scheme} to set $A_{[ab]}^{(M,0)}=0$, then we find
\begin{equation}
    A_{ab}^{(M,0)}=-\frac{1}{2}\left(h_{ab}^{(M,0)}+\mathcal{U}^{(M,0)}\right)\,.
\end{equation}
Thus, this proves that one can iteratively find higher-order corrections to the background tetrad, such that the orthogonality condition in Eq.~\eqref{eq:tetrad_ortho} is preserved.

Next, let us consider tetrad rotations. Inspecting the rotations we performed in Eqs.~\eqref{eq:rotate2_reduce_typeD} and \eqref{eq:rotate2_reduce}, one can immediately notice that, under any type II rotation [cf. Eq.~\eqref{eq:rotate2}] with rotation parameter $b^{(m,n)}$ at $\mathcal{O}(\zeta^m,\epsilon^n)$ with $m\geq0$, $n\geq1$, the Weyl scalars at $\mathcal{O}(\zeta^m,\epsilon^n)$ transform as
\begin{equation} \label{eq:rotate2_reduce_nonlinear}
	\begin{array}{l}
		\Psi_{0}^{(m,n)}\rightarrow\Psi_{0}^{(m,n)}+4b^{(m,n)}\Psi_{1}^{(0,0)}\,, \\
		\Psi_{1}^{(m,n)}\rightarrow\Psi_{1}^{(m,n)}+3b^{(m,n)}\Psi_{2}^{(0,0)}\,, \\
		\Psi_{2}^{(m,n)}\rightarrow\Psi_{2}^{(m,n)}+2b^{(m,n)}\Psi_{3}^{(0,0)}\,, \\
		\Psi_{3}^{(m,n)}\rightarrow\Psi_{3}^{(m,n)}+b^{(m,n)}\Psi_{4}^{(0,0)}\,, \\ 
		\Psi_{4}^{(m,n)}\rightarrow\Psi_{4}^{(m,n)}\,,
	\end{array} 
\end{equation}
where any terms beyond $\mathcal{O}(\zeta^m,\epsilon^n)$ are dropped. Since the background at  $\mathcal{O}(\zeta^0,\epsilon^0)$ is Petrov type D, where $\Psi_{0,1,3,4}^{(0,0)}=0$, if we pick $b^{(m,n)}=-\Psi_1^{(m,n)}/(3\Psi_2^{(0,0)})$, then
\begin{equation} \label{eq:rotate2_result_nonlinear}
	\begin{array}{l}
		\Psi_{1}^{(m,n)}\rightarrow0\,,\quad	
		\Psi_{0,2,3,4}^{(m,n)}\rightarrow\Psi_{0,2,3,4}^{(m,n)}\,.
	\end{array}
\end{equation}
Similarly, by performing a type I rotation with the rotation parameter $a^{(m,n)}=-\left[\Psi_3^{(m,n)}/(3\Psi_2^{(0,0)})\right]^{*}$, one can remove $\Psi_3^{(m,n)}$.

One may worry that a rotation at $\mathcal{O}(\zeta^{m_1},\epsilon^{n_1})$ will affect the Weyl scalars at $\mathcal{O}(\zeta^{m_2},\epsilon^{n_2})$, where $m_2>m_1\,,n_2>n_1$, since many Weyl scalars at  $\mathcal{O}(\zeta^{m_2-m_1},\epsilon^{n_2-n_1})$ might be nonzero. However, this problem can be avoided if one performs these rotations systematically from lower order to higher order. For example, one may consider the following procedures:
\begin{enumerate}
    \item Perform tetrad rotations step by step from $\mathcal{O}(\zeta^0,\epsilon^1)$ to $\mathcal{O}(\zeta^M,\epsilon^1)$ to remove $(\Psi_{1,3}^{(0,1)}\,,\cdots\,,\Psi_{1,3}^{(M,1)})$.
    \item Next, perform tetrad rotations step by step from $\mathcal{O}(\zeta^0,\epsilon^2)$ to $\mathcal{O}(\zeta^M,\epsilon^2)$ to remove $(\Psi_{1,3}^{(0,2)}\,,\cdots\,,\Psi_{1,3}^{(M,2)})$.
    \item $\cdots$\nonumber
    \item At the $N$-th step, perform tetrad rotations step by step from $\mathcal{O}(\zeta^0,\epsilon^N)$ to $\mathcal{O}(\zeta^M,\epsilon^N)$ to remove $(\Psi_{1,3}^{(0,N)}\,,\cdots\,,\Psi_{1,3}^{(M,N)})$.
\end{enumerate}
Following this sequence, any higher-order modifications to $\Psi_{1,3}$ due to lower-order rotations are removed at the corresponding step, and higher-order rotations do not affect the lower-order $\Psi_{1,3}$, which have been set to $0$. Thus, for any perturbation at $\mathcal{O}(\zeta^{M},\epsilon^{N})$ with $M\geq0$, $N\geq1$, we can consistently set 
\begin{equation} \label{eq:gauge_Psi1_nonlinear}
    \Psi_{1,3}^{(m,n)}=0\,,\quad0\leq m\leq M\,,\;1\leq n\leq N\,.
\end{equation}

Now, one can directly make an expansion of Eq.~\eqref{eq:3_to_1} similar to what we did at $\mathcal{O}(\zeta^1,\epsilon^1)$ in Sec.~\ref{sec:The_modified_Teukolsky_equation}. One direct consequence of the tetrad rotations above is that we can drop all $\Psi_1^{(m,n)}$, with $m\geq0$, $n\geq1$ [e.g., Eq.~\eqref{eq:gauge_Psi1_nonlinear}], so there is only the stationary part of $\Psi_1$ contributing to Eq.~\eqref{eq:3_to_1}. Then, following the same procedures as in Sec.~\ref{sec:The_modified_Teukolsky_equation}, for perturbations at $\mathcal{O}(\zeta^M,\epsilon^N)$, we find
\begin{align} \label{eq:master_eqn_non_typeD_Psi0_nonlinear}
    H_{0}^{\GR}\Psi_0^{(M,N)}
    =\mathcal{S}_{\geo}^{(M,N)}+\mathcal{S}^{(M,N)}\,,
\end{align}
where
\begin{align} \label{eq:sources_non_typeD_Psi0_nonlinear}
    & \mathcal{S}_{\geo}^{(M,N)}=\mathcal{S}_{0,I}^{(M,N)}
    +\mathcal{S}_{0,II}^{(M,N)}+\mathcal{S}_{1}^{(M,N)}\,, \nonumber\\
    & \mathcal{S}_{0,I}^{(M,N)}
    =\sum_{(m,n)=(0,1)}^{(m,n)<(M,N)}-H_0^{(M-m,N-n)}\Psi_0^{(m,n)}\,, \nonumber\\
    & \mathcal{S}_{0,II}^{(M,N)}
    =\sum_{m=1}^{M}-H_0^{(M-m,N)}\Psi_0^{(m,0)}\,, \nonumber\\
    & \mathcal{S}_{1}^{(M,N)}
    =\sum_{m=1}^{M}-H_1^{(M-m,N)}\Psi_1^{(m,0)}\,, \nonumber\\
    & \mathcal{S}^{(M,N)}
    =\sum_{m=1,n=0}^{(m,n)\leq(M,N)}
    \left[\mathcal{E}_{2}^{(M-m,N-n)}S_{2}^{(m,n)}\right. \nonumber\\
    & \left.-\mathcal{E}_{1}^{(M-m,N-n)}S_{1}^{(m,n)}\right]\,,
\end{align}
and where $(m,n)<(M,N)$ means $m\leq M,n<N$ or $m<M,n\leq N$. The equation for $\Psi_4$ can be found from the GHP transformation of Eqs.~\eqref{eq:master_eqn_non_typeD_Psi0_nonlinear}-\eqref{eq:sources_non_typeD_Psi0_nonlinear}. For the case of higher-order perturbations in GR, $\zeta=0$, so one can simply set $S_{0,II}^{(M,N)}=S_{1}^{(M,N)}=\mathcal{S}^{(M,N)}=0$, where the sum starts from $\mathcal{O}(\zeta^1)$. As discussed in Sec.~\ref{sec:second_order_Teuk_GR} and shown in detail in Appendix~\ref{appendix:consistency_check}, if one chooses the gauge in which $\Psi_{1,3}^{(0,n)}=0$, with $1\leq n\leq N$, then Eqs.~(7)-(10) of \cite{Campanelli:1998jv} are the same as the GHP transformation of Eqs.~\eqref{eq:master_eqn_non_typeD_Psi0_nonlinear}-\eqref{eq:sources_non_typeD_Psi0_nonlinear}. Thus, one can treat this higher-order extension of our formalism as a modified-gravity generalization of the higher-order Teukolsky formalism in \cite{Campanelli:1998jv}.

\subsection{Potential challenges}
\label{sec:potential_challenges}
In the previous subsection, we have successfully extended our formalism to higher order in both $\epsilon$ and $\zeta$. In this case, all NP quantities are decoupled at each perturbative order, and Weyl scalars $\Psi_{0,4}$ can be solved, given their solutions at lower orders. This shows that similar to any perturbation theory problem (e.g., solving the hydrogen atom in quantum mechanics), by working out the leading-order perturbation theory, one can iterate it to solve for higher-order perturbations. On the other hand, this procedure also inherits the same challenges of any perturbation theory solution. For example, the source terms made up of lower-order perturbations become complicated at very high order. However, developing a non-perturbative approach is outside the scope of this work, and one may have to rely on numerical relativity in the end. In this subsection, we will discuss other challenges and potential solutions when applying this higher-order modified Teukolsky formalism to the first few orders beyond $\mathcal{O}(\zeta^1,\epsilon^1)$ [e.g., $\mathcal{O}(\zeta^2,\epsilon^1)$ or $\mathcal{O}(\zeta^1,\epsilon^2)$], where perturbation theory is still tractable.

The major challenge of this higher-order modified Teukolsky formalism is the need of metric reconstruction in non-Ricci-flat backgrounds, since we need to evaluate NP quantities at $\mathcal{O}(\zeta^m,\epsilon^n)$ with $m>0$, $n\geq1$ in general. For example, at $\mathcal{O}(\zeta^2,\epsilon^1)$ or $\mathcal{O}(\zeta^1,\epsilon^2)$, one needs to reconstruct the perturbed metric at $\mathcal{O}(\zeta^1,\epsilon^1)$. At this order, we have taken advantage of the fact that the metric reconstruction procedure for $\mathcal{O}(\zeta^0,\epsilon^1)$ GW perturbations in GR is well developed~\cite{Chrzanowski:1975wv, Kegeles_Cohen_1979, Whiting_Price_2005, Yunes_Gonzalez_2006, Chandrasekhar_1983, Loutrel_Ripley_Giorgi_Pretorius_2020}. However, for general perturbations at $\mathcal{O}(\zeta^m,\epsilon^n)$, the metric reconstruction procedure is unknown. Moreover, when $m>0$, the correction to the Einstein-Hilbert action generates some effective stress-energy tensor [see Sec.~\ref{sec:beyondGR-theory}], so the traceless condition $g^{\mu\nu}h_{\mu\nu}=0$ in the radiation gauge used in these metric reconstruction procedures with a Hertz potential \cite{Chrzanowski:1975wv, Kegeles_Cohen_1979, Whiting_Price_2005, Yunes_Gonzalez_2006} is violated.

However, this issue is not just present in our modified Teukolsky formalism, but also in the higher-order Teukolsky formalism in GR, since lower-order perturbations become effective sources in the higher-order version of the Teukolsky equation. References~\cite{Green:2019nam, Toomani:2021jlo, Pound:2021qin} have shown that one can extend the Hertz potential approach by adding certain correction fields to the metric perturbation constructed from a usual Hertz potential. These correction fields can be obtained from certain decoupled ordinary differential equations, sourced by the effective stress-energy tensor. These references have proven that this procedure works for any smooth, compactly-supported source, which is unfortunately not satisfied by sources driven by non-linear couplings of gravitational fields. Thus, to apply their formalism to our non-linear Teukolsky formalism, additional work would have to be done. Besides an extension of the Hertz potential approach, there are also methods that do not rely on the radiation gauge, such as the approach of solving the remaining NP equations directly \cite{Chandrasekhar_1983, Loutrel_Ripley_Giorgi_Pretorius_2020, Ripley_Loutrel_Giorgi_Pretorius_2021}. This approach has been implemented for vacuum Petrov type D spacetimes \cite{Loutrel_Ripley_Giorgi_Pretorius_2020, Ripley_Loutrel_Giorgi_Pretorius_2021}, and it is worth exploring whether one can extend it to non-vacuum backgrounds.

Another challenge is the presence of extra fields. For the class A beyond GR theories mentioned in Sec.~\ref{sec:beyondGR-theory}, one has to solve the coupled equations of metric fields and extra fields at each perturbed order. In terms of solving the coupled equation itself, this will not be a huge challenge since similar problems have been solved in these approaches using metric perturbations \cite{Wagle_Yunes_Silva_2021, Srivastava_Chen_Shankaranarayanan_2021}. There might be numerical challenges when going to very high order since the source terms are complicated non-linear couplings of reconstructed NP quantities with extra fields at lower orders, which need to be solved together with the modified Teukolsky equation. Nonetheless, this is merely an unavoidable consequence of perturbation theory.

To summarize, the connection of our work to the second-order Teukolsky formalism in GR demonstrates the feasibility of the approach presented in this work. When applying our formalism to specific modified gravity theories, one should not expect more difficulties than when solving the second-order Teukolsky equation in GR, which has been widely studied. On the other hand, the formalism developed in this work aims to incorporate corrections from modified gravity, so it contains features unique to modified gravity and cannot be directly obtained from the second-order Teukolsky formalism in GR. The extension of our formalism to higher order naturally generalizes the higher-order Teukolsky formalism in \cite{Campanelli:1998jv} from GR to modified gravity. As a consistency check, we have studied the limiting case of $\zeta \to 0$, compared the results to those obtained in~\cite{Campanelli:1998jv}, and presented these concrete comparisons in Appendix~\ref{appendix:consistency_check}.

%%%%%%%%%%%%%%%%%%%%%%%%%%%%%%%%%%%%%%%

\section{Discussions}
\label{sec:discussions}

In this work, we extended the Teukolsky formalism to non-Ricci-flat, Petrov type D BH backgrounds, as well as to non-Ricci-flat, Petrov type \rom{1} BH backgrounds that can be treated as a linear perturbation of a Petrov type D background. We began by presenting a brief review of the derivation of the Teukolsky equation for a Ricci-flat and Petrov type D background in GR via the original approach in Teukolsky's paper~\cite{Teukolsky:1973ha}, as well as using an approach proposed by Chandrasekhar~\cite{Chandrasekhar_1983}. These two approaches differ in the method adopted to eliminate the $\Psi_1$ and $\Psi_3$ dependence from the two Bianchi identities and one Ricci identity [see e.g., Eq.~\eqref{eq:Bianchi_simplified}]. Teukolsky's approach makes use of additional Bianchi identities to obtain a commutation relation to eliminate $\Psi_1$ and $\Psi_3$. Chandrasekhar's approach uses the available gauge freedom to make a convenient gauge choice that eliminates $\Psi_1$ and $\Psi_3$ directly. One can then solve these equations to obtain a single decoupled differential equation for the perturbed Weyl scalars $\Psi_0$ and $\Psi_4$.

We first extended both approaches to obtain the modified Teukolsky equation in a generic modified gravity theory that allows BH backgrounds to be non-Ricci-flat and Petrov type D backgrounds. Since the background is now non-Ricci-flat, we have additional non-vanishing background NP quantities. We then used the two approaches described above to obtain decoupled differential equations for the perturbed Weyl scalars $\Psi_0$ and $\Psi_4$. We found that for non-Ricci-flat, Petrov type D BH backgrounds in modified gravity, the master equations for curvature perturbations acquire a source term [see e.g., Eqs.~\eqref{eq:nongr-teuk} and~\eqref{eq:nongr-teuk-psi4}]. In order to evaluate these source terms, we found that one needs to perform metric reconstruction from the GR curvature perturbations~\cite{Chrzanowski:1975wv, Kegeles_Cohen_1979, Whiting_Price_2005, Yunes_Gonzalez_2006, Chandrasekhar_1983, Loutrel_Ripley_Giorgi_Pretorius_2020} [i.e., to ${\cal{O}}(\zeta^0,\epsilon^1)$, where $\zeta$ labels the order of the GR deformation, and $\epsilon$ labels the order of the dynamic GW perturbation from the stationary background]. We showed that both the Teukolsky's approach and the Chandrasekhar's approach lead to the same modified Teukolsky equation, but the latter is algebraically simpler and thus more convenient. 

The algebraic simplicity of Chandrasekhar's approach makes this method ideal for the study of curvature perturbations of BH backgrounds that are non-Ricci-flat and Petrov type \rom{1}. We thus extended Chandrasekhar's approach to such BH backgrounds. The non-vanishing of the background NP Ricci scalars, the background NP spin coefficients, and the background Weyl scalars $\Psi_1$,$\Psi_2$, and $\Psi_3$ forces the NP equations [see e.g., Eq.~\eqref{eq:Bianchi_simplified}] to have additional non-vanishing NP quantities. However, when one requires the BH background to be a perturbation of a non-Ricci-flat, Petrov type D BH background at leading order in the GR deformation, the equations do decouple. This is achieved by rotating the tetrad such that the perturbed Weyl scalars $\Psi_1^{(1,1)}$ and $\Psi_3^{(1,1)}$ (at linear order in both the non-GR expansion parameter and the GW expansion parameter) vanish. With this, we then derived a single decoupled differential equation for $\Psi_0^{(1,1)}$ and $\Psi_4^{(1,1)}$.

The modified Teukolsky equation obtained in this way has the structure of the traditional Teukolsky equation but with certain source terms. The differential operator on the left-hand side of the modified Teukolsky equation acts on the perturbed Weyl scalar $\Psi_{0,4}$, and it has a functional form that is similar to the Teukolsky operators in GR~\cite{Teukolsky:1973ha}. The source terms on the right-hand side of the modified Teukolsky equation arise either because of either (i) modifications to the stationary BH background spacetime, or (ii) additional stress-tensor terms due to corrections to the Einstein-Hilbert action.

The first type of source terms comes from the homogeneous part of certain Bianchi and Ricci identities [see e.g., Eqs.~\eqref{eq:Bianchi_simplified}]. Some of these source terms can be directly evaluated using the modified background metric and the solution to the Teukolsky equation in GR. The rest are couplings of $\mathcal{O}(\zeta^1,\epsilon^0)$ corrections to the Weyl scalars with the $\mathcal{O}(\zeta^0,\epsilon^1)$ corrections to the metric due to GWs in GR. Thus, in order to evaluate these source terms, we need to reconstruct the metric for the curvature perturbations in GR \cite{Chrzanowski:1975wv, Kegeles_Cohen_1979, Whiting_Price_2005, Yunes_Gonzalez_2006, Chandrasekhar_1983, Loutrel_Ripley_Giorgi_Pretorius_2020}, just as in the case of non-Ricci-flat, Petrov type D backgrounds. 

The second type of source terms comes from the stress tensor due to corrections to the Einstein-Hilbert action. We have classified the modified gravity theories into two classes based on the presence or absence of extra non-gravitational dynamical fields. Class A beyond GR theories can have couplings to other dynamical scalar, vector or tensor fields (as is the case in dCS gravity \cite{Jackiw:2003pm,Alexander:2009tp}, EdGB gravity \cite{Gross_Sloan_1987, Kanti:1995vq, Moura:2006pz}, Horndeski theory \cite{Kobayashi:2019hrl}, scalar-tensor theories \cite{Sotiriou:2015lxa}, $f(R)$ gravity \cite{Sotiriou:2006hs, Sotiriou:2008rp}, Einstein-Aether theory \cite{Jacobson_2008}, and bi-gravity \cite{Schmidt-May:2015vnx}). Class B beyond GR theories depend only on the gravitational field and there are no additional dynamical fields (as is the case in certain effective field theory extensions of GR, such as higher-derivative gravity \cite{Burgess_2004, Donoghue_2012, Endlich_Gorbenko_Huang_Senatore_2017, Cano_Ruiperez_2019}). For class B beyond GR theories, these source terms can be directly evaluated with the background metric and the reconstructed metric. For class A beyond GR theories, one must solve the equations of motion for these extra fields to evaluate the stress tensor, and this can only be done on a theory-per-theory basis. The case-by-case treatment of these extra field equations is left to future work.

The major goal of this work was to simplify the perturbed gravitational equations in general for modified gravity theories that admit non-Ricci-flat and Petrov type \rom{1} or Petrov type D BH backgrounds such that all the curvature perturbations are packed into two fundamental variables $\Psi_0$ and $\Psi_4$. With this at hand, one can now in principle evaluate all source terms and separate the modified Teukolsky equation into radial and angular parts to solve for the QNM frequencies of perturbed BHs in modified gravity. It is important to realize that this was not possible until this work due to the inherently complicated nature of the perturbed field equations when working with metric perturbations. Indeed, up until now, the QNM spectrum of perturbed BHs in modified gravity had only been studied for non-rotating BHs  [e.g., in dCS gravity \cite{Cardoso:2009pk, Molina:2010fb, Pani_Cardoso_Gualtieri_2011}, EdGB theory \cite{Blazquez-Salcedo:2016enn, Blazquez-Salcedo_Khoo_Kunz_2017}, Einstein-Aether theory \cite{Konoplya_Zhidenko_2006, Konoplya_Zhidenko_2007, Ding_2017, Ding_2019, Churilova_2020}, higher-derivative gravity (quadratic \cite{Cardoso_Kimura_Maselli_Senatore_2018}, cubic \cite{deRham_Francfort_Zhang_2020}, and more generic \cite{Cardoso_Kimura_Maselli_Berti_Macedo_McManus_2019, McManus_Berti_Macedo_Kimura_Maselli_Cardoso_2019}), and Horndeski gravity \cite{Tattersall_Ferreira_2018}] or for slowly-rotating BHs (e.g., in EdGB theory \cite{Pierini:2021jxd}, dCS gravity \cite{Wagle_Yunes_Silva_2021,Srivastava_Chen_Shankaranarayanan_2021}, and higher-derivative gravity \cite{Cano:2020cao, Cano_Fransen_Hertog_Maenaut_2021}). The only study of QNM perturbations of rotating BHs was carried out in dCS gravity from numerical relativity simulations of BH mergers, but these suffer from secularly-growing uncontrolled remainders~\cite{Okounkova:2019dfo, Okounkova_Stein_Moxon_Scheel_Teukolsky_2020}. 

Our work creates a new path to directly calculate the corrections to the QNM frequencies of perturbed BHs with arbitrary spin in modified gravity and, more generally, any background spacetime that can be treated as a linear perturbation of a Petrov type D spacetime. One of our next major goals is to do a case-by-case study of all these well-motivated modified theories, using the formalism developed here, to then use GW observations to constrain these theories. For dCS gravity, we would like to compare the QNM frequencies obtained for arbitrarily rotating BHs to those found in the slow-rotation approximation to linear order in spin~\cite{Wagle_Yunes_Silva_2021}, as well as others that use metric perturbations \cite{Cardoso:2009pk, Molina:2010fb, Pani_Cardoso_Gualtieri_2011, Wagle_Yunes_Silva_2021, Srivastava_Chen_Shankaranarayanan_2021} and numerical relativity \cite{Okounkova_Stein_Scheel_Hemberger_2017, Okounkova_Scheel_Teukolsky_2019, Okounkova:2019dfo, Okounkova_Stein_Moxon_Scheel_Teukolsky_2020}.

By extending the Teukolsky formalism, we have also laid the foundation for studying gravitational perturbations other than QNMs around BHs in modified gravity. For example, the Teukolsky formalism has been applied to compute gravitational waveforms and energy/angular momentum fluxes sourced by a point particle orbiting around a BH in extreme mass-ratio binary inspirals (EMRI) \cite{Poisson_1993, Cutler_Finn_Poisson_Sussman_1993, Apostolatos_Kennefick_Ori_Poisson_1993, Poisson_Sasaki_1995, Poisson_1995, Tanaka_Tagoshi_Sasaki_1996}. The same procedure has been applied to a few modified gravity theories, e.g., in scalar-tensor theories \cite{Yunes:2011aa} and for a spinning horizonless compact object \cite{Maggio_vandeMeent_Pani_2021}, where the Teukolsky formalism in GR can be directly applied. With this extended Teukolsky formalism, we are now able to study EMRIs in a much wider class of modified gravity theories. These results can also be compared with those obtained using post-Newtoninan studies of EMRIs in GR and modified gravity~\cite{Kocsis:2011dr,Moore:2018kvz,Moore:2019xkm,Moore:2020rva,Sopuerta_Yunes_2009,Yagi:2011xp,Pani_Cardoso_Gualtieri_2011}.

Another example is the break of isospectrality (where even and odd parity modes have the same QNM frequencies) in certain modified gravity theories, e.g., dCS gravity \cite{Cardoso:2009pk, Molina:2010fb, Pani_Cardoso_Gualtieri_2011, Wagle_Yunes_Silva_2021, Srivastava_Chen_Shankaranarayanan_2021}, EdGB gravity \cite{Blazquez-Salcedo:2016enn, Blazquez-Salcedo_Khoo_Kunz_2017, Pierini:2021jxd}, and higher-derivative gravity \cite{Cano_Fransen_Hertog_Maenaut_2021}. The study of isospectrality is mostly done with metric perturbations since the Zerilli-Moncrief and the Regge-Wheeler functions naturally divide the metric perturbations into even and odd parity sectors~\cite{Zerilli:1971wd, Regge:PhysRev.108.1063}. For BHs with arbitrary spin, there are no known extensions of the Zerilli-Moncrief and the Regge-Wheeler functions, so we may have to use NP quantities in this extended Teukolsky formalism to study parity breaking. Since Teukolsky equation does not naturally classify its solutions into different parities, we will first need to understand better what even and odd parity modes mean in the Teukolsky formalism and their connections to the Zerilli-Moncrief and Regge-Wheeler functions even in GR. This, and much more, is now possible thanks to the derivation of a master evolution equation for curvature perturbations in modified gravity. 

In this work, we have focused on the formalism up to leading order in modified gravity corrections, i.e., at $\mathcal{O}(\zeta)$. This is mainly because the theories we have discussed in Sec.~\ref{sec:beyondGR-theory} are only presented to leading order in corrections since these are treated in an effective field theory approach, considering small deformations from GR. However, one can consider a modified theory of gravity different from the examples shown in Sec.~\ref{sec:beyondGR-theory}, where one can look at higher-order deformations from GR. As discussed in Sec.~\ref{sec:higher_order_perturbations}, our leading-order formalism can be extended to higher order [$\mathcal{O}(\zeta^m,\epsilon^n)$, $m\geq0$, $n\geq1$] by iterating the perturbation scheme in Sec.~\ref{sec:perturbation_scheme} and the procedure of finding the master equation in Sec.~\ref{sec:teuknonD}. However, utmost care needs to be taken when considering theories at higher than leading-order corrections to GR, as such theory may admit ghost modes~\cite{Yagi:2012ya}. Additionally, this formalism relies on the approximation that the theories mentioned in Sec.~\ref{sec:beyondGR-theory} are an effective field theory of GR. Therefore, the spacetimes we can probe using this formalism cannot deviate too much from their GR counterparts.

To present the feasibility of our formalism extending the Teukolsky equation to non-Ricci-flat Petrov type D and Petrov type \rom{1} spacetimes, our collaboration is already working on a series of calculations. The first in this planned series of works is the study of perturbations of a non-Ricci-flat vacuum Petrov type D BH spacetime representing a slowly-rotating BH to leading order in spin in dCS gravity~\cite{dcstyped1}. In~\cite{dcstyped1}, we will present the calculation of the perturbed field equations. These field equations, as expected from the results of this paper, are sourced equations which we will compute in the null basis. We will then implement the necessary metric reconstruction procedures and tetrad rotations. In the last step, we will convert all NP quantities to a coordinate basis to separate the master equation into radial and angular ordinary differential equations with couplings between the gravitational and scalar sectors. Then, in a follow-up work~\cite{dcstyped2}, we will make use of the EVP method to calculate the QNM frequencies of these BH spacetimes and verify our results with previously obtained frequencies computed in the slow-rotation limit~\cite{Wagle_Yunes_Silva_2021,Srivastava_Chen_Shankaranarayanan_2021}. We will then extend these calculations to arbitrarily spinning BHs in dCS gravity, which are described by non-Ricci-flat, vacuum, Petrov type \rom{1} BH metrics in~\cite{dcstype1}. This problem is more challenging due to the presence of additional theory-independent source terms (i.e., $\mathcal{S}_{\geo}^{(1,1)}$), which need metric reconstruction (e.g., $\mathcal{S}_{0,\nonD}^{(1,1)}$). However, it is much simpler to evaluate these additional terms than the theory-dependent source terms (i.e., $\mathcal{S}^{(1,1)}$) coupled to the pseudoscalar field, which we would have already computed in our previous work~\cite{dcstyped1} on Petrov type D BHs in dCS gravity mentioned above. We expect that through these extensions, we will acquire a deep knowledge of QNMs in modified gravity.

\vspace{0.2cm}
\noindent {\bf{Note added after completion:}} While writing up our analysis, we became aware of an equivalent and independent analysis of decoupled equations for gravitational perturbations around BHs in modified gravity~\cite{Hussain:2022ins}. Instead of using the NP formalism, Ref.~\cite{Hussain:2022ins} focuses mostly on the Einstein equations and shows how to partially decouple them, following the order-reduction scheme in~\cite{Okounkova_Stein_Scheel_Hemberger_2017}. To make the equations of gravitational perturbations separable, Ref.~\cite{Hussain:2022ins} uses Wald's formalism to project the Einstein equations onto a (modified) Teukolsky equation \cite{Wald_1978}. Although our work is independent of that of Ref.~\cite{Hussain:2022ins}, there are similarities in the general format of the final master equation. For example, both approaches require metric reconstruction of GWs in GR. Reference~\cite{Hussain:2022ins} also presents a direct derivation of the modified Teukolsky equation following Teukolsky's original approach~\cite{Teukolsky:1973ha}. Our work greatly simplifies the NP approach through the use of gauge freedom, following Chandrasekhar's approach \cite{Chandrasekhar_1983}. These two independent studies can be used to validate results when computing the shift of QNM frequencies in certain modified gravity theories. 

%%%%%%%%%%%%%%%%%%%%%%%%%%%%%%%%%%%%%%
\section{Acknowledgements}

We thank Aaron Zimmerman, Asad Hussain, Kwinten Fransen, and Adrian Chung for helpful discussions. We thank Yasmine Steele for creating the key image used by Physical Review X. N. Y. and P. W. acknowledge support from the Simons Foundation through Award No. 896696 and National Science Foundation (NSF) Grant No. PHY-2207650. Y. C. and D. L. acknowledge support from the Brinson Foundation, the Simons Foundation (Award No. 568762), and NSF Grants No. PHY-2011961, No. PHY-2011968, and No. PHY-1836809. We thank KITP, which is supported in part by NSF Grant No. PHY-1748958, for hosting and supporting the visit of one of us during the final stages of the completion of this manuscript. Some of our algebraic work used the package {\sc xAct}~\cite{xact} for Mathematica.

%%%%%%%%%%%%%%%%%%%%%%%%%%%%%%%%%%%%%%

\appendix

\section{NP formalism (continued)} \label{appendix:NPformalism_addition}

In Sec.~\ref{sec:NPformalism}, we have presented the orthogonality relations for the tetrad basis vectors in NP formalism. One can further compactly express the relation in Eq.~\eqref{eq:tetrad_ortho} as $g_{\mu\nu} = e^{a}_{\mu}e^{b}_{\nu}\eta_{ab}$ where,
\begin{align} \label{eq:ztetrad}
    e_{m}^{\mu} &= \left(l^{\mu},n^{\mu},m^{\mu},\Bar{m}^{\mu}\right) \,, \nonumber \\
    \eta_{ab} = \eta^{ab} &= \begin{pmatrix} 0 & -1 & 0 & 0 \\ -1 & 0 & 0 & 0 \\ 0 & 0 & 0 & 1 \\ 0 & 0 & 1 & 0 \end{pmatrix} \,,
\end{align}
where we have used Latin indices to denote the null tetrad indices whereas the Greek indices are the tensor indices. Further, using the metric and the null tetrad, we can define the quantity known as Ricci rotation coefficients, which are similar to Christoffel symbols. These are complex quantities in nature and defined as
\be
\label{eq:ricci_rotation_coeffs}
    \gamma_{cab}=e_{a\mu;\nu}e_{c}^{\mu}e_{b}^{\nu}
\ee
with the symmetry,
\be
    \gamma_{cab}=-\gamma_{acb} \,.
\ee
The commutation relations of the intrinsic derivatives are related to the Ricci rotation coefficients by
\begin{equation} \label{eq:commutation_relation}
    \left[e_{a}^{\mu},e_{b}^{\mu}\right]
    =\left(\gamma_{cba}-\gamma_{cab}\right)e^{c\mu}\,.
\end{equation}

The tetrad components of the Riemann tensor can then be defined by
\begin{align}
    R_{abcd}=R_{\alpha\beta\gamma\delta}
    e^\alpha_a e^\beta_b e^\gamma_c e^\delta_d \,.
\end{align}
Using a form of Eq.~\eqref{eq:ricci_rotation_coeffs}, the Riemann tensor can also be expressed in terms of the Ricci rotation coefficients,
\begin{align}
\label{eq:riemann_ricci_rot}
    R_{abcd}
    =& \;-\gamma_{abc,d}+\gamma_{abd,c}
    +\gamma_{abf}\left(\gamma^f{}_{cd}-\gamma^{f}{}_{dc}\right) \nonumber\\
    & \;+\gamma^{f}{}_{ac}\gamma_{bfd}-\gamma^{f}{}_{ad}\gamma_{bfc}\,,
\end{align}
where $\gamma_{abc,d}\equiv\gamma_{abc,\mu}e_{d}^{\mu}$. The relationship among the Riemann tensor, Weyl tensor $C_{\alpha\beta\gamma\delta}$, and Ricci tensor $R_{\alpha\beta}$ remains unchanged in tetrad notation.
\begin{align}
\label{eq:Weyl_Riemann}
    R_{abcd}=C_{abcd}&-\frac{1}{2}
    \left(\eta_{ac}R_{bd}-\eta_{bc}R_{ad}-\eta_{ad}R_{bc}
    +\eta_{bd}R_{ac}\right) \nonumber\\  
    & +\frac{1}{6}\left(\eta_{ac}\eta_{bd}-\eta_{ad}\eta_{bc}\right)R \,.
\end{align}
In tetrad notation, Bianchi identities ($R_{\alpha\beta[\gamma\delta;\mu]}=0$) take the form,
\begin{align}
\label{eq:Bianchi_gen}
    R_{ab[cd;f]}
    =& \;\frac{1}{6}\sum_{[cdf]}
    \left[R_{abcd,f}-\eta^{nm}\left(\gamma_{naf}R_{mbcd}
    \right.\right. \nonumber\\
    & \;\left.\left.+\gamma_{nbf}R_{amcd}+\gamma_{ncf}R_{abmd}
    +\gamma_{ndf}R_{abcm}\right)\right]\,.
\end{align}

\subsection{NP quantities}
\label{appendix:NPquantity}

With the formalism developed above, Newman and Penrose defined twelve complex functions known as the spin coefficients which can be defined in terms of the Ricci rotation coefficients (and thus the tetrad). The spin coefficients are as follows:
\begin{align} \label{eq:spin-coeff}
\kappa & = \gamma_{131} = l_{\mu;\nu} m^\mu l^\nu \,, \nonumber \\
\pi & = - \gamma_{241} = - n_{\mu;\nu} \Bar{m}^\mu l^\nu \,, \nonumber \\
\varepsilon & = \frac{1}{2}(\gamma_{121} - \gamma_{341}) = \frac{1}{2} (l_{\mu;\nu} n^\mu l^\nu - m_{\mu;\nu} \Bar{m}^\mu l^\nu) \,, \nonumber \\
\rho & = \gamma_{134} = l_{\mu;\nu} m^\mu \Bar{m}^\nu \,, \nonumber \\
\lambda & = - \gamma_{244} = -n_{\mu;\nu} \Bar{m}^\mu \Bar{m}^\nu \,, \nonumber \\
\alpha & = \frac{1}{2}(\gamma_{124} - \gamma_{344}) = \frac{1}{2} (l_{\mu;\nu} n^\mu  \Bar{m}^\nu - m_{\mu;\nu} \Bar{m}^\mu  \Bar{m}^\nu) \,, \nonumber \\
\sigma & = \gamma_{133} = l_{\mu;\nu} m^\mu m^\nu \,, \nonumber \\
\mu & = - \gamma_{243} = -n_{\mu;\nu}  \Bar{m}^\mu m^\nu \,, \nonumber \\
\beta & = \frac{1}{2}(\gamma_{123} - \gamma_{343}) = \frac{1}{2} (l_{\mu;\nu} n^\mu  m^\nu - m_{\mu;\nu} \Bar{m}^\mu  m^\nu) \,, \nonumber \\
\nu & = -\gamma_{242} = -n_{\mu;\nu} \Bar{m}^\mu n^\nu \,, \nonumber \\
\gamma & = \frac{1}{2}(\gamma_{122} - \gamma_{342}) = \frac{1}{2} (l_{\mu;\nu} n^\mu  n^\nu - m_{\mu;\nu} \Bar{m}^\mu  n^\nu) \,, \nonumber \\
\tau & = \gamma_{132} = l_{\mu;\nu} m^\mu n^\nu \,.
\end{align}

Using Eq.~\eqref{eq:Weyl_Riemann}, one can decompose the Riemann tensor into the Weyl tensor, completely determined by $5$ complex Weyl scalars,
\begin{align} \label{eq:Weylscalar_NP}
    \Psi_0 &=C_{1313}=C_{\alpha\beta\gamma\delta}
    l^\alpha m^\beta l^\gamma m^\delta \,, \nonumber \\
    \Psi_1 &=C_{1213}=C_{\alpha\beta\gamma\delta}
    l^\alpha n^\beta l^\gamma m^\delta \,, \nonumber \\
    \Psi_2 &=C_{1342}=C_{\alpha\beta\gamma\delta}
    l^\alpha m^\beta \Bar{m}^\gamma n^\delta\,, \nonumber \\
    \Psi_3 &=C_{1242}=C_{\alpha\beta\gamma\delta}
    l^\alpha n^\beta \Bar{m}^\gamma n^\delta \,, \nonumber \\
    \Psi_4 &=C_{2424}=C_{\alpha\beta\gamma\delta}
    n^\alpha \Bar{m}^\beta n^\gamma \Bar{m}^\delta \,,
\end{align}
the Ricci tensor, and the Ricci scalar, characterized by $10$ NP Ricci scalars,
\begin{equation} \label{eq:Ricci_NP}
    \begin{aligned}
        & \Phi_{00}=\frac{1}{2}R_{11}
        =\frac{1}{2}R_{\mu\nu}l^{\mu}l^{\nu}\,, \\
        & \Phi_{01}=\frac{1}{2}R_{13}
        =\frac{1}{2}R_{\mu\nu}l^{\mu}m^{\nu}\,,\;
        \Phi_{10}=\frac{1}{2}R_{14}
        =\frac{1}{2}R_{\mu\nu}l^{\mu}\bar{m}^{\nu}\,,\\
        & \Phi_{11}=\frac{1}{4}(R_{12}+R_{34})
        =\frac{1}{2}R_{\mu\nu}(l^{\mu}n^{\nu}+m^{\mu}\bar{m}^{\nu})\,, \\
        & \Phi_{02}=\frac{1}{2}R_{33}
        =\frac{1}{2}R_{\mu\nu}m^{\mu}m^{\nu}\,,\;
        \Phi_{12}=\frac{1}{2}R_{23}
        =\frac{1}{2}R_{\mu\nu}n^{\mu}m^{\nu}\,, \\
        & \Phi_{20}=\frac{1}{2}R_{44}
        =\frac{1}{2}R_{\mu\nu}\bar{m}^{\mu}\bar{m}^{\nu}\,,\;
        \Phi_{21}=\frac{1}{2}R_{24}
        =\frac{1}{2}R_{\mu\nu}n^{\mu}\bar{m}^{\nu}\,, \\
        & \Phi_{22}=\frac{1}{2}R_{22}
        =\frac{1}{2}R_{\mu\nu}n^{\mu}n^{\nu}\,,\;
        \Lambda=R/24 \,. 
    \end{aligned}
\end{equation}

\subsection{NP equations}
\label{appendix:NPeqns}

Using the NP quantities defined above, one can consider appropriate linear combinations of Eq.~\eqref{eq:riemann_ricci_rot} and rewrite the equations in terms of the NP quantities. The resulting equations are called Ricci identities in \cite{Chandrasekhar_1983} and given by
\begin{subequations} \label{eq:Ricci_ids_all}
\allowdisplaybreaks
\begin{align} 
D\rho-\delta^{*}\kappa
=& \;\left(\rho^{2}+\sigma\sigma^{*}\right)
+(\varepsilon+\varepsilon^{*})\rho-\kappa^{*}\tau \nonumber\\
& \;-\kappa(3\alpha+\beta^{*}-\pi)+\Phi_{00} \,,
\\
D\sigma-\delta\kappa
=& \;(\rho+\rho^{*})\sigma+(3\varepsilon-\varepsilon^{*})\sigma \nonumber\\
& \;-(\tau-\pi^{*}+\alpha^{*}+3\beta)\kappa+\Psi_{0} \,,
\\
D\tau-\Delta\kappa
=& \;(\tau+\pi^{*})\rho+(\tau^{*}+\pi)\sigma
+(\varepsilon-\varepsilon^{*})\tau \nonumber\\
& \;-(3\gamma+\gamma^{*}) \kappa+\Psi_{1}+\Phi_{01} \,,
\\
D\alpha-\delta^{*}\varepsilon
=& \;(\rho+\varepsilon^{*}-2\varepsilon)\alpha+\beta\sigma^{*}
-\beta^{*}\varepsilon \nonumber\\
& \;-\kappa\lambda-\kappa^{*}\gamma+(\varepsilon+\rho) \pi+\Phi_{10} \,,
\\
D\beta-\delta\varepsilon
=& \;(\alpha+\pi)\sigma+(\rho^{*}-\varepsilon^{*})\beta \nonumber\\
& \;-(\mu+\gamma)\kappa-(\alpha^{*}-\pi^{*})\varepsilon+\Psi_{1} \,,
\\
D\gamma-\Delta\varepsilon
=& \;(\tau+\pi^{*})\alpha+(\tau^{*}+\pi)\beta
-(\varepsilon+\varepsilon^{*})\gamma \nonumber\\
& \;-(\gamma+\gamma^{*})\varepsilon+\tau\pi-\nu\kappa
+\Psi_{2}-\Lambda+\Phi_{11} \,,
\\
D\lambda-\delta^{*}\pi
=& \;(\rho\lambda+\sigma^{*}\mu)+\pi^{2}
+(\alpha-\beta^{*})\pi \nonumber\\
& \;-\nu\kappa^{*}-(3\varepsilon-\varepsilon^{*})\lambda
+\Phi_{20} \,,
\\
D\mu-\delta \pi
=& \;(\rho^{*}\mu+\sigma\lambda)+\pi\pi^{*}
-(\varepsilon+\varepsilon^{*})\mu \nonumber\\
& \;-\pi(\alpha^{*}-\beta)-\nu\kappa+\Psi_{2}+2\Lambda \,,
\\
D\nu-\Delta\pi
=& \;(\pi+\tau^{*})\mu+(\pi^{*}+\tau)\lambda
+(\gamma-\gamma^{*})\pi \nonumber\\
& \;-(3\varepsilon+\varepsilon^{*})\nu
+\Psi_{3}+\Phi_{21} \,,
\\
\Delta\lambda-\delta^{*}\nu
=& \;-(\mu+\mu^{*})\lambda-(3\gamma-\gamma^{*})\lambda \nonumber\\
& \;+(3\alpha+\beta^{*}+\pi-\tau^{*})\nu-\Psi_{4} \,,
\\
\delta\rho-\delta^{*}\sigma
=& \;\rho(\alpha^{*}+\beta)-\sigma(3\alpha-\beta^{*}) 
+(\rho-\rho^{*})\tau\nonumber\\
& \;+(\mu-\mu^{*})\kappa
-\Psi_{1}+\Phi_{01} \,,
\\
\delta\alpha-\delta^{*}\beta
=& \;(\mu\rho-\lambda\sigma)+\alpha\alpha^{*}+\beta\beta^{*}
-2\alpha\beta \nonumber\\
& \;+\gamma(\rho-\rho^{*})+\varepsilon(\mu-\mu^{*})-\Psi_{2}
+\Lambda+\Phi_{11} \,,
\\
\delta\lambda-\delta^{*}\mu
=& \;(\rho-\rho^{*})\nu+(\mu-\mu^{*})\pi
+\mu(\alpha+\beta^{*}) \nonumber\\
& \;+\lambda(\alpha^{*}-3\beta)-\Psi_{3}+\Phi_{21} \,,
\\
\delta\nu-\Delta\mu
=& \;\left(\mu^{2}+\lambda\lambda^{*}\right)
+(\gamma+\gamma^{*})\mu \nonumber\\
& \;-\nu^{*}\pi+(\tau-3\beta-\alpha^{*})\nu+\Phi_{22} \,,
\\
\delta\gamma-\Delta\beta
=& \;(\tau-\alpha^{*}-\beta)\gamma+\mu\tau
-\sigma\nu-\varepsilon\nu^{*} \nonumber\\
& \;-\beta(\gamma-\gamma^{*}-\mu)+\alpha\lambda^{*}+\Phi_{12} \,,
\\
\delta\tau-\Delta\sigma
=& \;\left(\mu\sigma+\lambda^{*}{\rho}\right)
+(\tau+\beta-\alpha^{*})\tau \nonumber\\
& \;-(3\gamma-\gamma^{*})\sigma-\kappa\nu^{*}+\Phi_{02} \,,
\\
\Delta\rho-\delta^{*}\tau
=& \;-(\rho\mu^{*}+\sigma\lambda)
+(\beta^{*}-\alpha-\tau^{*})\tau \nonumber\\
& \;+(\gamma+\gamma^{*})\rho+\nu\kappa-\Psi_{2}-2\Lambda \,,
\\
\Delta\alpha-\delta^{*}\gamma
=& \;(\rho+\varepsilon)\nu-(\tau+\beta)\lambda\nonumber\\
& \;+(\gamma^{*}-\mu^{*})\alpha+(\beta^{*}-\tau^{*})\gamma-\Psi_{3} \,.
\end{align}
\end{subequations}

Similarly, rewriting Eq.~\eqref{eq:Bianchi_gen} in terms of the NP quantities, one gets a set of equations called Bianchi identities in \cite{Chandrasekhar_1983}. These equations are given by
\begin{subequations}
\label{eq:bianchi_iden}
\allowdisplaybreaks
\begin{align}
& (\delta^{*}-4\alpha+\pi)\Psi_{0}
-(D-4\rho-2\varepsilon)\Psi_{1}-3\kappa\Psi_{2}=S_1\,, \\
& (\Delta-4\gamma+\mu)\Psi_{0}-(\delta-4\tau-2\beta)\Psi_{1}
-3\sigma\Psi_{2}=S_2\,, \\
& (\delta+4\beta-\tau)\Psi_{4}-(\Delta+2\gamma+4\mu)\Psi_{3}
+3\nu\Psi_{2}=S_3\,, \\
& (D+4\varepsilon-\rho)\Psi_{4}-(\delta^{*}+4\pi+2\alpha)\Psi_{3}
+3\lambda\Psi_{2}=S_4\,, \\
& (\delta^{*}+3\pi)\Psi_{2}-(D+2\varepsilon-2\rho)\Psi_{3}
-2\lambda\Psi_{1}-\kappa\Psi_{4}=S_5\,, \\
& (\Delta+3\mu)\Psi_{2}-(\delta+2\beta-2\tau)\Psi_{3}
-2\nu\Psi_{1}-\sigma\Psi_{4}=S_6\,, \\
\label{eq:bianchi_delpsi2}
& (\delta-3\tau)\Psi_{2}-(\Delta-2\gamma+2\mu)\Psi_{1}
+\nu\Psi_{0}+2\sigma\Psi_{3}=S_7\,, \\
\label{eq:bianchi_dpsi2}
& (D-3\rho)\Psi_{2}-(\delta^{*}+2\pi-2\alpha)\Psi_{1}
+\lambda\Psi_{0}+2\kappa\Psi_{3}=S_8\,, \\
& \delta^{*}\Phi_{01}+\delta\Phi_{10}
-D(\Phi_{11}+3\Lambda)-\Delta\Phi_{00} \nonumber\\
& =\kappa^{*}\Phi_{12}+\kappa\Phi_{21}
+(2\alpha+2\tau^{*}-\pi)\Phi_{01} \nonumber\\
& +(2\alpha^{*}+2\tau-\pi^{*})\Phi_{10}
-2(\rho+\rho^{*})\Phi_{11} \nonumber\\
& -\sigma^{*}\Phi_{02}-\sigma\Phi_{20} +\left[\mu+\mu^{*}-2(\gamma+\gamma^{*})\right]\Phi_{00}\,, \\
& \delta^{*}\Phi_{12}+\delta\Phi_{21}
-\Delta(\Phi_{11}+3\Lambda)-D\Phi_{22} \nonumber\\
& =-\nu\Phi_{01}-v^{*}\Phi_{10}
+(\tau^{*}-2\beta^{*}-2\pi)\Phi_{12} \nonumber\\
& +(\tau-2\beta-2\pi^{*})\Phi_{21}
+2(\mu+\mu^{*})\Phi_{11} \nonumber\\
& -(\rho+\rho^{*}-2\varepsilon-2\varepsilon^{*})\Phi_{22}
+\lambda\Phi_{02}+\lambda^{*}\Phi_{20}\,, \\ \nonumber\\
& \delta\left(\Phi_{11}-3\Lambda\right)
-D\Phi_{12}-\Delta\Phi_{01}+\delta^{*}\Phi_{02} \nonumber\\
& =\Phi_{22}-\nu^{*}\Phi_{00}
+(\tau^{*}-\pi+2\alpha-2\beta^{*})\Phi_{02} \nonumber\\
& -\sigma\Phi_{21}+\lambda^{*}\Phi_{10}
+2(\tau-\pi^{*})\Phi_{11} \nonumber\\
& -(2\rho+\rho^{*}-2\varepsilon^{*})\Phi_{12}
+(2\mu^{*}+\mu-2\gamma)\Phi_{01}\,,
\end{align}
\end{subequations}
where $S_i$ are related to the Ricci tensor and defined to be
\begin{subequations}
\begin{align}
S_1
\equiv& \;\left(\delta+\pi^{*}-2\alpha^{*}-2\beta\right)\Phi_{00}
-\left(D-2\varepsilon-2\rho^{*}\right)\Phi_{01} \nonumber\\
& \;+2\sigma\Phi_{10}-2\kappa\Phi_{11}-\kappa^{*}\Phi_{02}\,, \\
S_2
\equiv& \;\left(\delta+2\pi^{*}-2\beta\right)\Phi_{01}
-\left(D-2\varepsilon+2\varepsilon^{*}-\rho^{*}\right)\Phi_{02} \nonumber\\
& \;-\lambda^{*}\Phi_{00}+2\sigma\Phi_{11}-2\kappa\Phi_{12}\,, \\
S_3
\equiv& \;-\left(\Delta+2\mu^{*}+2\gamma\right)\Phi_{21}
+\left(\delta^{*}-\tau^{*}+2\alpha+2\beta^{*}\right)\Phi_{22} \nonumber\\
& \;+2\nu\Phi_{11}+\nu^{*}\Phi_{20}-2\lambda\Phi_{12}\,, \\
S_4
\equiv& \;-\left(\Delta+\mu^{*}+2\gamma-2\gamma^{*}\right)\Phi_{20}
+\left(\delta^{*}+2\alpha-2\tau^{*}\right)\Phi_{21} \nonumber\\
& \;+2\nu\Phi_{10}-2\lambda\Phi_{11}+\sigma^{*}\Phi_{22}\,, \\
S_5
\equiv& \;(\delta-2\alpha^{*}+2\beta+\pi^{*})\Phi_{20}
-(D-2\rho^{*}+2\varepsilon)\Phi_{21} \nonumber\\
& \;-2\mu\Phi_{10}+2\pi\Phi_{11}-\kappa^{*}\Phi_{22}
-2\delta^{*}\Lambda\,, \\
S_6
\equiv& \;(\delta+2\pi^{*}+2\beta)\Phi_{21}
-(D-\rho^{*}+2\varepsilon+2 \varepsilon^{*})\Phi_{22} \nonumber\\
& \;-2\mu\Phi_{11}-\lambda^{*}\Phi_{20}+2\pi\Phi_{12}
-2\Delta\Lambda\,, \\
S_7
\equiv& \;-(\Delta+2\mu^{*}-2\gamma)\Phi_{01}
+(\delta^{*}-\tau^{*}+2\beta^{*}-2\alpha)\Phi_{02} \nonumber\\
& \;+2\rho\Phi_{12}+\nu^{*}\Phi_{00}-2\tau\Phi_{11}
-2\delta\Lambda\,, \\
S_8
\equiv& \;-(\Delta+\mu^{*}-2\gamma-2\gamma^{*})\Phi_{00}
+(\delta^{*}-2\alpha-2\tau^{*})\Phi_{01} \nonumber\\
& \;+2\rho\Phi_{11}+\sigma^{*}\Phi_{02}-2\tau\Phi_{10}
-2D\Lambda\,.
\end{align}
\end{subequations}
For the Bianchi identities, we have re-organized the terms and shuffled the sequence of equations in comparison to the one in \cite{Chandrasekhar_1983}, so our equations here are consistent with the equations in Sec.~\ref{sec:NPformalism}.

Finally, the commutation relation in Eq.~\eqref{eq:commutation_relation} can be written as
\begin{subequations}
    \begin{align}
        [\Delta,D]
        =& \;\left(\gamma+\gamma^{*}\right)D
        +\left(\varepsilon+\varepsilon^{*}\right)\Delta
        -\left(\tau^{*}+\pi\right)\delta \nonumber\\
        & \;-\left(\tau+\pi^{*}\right)\delta^{*}\,, \\
        [\delta,D] 
        =& \;\left(\alpha^{*}+\beta-\pi^{*}\right)D+\kappa\Delta
        -\left(\rho^{*}+\varepsilon-\varepsilon^{*}\right)\delta
        \nonumber \\
        & -\sigma\delta^{*}\,, \\
        [\delta,\Delta] 
        =& \;-\nu^{*}D+\left(\tau-\alpha^{*}-\beta\right) \Delta+\left(\mu-\gamma+\gamma^{*}\right)\delta \nonumber\\
        & \;+\lambda^{*}\delta^{*}\,, \\
        [\delta^{*},\delta]
        =& \;\left(\mu^{*}-\mu\right)D+\left(\rho^{*}-\rho\right) \Delta+\left(\alpha-\beta^{*}\right)\delta \nonumber\\
        & \;+\left(\beta-\alpha^{*}\right)\delta^{*}\,.
    \end{align}
\end{subequations}

\subsection{Tetrad rotations}
\label{appendix:rotations}

In Sec.~\ref{sec:NPformalism}, we have mentioned that the tetrad basis vectors can be rotated in certain ways such that the orthogonality conditions in Eq.~\eqref{eq:tetrad_ortho} are still preserved. As discussed in \cite{Chandrasekhar_1983}, all these tetrad rotations can be classified into three types,
\begin{subequations} \label{eq:tetrad_rotations}
    \begin{align}
        \begin{split} \label{eq:rotate1}
	    	\text{\rom{1}}: 
		    & \;l\rightarrow l\,,\; m\rightarrow m+al\,,\;
		    \bar{m}\rightarrow \bar{m}+a^*l\,,\; \\
		    & \;n\rightarrow n+a^*m+a\bar{m}+aa^*l\,.
	    \end{split} \\
	    \begin{split} \label{eq:rotate2}
	    	\text{\rom{2}}: 
		    & \;n\rightarrow n\,,\; m\rightarrow m+bn\,,\;
		    \bar{m}\rightarrow \bar{m}+b^*n\,,\; \\
		    & \;l\rightarrow l+b^*m+b\bar{m}+bb^*n\,.
	    \end{split} \\
	    \begin{split} \label{eq:rotate3}
		    \text{\rom{3}}: 
		    & \;l\rightarrow A^{-1}l\,,\; n\rightarrow An\,,\;
		    m\rightarrow e^{i\theta}m\,,\; \\
	    	& \bar{m}\rightarrow e^{-i\theta}\bar{m}\,.
	    \end{split}
    \end{align}		
\end{subequations}
Here, $a$, $b$ are complex functions, and $A$, $\theta$ are real functions. Under these rotations, the Weyl scalars transform in the following way,
\begin{subequations} \label{eq:rotate_Weyl_scalars}
\allowdisplaybreaks
    \begin{align}
	    \begin{split} \label{eq:change1}
		    \text{\rom{1}}: 
		    & \begin{array}{l}
			    \Psi_{0}\rightarrow\Psi_{0}\,,\;
			    \Psi_{1}\rightarrow\Psi_{1}+a^{*}\Psi_{0}\,, \\ \Psi_{2}\rightarrow\Psi_{2}+2a^{*}\Psi_{1}
			    +\left(a^{*}\right)^{2}\Psi_{0}\,, \\
			    \Psi_{3}\rightarrow\Psi_{3}+3a^{*}\Psi_{2}+3\left(a^{*}\right)^{2} \Psi_{1}+\left(a^{*}\right)^{3}\Psi_{0}\,, \\
			    \Psi_{4}\rightarrow\Psi_{4}
			    +4 a^{*}\Psi_{3}+6\left(a^{*}\right)^{2}\Psi_{2}
			    +4\left(a^{*}\right)^{3}\Psi_{1} \\
			    +\left(a^{*}\right)^{4}\Psi_{4}\,.
		    \end{array}
	    \end{split} \\
	    \begin{split} \label{eq:change2}
		    \text{\rom{2}}: 
		    & \begin{array}{l}
			    \Psi_{0}\rightarrow\Psi_{0}+4b\Psi_{1}+6b^{2}\Psi_{2}
			    +4b^{3}\Psi_{3}+b^{4}\Psi_{4}\,, \\
			    \Psi_{1}\rightarrow\Psi_{1}+3b\Psi_{2}+3b^{2}\Psi_{3}
			    +b^{3}\Psi_{4}\,, \\
			    \Psi_{2}\rightarrow\Psi_{2}+2b\Psi_{3}+b^{2}\Psi_{4}\,,\; 
			    \Psi_{3}\rightarrow\Psi_{3}+b\Psi_{4}\,, \\ 
			    \Psi_{4}\rightarrow\Psi_{4}\,.
		    \end{array}
	    \end{split} \\
	    \begin{split} \label{eq:change3}
	    	\text{\rom{3}}:
		    & \begin{array}{l}
		    	\Psi_{0}\rightarrow A^{-2}e^{2i\theta}\Psi_{0}\,,\; 
		    	\Psi_{1}\rightarrow A^{-1}e^{i\theta}\Psi_{1}\,, \\
		    	\Psi_{2}\rightarrow \Psi_{2}\,,\;
		    	\Psi_{3}\rightarrow Ae^{-i\theta}\Psi_{3}\,, \\
		    	\Psi_{4}\rightarrow A^{2}e^{-2i\theta}\Psi_{4}\,.
	    	\end{array} 
	    \end{split}
    \end{align}
\end{subequations}
For the transformations of the spin-coefficients under the tetrad rotations, since we haven't used them explicitly in our calculations, we refer the readers to \cite{Chandrasekhar_1983} for all the details.
\vspace{4cm}

\onecolumngrid
\section{Modified Teukolsky Equation in One Place} \label{appendix:eqns_in_one_place}
For convenience of the reader, we organize the modified Teukolsky equation in one place. For $\Psi_0$, we have
\begin{equation}
    H_0^{(0,0)}\Psi_0^{(1,1)} +
    H_0^{(1,0)}\Psi_0^{(0,1)}+
    H_0^{(0,1)}\Psi_0^{(1,0)}-
    H_1^{(0,1)}\Psi_1^{(1,0)}
    =\mathcal{E}_2^{ (0,0)} S_2^{(1,1)}+
    \mathcal{E}_2^{ (0,1)} S_2^{(1,0)}
    -\mathcal{E}_1^{ (0,0)} S_1^{(1,1)}
    -\mathcal{E}_1^{ (0,1)} S_1^{(1,0)}\,.
\end{equation}
Here we have
\begin{equation}
    H_0 = \mathcal{E}_2F_2-\mathcal{E}_1 F_1-3\Psi_2\,,\quad
    H_1 = \mathcal{E}_2J_2-\mathcal{E}_1 J_1\,, 
\end{equation}
and
\begin{align}
    & \mathcal{E}_1=\delta-\tau+\pi^{*}-\alpha^{*}
    -3\beta-\Psi_2^{-1}\delta\Psi_2\,,\quad
    F_1\equiv\delta^{*}-4\alpha+\pi\,,\quad
    J_1\equiv D-2\varepsilon-4\rho\,, \nonumber\\
    & \mathcal{E}_2=D-\rho-\rho^{*}-3\varepsilon
    +\varepsilon^{*}-\Psi_2^{-1}D\Psi_2\,,\quad
    F_2\equiv\Delta-4\gamma+\mu\,,\quad
    J_2\equiv\delta-4\tau-2\beta\,,
\end{align}
with 
\begin{align}
    & S_1=\left(\delta+\pi^{*}-2\alpha^{*}-2\beta\right)\Phi_{00}
	-\left(D-2\varepsilon-2\rho^{*}\right)\Phi_{01} +2\sigma\Phi_{10}-2\kappa\Phi_{11}-\kappa^{*}\Phi_{02}\,, \nonumber\\
	& S_2=\left(\delta+2\pi^{*}-2\beta\right)\Phi_{01}
	-\left(D-2\varepsilon+2\varepsilon^{*}-\rho^{*}\right)\Phi_{02} -\lambda^{*}\Phi_{00}+2\sigma\Phi_{11}-2\kappa\Phi_{12}\,. 
\end{align}
For $\Psi_4$, we have
\begin{equation}
    H_4^{(0,0)}\Psi_4^{(1,1)}+
    H_4^{(1,0)}\Psi_4^{(0,1)}+
    H_4^{(0,1)}\Psi_4^{(1,0)}-
    H_3^{(0,1)}\Psi_3^{(1,0)}
    =\mathcal{E}_4^{(0,0)}S_4^{(1,1)}+
    \mathcal{E}_4^{(0,1)}S_4^{(1,0)}
    -\mathcal{E}_3^{(0,0)}S_3^{(1,1)}
    -\mathcal{E}_3^{(0,1)}S_3^{(1,0)}\,.
\end{equation}
Here we have
\begin{equation}
    H_4=\mathcal{E}_4F_4-\mathcal{E}_3F_3-3\Psi_2\,,\quad
    H_3=\mathcal{E}_4J_4-\mathcal{E}_3J_3\,, 
\end{equation}
and
\begin{align}
    & \mathcal{E}_3=\delta^{*}+3\alpha+\beta^{*} +\pi-\tau^{*}-\Psi_2^{-1}\delta^*\Psi_2\,,\quad
    F_3\equiv \delta+4\beta-\tau\,,\quad
	J_3\equiv \Delta+2\gamma+4\mu\,, \nonumber\\
    & \mathcal{E}_4=\Delta+\mu+\mu^{*}+3\gamma-\gamma^*
    -\Psi_2^{-1}\Delta\Psi_2\,,\quad
    F_4\equiv D+4\varepsilon-\rho\,,\quad
    J_4\equiv \delta^{*}+4\pi+2\alpha \,,
\end{align}
with 
\begin{align}
	& S_3=-\left(\Delta+2\mu^{*}+2\gamma\right)\Phi_{21}
	+\left(\delta^{*}-\tau^{*}+2\alpha+2\beta^{*}\right)\Phi_{22}
    +2\nu\Phi_{11}+\nu^{*}\Phi_{20}-2\lambda\Phi_{12}\,, \nonumber\\
	& S_4=-\left(\Delta+\mu^{*}+2\gamma-2\gamma^{*}\right)\Phi_{20}
	+\left(\delta^{*}+2\alpha-2\tau^{*}\right)\Phi_{21}
    +2\nu\Phi_{10}-2\lambda\Phi_{11}+\sigma^{*}\Phi_{22}\,.
\end{align}

\twocolumngrid

\section{Consistency check with previous higher-order Teukolsky formalism} 
\label{appendix:consistency_check}
In this appendix, we show that the GHP transformation of Eqs.~\eqref{eq:master_eqn_non_typeD_Psi0_nonlinear}-\eqref{eq:sources_non_typeD_Psi0_nonlinear} when $\zeta=0$ are consistent with Eqs.~(7)-(10) of \cite{Campanelli:1998jv} when we are in the same gauge as in Eq.~\eqref{eq:gauge_Psi1_nonlinear}.

First, let us write down the GHP transformation of Eqs.~\eqref{eq:master_eqn_non_typeD_Psi0_nonlinear}-\eqref{eq:sources_non_typeD_Psi0_nonlinear} when $\zeta=0$, 
\begin{align} \label{eq:Teuk_nonlinear_Psi4}
    & H_{4}^{\GR}\Psi_4^{(N)}
    =\mathcal{T}_{\geo}^{(N)}\,, \nonumber \\
    & \mathcal{T}_{\geo}^{(N)}=\sum_{n=1}^{N-1}-H_4^{(N-n)}\Psi_4^{(n)}\,,
\end{align}
where we have used the single superscript notation since there is only one expansion parameter, $\epsilon$. In comparison, Ref.~\cite{Campanelli:1998jv} found
\begin{align} \label{eq:Teuk_nonlinear_Psi4_Campanelli}
    & H_{4}^{\GR}\Psi_4^{(N)}
    =\mathcal{T'}_{\geo}^{(N)}\,, \nonumber\\
    & \begin{aligned}
        \mathcal{T'}_{\geo}^{(N)}
        =& \;\sum_{n=1}^{N-1}\left[\left(\mathcal{E}_3^{(0)}F_3^{(N-n)}
        -\mathcal{E}_4^{(0)}F_4^{(N-n)}\right)\Psi_4^{(n)}\right. \\
        & \;+3\mathcal{E}_3^{(0)}\left(\Psi_2^{(n)}\nu^{(N-n)}\right)
        -3\mathcal{E}_4^{(0)}\left(\Psi_2^{(n)}\lambda^{(N-n)}\right) \\
        & \;\left.-3\Psi_2^{(0)}
        \left(E_3^{(N-n)}\nu^{(n)}-E_4^{(N-n)}\lambda^{(n)}\right)\right]\,,
    \end{aligned}
\end{align}
where we have set all the terms containing $\Psi_3^{(0,n)}$ for $n>0$ to zero and replaced the operators $\bar{d}_{3,4}$ in \cite{Campanelli:1998jv} with the operators $E_{3,4}$ by observing that
\begin{align}
    & \bar{d}_3=E_3+3\pi\,,\quad\bar{d}_4=E_4+3\mu\,, \\
    & \bar{d}_3^{(0)}=E_3^{\GR}=\mathcal{E}_3^{(0)}\,,\quad
    \bar{d}_4^{(0)}=E_4^{\GR}=\mathcal{E}_4^{(0)}\,.
\end{align}

As discussed in Sec.~\ref{sec:second_order_Teuk_GR}, to show that Eq.~\eqref{eq:Teuk_nonlinear_Psi4_Campanelli} is the same as Eq.~\eqref{eq:Teuk_nonlinear_Psi4}, one needs to use Bianchi identities to express $\lambda$ and $\nu$ in terms of $\Psi_4$ or vice versa. Since $\Psi_3^{(0)}=0$ for Petrov type D spacetimes, and we have chosen a gauge in which $\Psi_3^{(n)}=0$ for all $n\geq1$, we can set $\Psi_3=0$ in Eq.~\eqref{eq:Bianchi_simplified_psi4}, such that
\begin{equation} \label{eq:nu_lambda}
    F_3\Psi_4=-3\Psi_2\nu\,,\quad
    F_4\Psi_4=-3\Psi_2\lambda\,,
\end{equation}
where we have also set $S_{3}=S_{4}=0$ since we focus on vacuum spacetimes. Notice that Eq.~\eqref{eq:nu_lambda} is true at all orders in $\epsilon$.

Expressing $\Psi_4$ in terms of $\lambda$ and $\nu$ is easier when comparing Eq.~\eqref{eq:Teuk_nonlinear_Psi4_Campanelli} with Eq.~\eqref{eq:Teuk_nonlinear_Psi4}. Let us first perform this transformation on Eq.~\eqref{eq:Teuk_nonlinear_Psi4}. From the definition in Eqs.~\eqref{eq:simplify_operators_4} and \eqref{eq:tile_E_operators}, we know that
\begin{equation} \label{eq:E_tildeE}
    \mathcal{E}_{3}=E_3-\Psi_2^{-1}\delta^{*}\Psi_2\,,\quad
    \mathcal{E}_{4}=E_4-\Psi_2^{-1}\Delta\Psi_2\,.
\end{equation}
Inserting Eqs.~\eqref{eq:nu_lambda}-\eqref{eq:E_tildeE} into Eq.~\eqref{eq:Teuk_nonlinear_Psi4}, we find
\begin{equation}
    \begin{aligned}
        H_4\Psi_4
        =& \;(\mathcal{E}_4F_4-\mathcal{E}_3F_3-3\Psi_2)\Psi_4 \\
        =& \;-3\left[E_4(\Psi_2\lambda)-\Delta\Psi_2
        -E_3(\Psi_2\nu)+\delta^{*}\Psi_2+\Psi_2\Psi_4\right] \\
        =& \;-3\Psi_2\left(E_4\lambda-E_3\nu+\Psi_4\right)\,,
    \end{aligned}
\end{equation}
which is simply $-3\Psi_2$ times the Ricci identity in Eq.~\eqref{eq:RicciId_Psi4_simplify}. Since Eq.~\eqref{eq:Teuk_nonlinear_Psi4} is essentially the $N$-th order expansion of $H_4\Psi_4$, we find
\begin{equation} \label{eq:nonlinear_transformed}
    \left[-3\Psi_2\left(E_4\lambda-E_3\nu+\Psi_4\right)\right]^{(N)}=0\,.
\end{equation}
Equation~\eqref{eq:nonlinear_transformed} is consistent with our procedures to derive the master equation in Secs.~\ref{sec:The_modified_Teukolsky_equation} and \ref{sec:teuknonD_nonlinear}. The equation we used is indeed $3\Psi_2$ multiplying the Ricci identity Eq.~\eqref{eq:RicciId_Psi4_simplify} with $\lambda$ and $\nu$ replaced by the Bianchi identities Eqs.~\eqref{eq:BianchiId_Psi4_1_simplify}-\eqref{eq:BianchiId_Psi4_2_simplify}. Since the Teukolsky equations have to be consistent with all the Bianchi identities and Ricci identities, one also expects that starting from a Teukolsky equation and simplifying it using Bianchi identities, one will get back the original Ricci identity.

Now, let us transform Eq.~\eqref{eq:Teuk_nonlinear_Psi4_Campanelli}. We first move the first line of $\mathcal{T'}_{\geo}^{(N)}$ in Eq.~\eqref{eq:Teuk_nonlinear_Psi4_Campanelli} to the left-hand side of the equation, so it becomes
\begin{align} \label{eq:psi_4_nonlinear_Campanelli_first}
    & \sum_{n=1}^{N}\left(\mathcal{E}_4^{(0)}F_4^{(N-n)}
    -\mathcal{E}_3^{(0)}F_3^{(N-n)}\right)\Psi_4^{(n)}
    -3\Psi_2^{(0)}\Psi_4^{(N)}\nonumber\\
    & =\mathcal{E}_4^{(0)}(F_4\Psi_4)^{(N)}
    -\mathcal{E}_3^{(0)}(F_3\Psi_4)^{(N)}
    -3\Psi_2^{(0)}\Psi_4^{(N)} \nonumber \\
    & =-3\left[\mathcal{E}_4^{(0)}(\Psi_2\lambda)^{(N)}
    -\mathcal{E}_3^{(0)}(\Psi_2\nu)^{(N)}\right]
    -3\Psi_2^{(0)}\Psi_4^{(N)}\,.
\end{align}
Next, subtracting off the second line of $\mathcal{T'}_{\geo}^{(N)}$ in Eq.~\eqref{eq:Teuk_nonlinear_Psi4_Campanelli} from Eq.~\eqref{eq:psi_4_nonlinear_Campanelli_first}, we find
\begin{align}
    & -3\left[\mathcal{E}_4^{(0)}\left(\Psi_2^{(0)}\lambda^{(N)}\right)
    -\mathcal{E}_3^{(0)}\left(\Psi_2^{(0)}\nu^{(N)}\right)\right]
    -3\Psi_2^{(0)}\Psi_4^{(N)} \nonumber \\
    & =-3\Psi_2^{(0)}\left(E_4^{(0)}\lambda^{(N)}
    -E_3^{(0)}\nu^{(N)}+\Psi_4^{(N)}\right)\,,
\end{align}
which, with the last line of $\mathcal{T'}_{\geo}^{(N)}$ in Eq.~\eqref{eq:Teuk_nonlinear_Psi4_Campanelli}, gives us
\begin{equation} \label{eq:Campanelli_transformed}
    -3\Psi_2^{(0)}\left[\left(E_4\lambda-E_3\nu+\Psi_4\right)\right]^{(N)}
    =0\,.
\end{equation}
As discussed above, Eq.~\eqref{eq:Campanelli_transformed} is expected since the Teukolsky equations are consistent with the Ricci identities.

Comparing Eq.~\eqref{eq:Campanelli_transformed} to Eq.~\eqref{eq:nonlinear_transformed}, one can notice that the only difference is the overall normalization factor. In Eq.~\eqref{eq:Campanelli_transformed}, this normalization factor is $-3\Psi_2^{(0)}$, while in Eq.~\eqref{eq:nonlinear_transformed}, a normalization factor of $-3\Psi_2$ appears before the expansion. Then, when expanding Eq.~\eqref{eq:nonlinear_transformed}, we also mix lower-order Ricci identities in the equation. For example, we can get the term $-3\Psi_2^{(1)}(E_4\lambda-E_3\nu+\Psi_4)^{(N-1)}$. Nonetheless, after inserting in all the lower-order NP quantities into the equation, these lower-order Ricci identities vanish, since they are automatically satisfied by the lower-order Teukolsky solutions in the previous steps. On the other hand, before inserting lower-order Teukolsky solutions, Eq.~\eqref{eq:nonlinear_transformed} might be more complicated than Eq.~\eqref{eq:Campanelli_transformed} due to these lower-order equations.

One can easily remove this difference by replacing the normalization factor $3\Psi_2$ in Eq.~\eqref{eq:3_to_1} with $3\Psi_2^{(0,0)}$. The reason  we inserted $3\Psi_2$ in Eq.~\eqref{eq:3_to_1} is that the $\mathcal{O}(\zeta^0,\epsilon^1)$ expansion of the equation reproduces the original Teukolsky equation in GR \cite{Teukolsky:1973ha}, which is also true if we instead insert $3\Psi_2^{(0,0)}$. Moreover, we can absorb the factors of $\Psi_2$ and $\Psi_2^{-1}$ in Eq.~\eqref{eq:3_to_1} nicely into the operators $\mathcal{E}_i$. If we instead use $3\Psi_2^{(0,0)}$, we can alternatively define the operators $\mathcal{E}_{i}$ as
\begin{equation}
    \mathcal{E}_i=\Psi_2^{(0,0)}E_i\Psi_2^{-1}
\end{equation}
in comparison to the original definition in Eqs.~\eqref{eq:def_tile_E_operators} and \eqref{eq:tile_E_operators}. For the goals of this paper, finding the $\mathcal{O}(\zeta^1,\epsilon^1)$ corrections to the Teukolsky equation, both ways of normalizing the equation are fine and make little difference.

\bibliography{refs}
\end{document}